\documentclass[
 reprint,
 amsmath,amssymb,
 aps,
]{revtex4-2}

\usepackage{placeins}
\usepackage{lipsum}
\usepackage{graphicx}
\usepackage{epstopdf}

\usepackage{dcolumn}
\usepackage{bm}
\usepackage[colorlinks=true, citecolor={blue!80!black}, urlcolor={blue!50!black}, linkcolor = {blue!80!black}]{hyperref}
\usepackage[colorinlistoftodos]{todonotes}
\usepackage[capitalize]{cleveref}
\usepackage[english]{babel}
\usepackage{braket}

\usepackage{siunitx}
\DeclareSIUnit{\belmilliwatt}{Bm}
\DeclareSIUnit{\dBm}{\deci\belmilliwatt}

\newcommand{\Sa}{$S_{a}$}
\newcommand{\Satilde}{$\tilde{S}_{a}$}
\newcommand{\Sb}{$S_{b}$}
\newcommand{\Iab}{$I_{ab}$}
\newcommand{\Da}{$D_{a}$}
\newcommand{\Db}{$D_{b}$}
\newcommand{\Tzero}{$T_{0}$}
\newcommand{\Tzerotilde}{$\tilde{T}_{0}$}
\newcommand{\Tplus}{$T_{+}$}
\newcommand{\Tmin}{$T_{-}$}
\newcommand{\Sab}{$S_{ab}$}

\newcommand{\ketSa}{$\ket{S_{a}}$}
\newcommand{\ketSatilde}{$\ket{\tilde{S}_{a}}$}
\newcommand{\ketSb}{$\ket{S_{b}}$}
\newcommand{\ketTzero}{$\ket{T_{0}}$}
\newcommand{\ketTzerotilde}{$\ket{\tilde{T}_{0}}$}
\newcommand{\ketTplus}{$\ket{T_{+}}$}
\newcommand{\ketTmin}{$\ket{T_{-}}$}
\newcommand{\ketSab}{$\ket{S_{ab}}$}

\newcommand{\By}{$B_{y}$}
\newcommand{\Bx}{$B_{x}$}
\newcommand{\Bz}{$B_{z}$}

\newcommand{\fr}{$f_\mathrm{r}$}
\newcommand{\fnull}{$f_0$}
\newcommand{\fd}{$f_\mathrm{d}$}
\newcommand{\LJ}{$L_\mathrm{J}$}
\newcommand{\Vg}{$V_\mathrm{g}$}
\newcommand{\Lr}{$L_\mathrm{r}$}
\newcommand{\Ls}{$L_\mathrm{s}$}
\newcommand{\Cc}{$C_\mathrm{c}$}
\newcommand{\Cr}{$C_\mathrm{r}$}

\begin{document}

\preprint{APS/123-QED}

\title{Microwave spectroscopy of interacting Andreev spins}

\author{J.\,J. Wesdorp\textsuperscript{1}}\email{j.j.wesdorp@tudelft.nl}
\author{A. Vaartjes\textsuperscript{1}}
\author{L. Gr{\"u}nhaupt\textsuperscript{1}}
\author{T. Laeven\textsuperscript{2}}
\author{S. Roelofs\textsuperscript{1}}
\author{L.\,J. Splitthoff\textsuperscript{1}}
\author{M. Pita-Vidal\textsuperscript{1}}
\author{A. Bargerbos\textsuperscript{1}}
\author{D. J. van Woerkom\textsuperscript{2}}
\author{L. P. Kouwenhoven\textsuperscript{1}}
\author{C. K. Andersen\textsuperscript{1}}
\author{B. van Heck\textsuperscript{2}}
\author{G. de Lange\textsuperscript{2}}\email{gijs.delange@microsoft.com}
\affiliation{\textsuperscript{1}QuTech and Kavli Institute of Nanoscience, Delft University of Technology, 2628 CJ, Delft, The Netherlands \\
\textsuperscript{2}Microsoft Quantum Lab Delft, 2628 CJ, Delft, The Netherlands}
\author{F. J. Matute-Cañadas\textsuperscript{4}}
\author{A. Levy Yeyati\textsuperscript{4}}
\affiliation{\textsuperscript{4} Departamento de Física Teórica de la Materia Condensada, Condensed Matter Physics Center (IFIMAC) and Instituto Nicolás Cabrera, Universidad Autónoma de Madrid, 28049 Madrid, Spain}
\author{P. Krogstrup\textsuperscript{3}}
\affiliation{\textsuperscript{3} Center for Quantum Devices, Niels Bohr Institute, University of Copenhagen
and Microsoft Quantum Materials Lab Copenhagen, Denmark}
\date{\today}

\begin{abstract}
Andreev bound states are fermionic states localized in weak links between superconductors which can be occupied with spinful quasiparticles. 
Microwave experiments using superconducting circuits with InAs/Al nanowire Josephson junctions have recently enabled probing and coherent manipulation of Andreev states but have remained limited to zero or small fields.
Here we use a flux-tunable superconducting circuit in external magnetic fields up to \SI{1}{\tesla} to perform spectroscopy of spin-polarized Andreev states up to $\sim$ \SI{250}{\milli\tesla}, beyond which the spectrum becomes gapless.
We identify singlet and triplet states of two quasiparticles occupying different Andreev states through their dispersion in magnetic field.
These states are split by exchange interaction and couple via spin-orbit coupling, analogously to two-electron states in quantum dots.
We also show that the magnetic field allows to drive a direct spin-flip transition of a single quasiparticle trapped in the junction.
Finally, we measure a gate- and field-dependent anomalous phase shift of the Andreev spectrum, of magnitude up to  approximately $0.7\pi$.
Our observations demonstrate new ways to manipulate Andreev states in a magnetic field and reveal  spin-polarized triplet states that carry supercurrent.
\end{abstract}

\maketitle

\section{\label{sec:introduction}Introduction}

Experimental results in recent years have advanced our understanding of the Josephson effect in terms of Andreev bound states (ABS)~\cite{kulik_macroscopic_1969, beenakker_universal_1991, klapwijk_proximity_2004}.
When two superconductors (S) are separated by a normal (N) material, the transport of Cooper pairs between them is mediated by Andreev reflections at the N-S interfaces.
The consequent formation of current-carrying, discrete Andreev states in SNS junctions can be observed with microwave spectroscopy~\cite{bretheau_exciting_2013, janvier_coherent_2015, bretheau_supercurrent_2013-1, van_woerkom_microwave_2017-1, hays_direct_2018, tosi_spin-orbit_2019, hays_continuous_2020, hays_coherent_2021,metzger_circuit-qed_2021-1,wesdorp_dynamical_2021, fatemi_microwave_2021-3, matute-canadas_signatures_2022}. 

In $s$-wave superconductors, which preserve time-reversal symmetry, Cooper pairs are formed with opposite spins in singlet states with zero total spin.
On the other hand, in semiconductors with strong spin-orbit coupling that are proximitized by an $s$-wave superconductor~\cite{gorkov_superconducting_2001, reeg_proximity-induced_2015}, a parallel magnetic field can induce a triplet $p$-wave component in the superconducting pairing due  to the competition of the spin-orbit interaction and the Zeeman effect~\cite{lutchyn_majorana_2010,oreg_helical_2010, alicea_majorana_2010,potter_majorana_2011}.
Such triplet pairing is of fundamental interest, in part because it is a key ingredient to create topological superconducting phases with Majorana zero modes~\cite{read_paired_2000-1,ivanov_non-abelian_2001, kitaev_unpaired_2001}.

The consequences of triplet pairing on the Josephson effect have been widely investigated theoretically and include the occurrence of the anomalous Josephson effect and of spin-polarized supercurrents~\cite{krive_chiral_2004,buzdin_proximity_2005,reynoso_anomalous_2008,yokoyama_anomalous_2014,konschelle_theory_2015}.
The experimental detection has, however, proven more challenging.
Early signatures of triplet supercurrent have been reported in Josephson junctions with magnetic materials~\cite{khaire_observation_2010, robinson_controlled_2010, sprungmann_evidence_2010, linder_superconducting_2015-1} and more recently in experiments making use of materials with spin-orbit coupling to induce spin-mixing~\cite{jeon_tunable_2020, cai_evidence_2021,yang_boosting_2021, ahmad_coexistence_2022}.
In hybrid semiconductor-superconductor systems, evidence of triplet pairing stems from the observation of the anomalous Josephson effect in InAs/Al nanowires~\cite{szombati_josephson_2016,strambini_josephson_2020} and in 2-dimensional electron gases (2DEGs)~\cite{mayer_gate_2020}. Additionally, there are indications of triplet pairing from microwave susceptibility measurements of resonators made out of InAs/Al 2DEGs~\cite{phan_detecting_2022} and from spin-polarized crossed Andreev reflection in InSb/Al nanowires~\cite{wang_singlet_2022}.
Evidence of spin-polarized triplet pairs based on microwave absorption and their associated supercurrent has, however, been elusive.

Embedding nanowire Josephson junctions in microwave superconducting circuits allows for probing of individual Andreev states with a remarkable energy resolution of $\sim\SI{100}{\mega\hertz}$ (i.e., $\sim$ \SI{0.5}{\micro\electronvolt})~\cite{hays_direct_2018,hays_continuous_2020,tosi_spin-orbit_2019,hays_coherent_2021,metzger_circuit-qed_2021-1, matute-canadas_signatures_2022,fatemi_microwave_2021-3,wesdorp_dynamical_2021} and with potential spin-sensitivity~\cite{tosi_spin-orbit_2019, hays_continuous_2020, hays_coherent_2021,metzger_circuit-qed_2021-1,wesdorp_dynamical_2021}.
Thus, such circuits provide an excellent platform to study the (spin) properties of Andreev bound states.
In fact, microwave spectroscopy has already revealed that spin-orbit coupling~\cite{tosi_spin-orbit_2019, hays_continuous_2020} and electron-electron interactions~\cite{matute-canadas_signatures_2022, fatemi_microwave_2021-3} are crucial ingredients that determine the many-body Andreev spectrum of hybrid nanowire Josephson junctions.
However, so far, such experiments using superconducting circuits have been limited to zero or small magnetic fields.

In this work, we demonstrate measurements of the Andreev spectra of an InAs/Al nanowire Josephson junction embedded in a superconducting circuit with magnetic fields up to $\sim\SI{250}{\milli\tesla}$.
The magnetic field dependence of the microwave absorption spectrum shows clear signatures of excited Andreev levels in a triplet state.
The spectrum can be well understood based on a minimal model which includes spin-orbit coupling, the Zeeman effect, and ferromagnetic exchange interaction between Andreev bound states, originating from electron-electron interactions in the junction.
A particularly interesting feature of the data is the presence of a singlet-triplet avoided crossing.
Due to quasiparticle poisoning~\cite{glazman_bogoliubov_2021}, the microwave absorption spectra also reveal transitions between odd-parity states, which were recently used to realize Andreev spin qubits~\cite{hays_coherent_2021}.
Here, we detect the direct driving of the spin-flip transition of an Andreev bound state, activated by the magnetic field.
Finally, at high fields we observe a gate-tunable anomalous Josephson effect and resolve the individual contributions of Andreev bound states to the anomalous phase shift.
In the next Section, we kick-off the presentation of our results by discussing the experimental setup and the ingredients that made these measurements possible.

\begin{figure}
\includegraphics{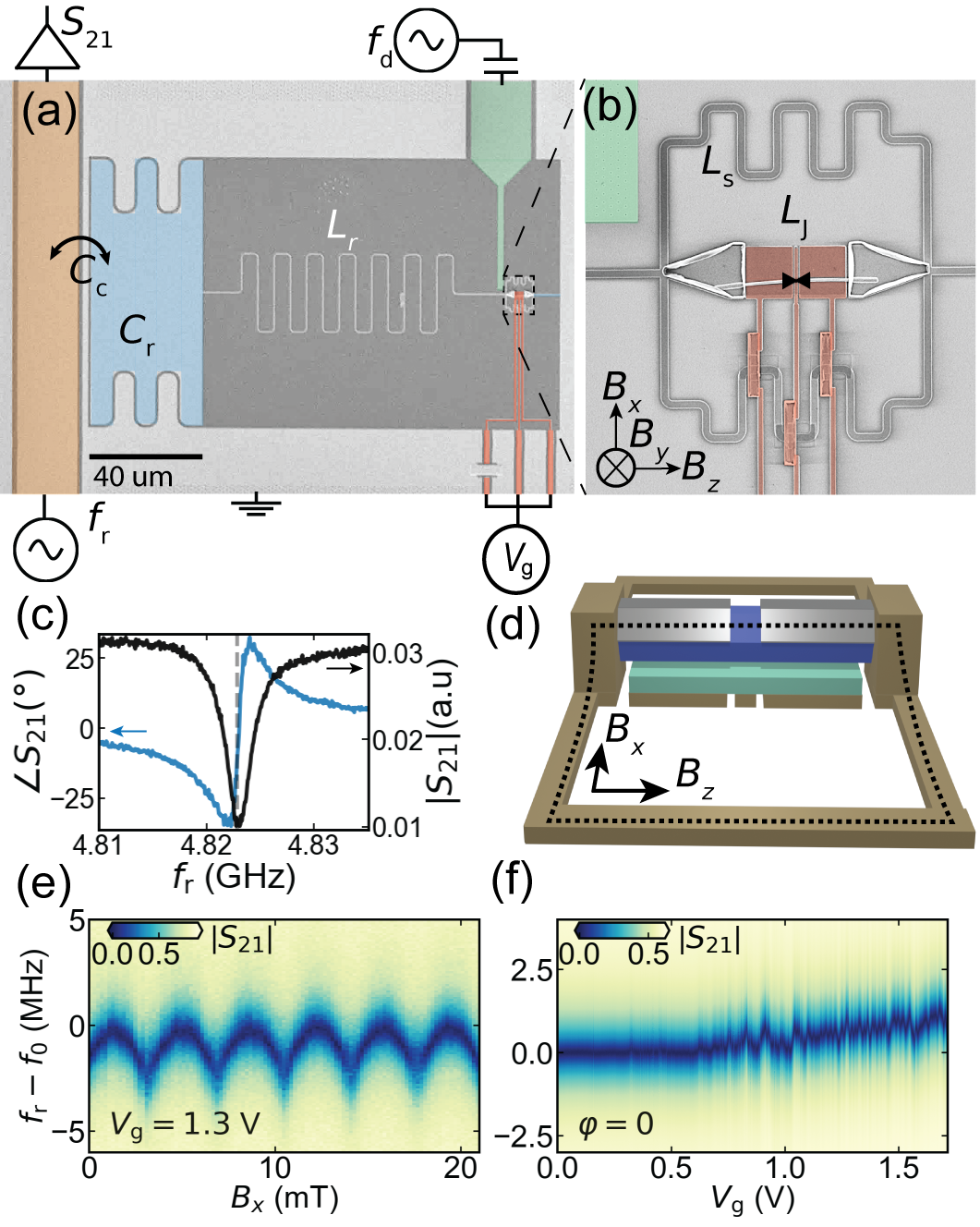}
\caption{\label{fig:f1} Field compatible circuit design and operation principle. \textbf{(a)} Device image and circuit schematic. A lumped element resonator is capacitively coupled (\Cc) to a transmission line (orange). The resonator consists of a capacitor (\Cr, blue) and inductor (\Lr, white) connected to ground via a gradiometric RF-SQUID that modulates the total inductance. \textbf{(b)}  The SQUID consists of two loops of inductance $L_\mathrm{s}$ that shunt a nanowire Josephson junction with gate-tunable Josephson inductance \LJ.
The gradiometric design reduces sensitivity to perpendicular field \By\ and the shunt-inductance determine the coupling strength to Andreev bound states in the junction.
The magnetic field coordinate system aligned to the nanowire and used throughout the text is indicated~\cite{wesdorp_supplementary_2022}.
\textbf{(c)} Amplitude and phase response of the resonator when the junction is pinched-off.
\textbf{(d)} 3D sketch of the SQUID loop.
An InAs nanowire (blue) with Al shell (silver) and a \SI{144}{\nano\meter} junction~\cite{wesdorp_dynamical_2021} is suspended on gate dielectric (teal) above bottom gates (gold).
By applying an in-plane field \Bx, we can thread a flux through a vertically defined loop (dashed dots).
\textbf{(e)} SQUID oscillations when applying \Bx. \textbf{(f)} Gate dependence of the junction without applying flux. \fnull\ increases as the critical current (inductance) of the junction increases (decreases).}
\end{figure}

\section{\label{sec:field_compatible_design} Field compatible design and operation}

Previous microwave experiments probing Andreev states with superconducting circuits have traditionally used thick (\SI{150}{\nano\meter}) coplanar-waveguides~\cite{tosi_spin-orbit_2019, janvier_coherent_2015, metzger_circuit-qed_2021-1, matute-canadas_signatures_2022} or coplanar stripline resonators~\cite{hays_direct_2018, hays_continuous_2020, hays_coherent_2021, fatemi_microwave_2021-3}.
Here we use thin-film (\SI{20}{\nano\meter}) lumped-element resonators due to their proven resilience to parallel fields shown earlier in fluxonium devices~\cite{pita-vidal_gate-tunable_2020-1}.
Additionally, the second harmonic of the resonator is expected to be at higher frequencies (\SI{28.5}{\giga\hertz}, see Supplement~\cite{wesdorp_supplementary_2022}) relative to the lowest mode compared to a coplanar geometry of equal fundamental frequency. 
This helps with spectroscopic measurements at frequencies up to the superconducting gap $\Delta\approx\SI{44}{\giga\hertz}$. 

We fabricate multiple resonators on a chip, one of which is shown in~\cref{fig:f1}(a). The resonator is coupled to a common feedline that is used for microwave readout.
The lumped-element resonator, with resonance frequency \fnull=\SI{4.823}{\giga\hertz}, consists of a capacitor ($C_\mathrm{r}\approx \SI{47}{\femto\farad}$) that is connected to the ground plane via an inductor~($L_\mathrm{r}\approx\SI{22}{\nano\henry}$).
The inductance is dominated by the kinetic inductance of the thin-film NbTiN~\cite{annunziata_tunable_2010}.
The inductor has a width of \SI{300}{\nano\meter}, such that the required perpendicular field for vortex generation is $>\SI{20}{\milli\tesla}$ in locations where the current is strongest -- well above the tolerance for misalignment using a vector magnet.
We patterned vortex traps with a diameter of \SI{80}{\nano\meter} within a \SI{8}{\micro\meter} radius in the capacitor and surrounding ground planes with a $\SI{200}{\nano\meter}$ gap to prevent flux jumps due to moving vortices~\cite{kroll_magnetic_2019-2}
\footnote{We found that holes closer spaced to the edges reduced the flux jumps significantly, compared to a $\SI{1}{\micro\meter}$ spacing used in Ref.~\cite{kroll_magnetic_2019-2}}.
The inductor is connected to ground via a gradiometric radio-frequency superconducting quantum interference device (RF-SQUID)~[\cref{fig:f1}(b)]~\cite{pita-vidal_gate-tunable_2020-1}, which consists of a nanowire Josephson junction shunted on two sides by an inductance ($L_\mathrm{s}=\SI{0.7}{\nano\henry}$) forming two nearly equal sized loops. 
We define a Josephson junction by selectively etching a $\SI{144}{\nano\meter}$ section of a $\sim \SI{6}{\nano\meter}$ thick aluminum shell that covers two facets of a hexagonal InAs nanowire of $\sim$\SI{80}{\nano\meter} diameter~\cite{chang_hard_2015}. The nanowire is placed on bottom gates defined in the NbTiN layer, which are covered with a \SI{28}{\nano\meter} $\mathrm{Si_3N_4}$ dielectric before nanowire placement. To each resonator, we add capacitive coupling to an additional transmission line to drive transitions in the junction and perform spectroscopy.

The specific gradiometric loop design~[\cref{fig:f1}(b, d)] was optimized to allow for flux-biased measurements in high magnetic field.
In a gradiometric geometry, the two loops create opposite circulating currents through the nanowire Josephson junction under applied flux (see Supplement~\cite{wesdorp_supplementary_2022}).
The effective loop area is therefore proportional to the area difference between the loops, which here is determined by the inaccuracy of the nanowire placement with respect to center axis of the two loops ($\sim \SI{300}{\nano\meter})$. 
The resulting effective loop area ($\approx\SI{0.8}{\micro\meter\squared}$) is much smaller than the individual patterned loop areas ($\approx \SI{50}{\micro\meter\squared}$). 
A small effective loop is desired to render the SQUID insensitive to flux from out-of-plane field (\By), reducing flux noise in presence of strong external fields.
The gradiometric design also allows for picking a shunt inductance $L_\mathrm{s}$ -- which determines the coupling strength to the Josephson junction -- nearly independent of the loop size, which makes for easier design and fabrication.
Additionally, by placing the nanowire on top of the bottom gates, we lift the nanowire and thus elevate part of the loop vertically in the $z$-$y$ plane~[\cref{fig:f1}(d)]. 
The orientation of the loop thus allows flux biasing the SQUID with an in-plane field \Bx\ with respect to the superconducting circuit. In this case the induced circulating currents add, so the effective flux is proportional to twice the loop area ($A=\SI{0.28}{\micro\meter\squared}$, $\Phi_0 \sim \SI{3.65}{\milli\tesla}$).
Due to the thin-film NbTiN, the area of superconducting film that is exposed to parallel field \Bx\ is much smaller compared to the area exposed to perpendicular field \By.
This is essential for flux biasing without flux jumps because vortex nucleation and circulating currents are proportional to the total area of superconducting film exposed to magnetic field~\cite{benfenati_vortex_2020, tinkham_introduction_2015}. 
Resulting SQUID oscillations are shown in~\cref{fig:f1}(e) over a range of \SI{20}{\milli\tesla}.

We operate the devices by sending a near resonant probe tone at frequency \fr\ through the feedline and monitoring the transmitted complex scattering parameter $S_\mathrm{21}$ using a vector network analyser.
Out of the four resonators we focus on the only one in which the junction showed considerable gate response (\fnull~$=\SI{4.823}{\giga\hertz}$). 
At \fr~=~\fnull\ there is a dip in the magnitude $|S_\mathrm{21}|$ and a $\sim 60^{\circ}$ shift in the phase $\angle S_\mathrm{21}$~[\cref{fig:f1}(c)]. 
The Josephson junction then acts as a gate-tunable inductor \LJ~[\cref{fig:f1}(b)] that changes $f_0$ via 
$$
f_0 = \frac{1}{\sqrt{\left(L_\mathrm{r}+L_\mathrm{squid}\right)C}}$$
where $L_\mathrm{squid}^{-1}=L_\mathrm{J}^{-1} + 2L_\mathrm{s}^{-1}$.
Thus, by monitoring changes in $f_0$ we get access to \LJ, which is related to the Andreev bound state energies and their occupation~\cite{ zazunov_andreev_2003, bretheau_localized_2013, park_adiabatic_2020-1}. 
As we increase the gate voltage \Vg\ on the bottom gates, we observe a trend that more current carrying channels start to conduct in the junction, which decreases \LJ\ and increases $f_0$~[\cref{fig:f1}(f)].
The smaller modulations on top of the general trend can be attributed to mesoscopic fluctuations of the transparency of individual Andreev states~\cite{doh_tunable_2005, goffman_conduction_2017}. As shown later, we use this to tune the Andreev energies over a large range within small \SI{}{\milli\volt} gate ranges. 
From the change in inductance at $\varphi=0$ between the junction being in an open configuration (\Vg~=~\SI{1.68}{\volt}) and pinched-off  ($V_\mathrm{g}=\SI{0}{\volt}$, $L_\mathrm{J}=\infty$), we estimate \LJ~=~\SI{38}{\nano\henry} at \Vg~=~\SI{1.68}{\volt}, resulting in an estimate for the maximal critical current $I_\mathrm{c}\approx \varphi_0/L_\mathrm{J}=\SI{8.5}{\nano\ampere}$.
In general, the Andreev states induce a state-dependent frequency shift~\cite{metzger_circuit-qed_2021-1} which generates changes in $\angle S_{21}$ monitored at \fr. 
This allows us to perform spectroscopy by sweeping a drive tone \fd\ via the drive line, which results in changes in $\angle S_{21}$ when \fd\ is equal to an energy difference between Andreev levels of the same parity.  

\section{Andreev bound state spectrum}

\begin{figure}[t]
\includegraphics{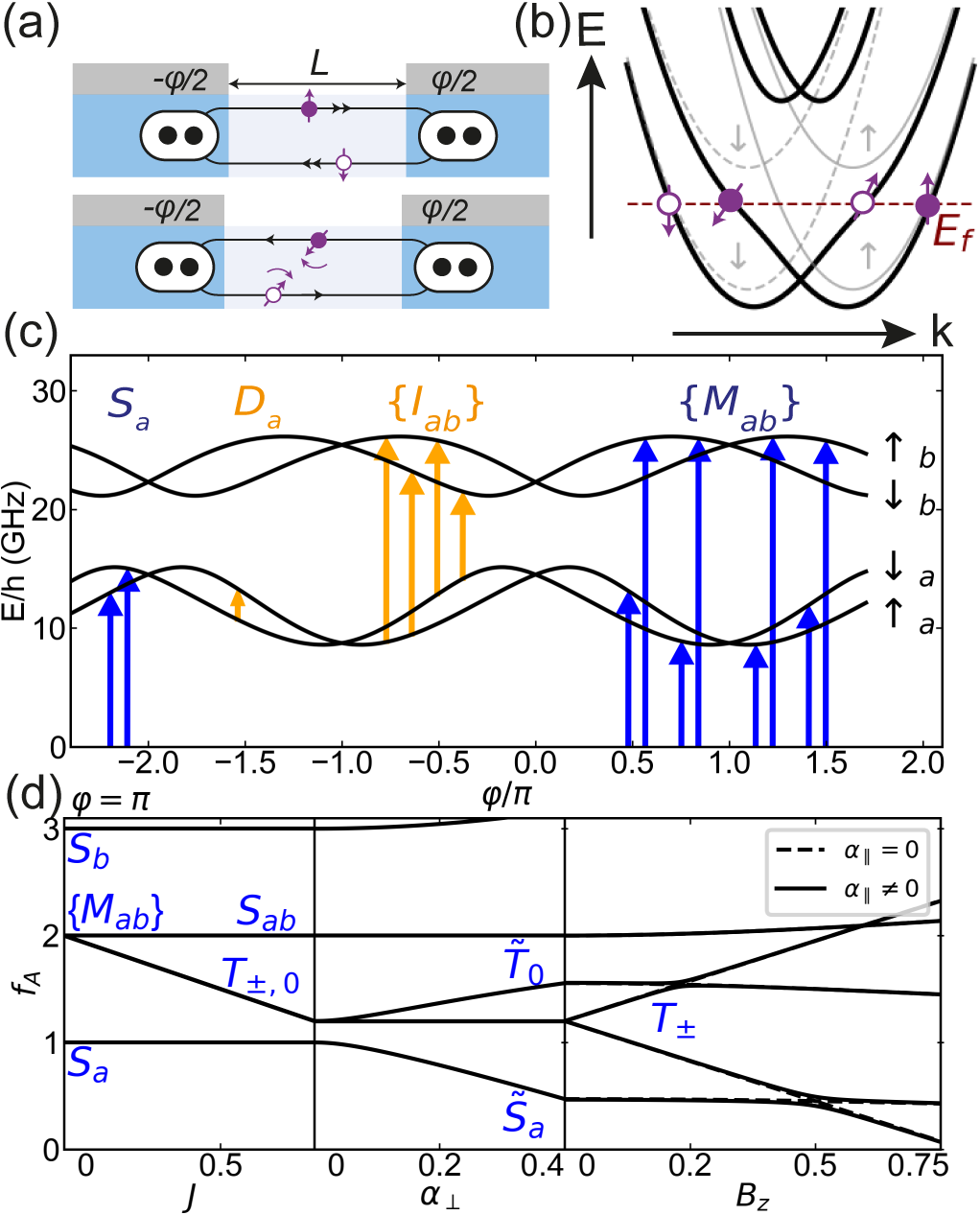}
\caption{\label{fig:f2}Hybrid nanowire Josephson junction hosting spin-split Andreev bound states at zero field. \textbf{(a)} Schematic of two Andreev reflection processes in the nanowire junction. Spin-orbit induced sub-band hybridization rotates the spins of the bottom Andreev states and lower the Fermi velocity ($v_\mathrm{F}$). Note that the time reversed processes are also possible (not shown), and that the Andreev states are generally superpositions of these four Andreev reflection processes. \textbf{(b)} Electron band structure indicating hybridized sub-bands due to spin orbit interaction. The anti-crossings lead to a rotated spin of the inner Andreev mode and a spin-dependent $v_\mathrm{F}$. \textbf{(c)} Phase dependence of two low lying spinful Andreev manifolds $(a, b)$ in the non-interacting picture~\cite{tosi_spin-orbit_2019}. Arrows denote possible parity conserving microwave transitions. Pairs of blue arrows indicate even-parity transitions starting from the ground state. Yellow arrows indicate transitions starting from one of the two lowest levels occupied with a quasiparticle. \textbf{(d)} Evolution of even parity transitions at phase difference $\varphi=\pi$ using Eq.~\eqref{eq:model}, illustrating the effect of exchange interaction $J$, spin-orbit interaction and Zeeman energy using $\alpha_{||}<J<\Delta$, which resembles the experiment. $J$ splits the four mixed states $\{M_{ab}\}$ into a singlet \Sab\ and three triplet transitions $\{T\}=\{$\Tzero,\Tplus, \Tmin\}. Spin-orbit interaction hybridizes \Tzero\ and \Sa, moving \Tzerotilde\ up and \Satilde\ down in energy. Finally, a magnetic field splits $T_\pm$.}
\end{figure}

In a nanowire Josephson junction, Andreev states arise due to constructive interference after consecutive Andreev reflections from the hybrid superconducting leads~\cite{kulik_macroscopic_1969, beenakker_universal_1991}~[\cref{fig:f2}(a)]. 
The energy of an Andreev state depends on an energy-dependent phase gained while Andreev reflection occurs, as well as, in certain conditions, a phase gained while traversing the junction.
In the presence of time-reversal symmetry, which holds at $\varphi=0$ or $\varphi=\pi$ when the magnetic field is zero, the Andreev energies are two-fold degenerate because of Kramers' theorem.
The number of Kramers doublets (manifolds) present below the gap depends on the number of the occupied sub-bands in the leads, and on the length of the junction.
In what follows, we restrict our attention to the two lowest manifolds of Andreev levels, labeled $a$ and $b$~\footnote{Note that the labels can refer either to manifolds that originate from the same transverse sub-band, due to finite-length effects, or to orbitals from different transverse sub-bands}.

Recent works have highlighted the importance of both spin-orbit interaction~\cite{park_andreev_2017, hays_continuous_2020, tosi_spin-orbit_2019,wesdorp_dynamical_2021} and electron-electron interaction~\cite{matute-canadas_signatures_2022,fatemi_microwave_2021-3} to understand the Andreev spectrum of nanowire Josephson junctions.
While Andreev bound states are spin-degenerate at all phases in the absence of spin-orbit interaction, the latter may lift the degeneracy away from $\varphi=0$ and $\varphi=\pi$.
This occurs in junctions of finite length such that a phase shift accumulated due to a spin-dependent Fermi velocity becomes relevant~\cite{governale_spin_2002,chtchelkatchev_andreev_2003,
krive_chiral_2004,beri2008,yokoyama_josephson_2013-1, yokoyama_anomalous_2014, konschelle_semiclassical_2016, park_andreev_2017}, see~\cref{fig:f2}(b).
The typical phase dispersion of the resulting spin-split manifolds is illustrated in~\cref{fig:f2}(c).
As inferred in Refs.~\cite{matute-canadas_signatures_2022, kurland_mesoscopic_2000}, electron-electron interaction manifests itself via a ferromagnetic exchange interaction $-J \vec{S}^2$ between two quasiparticles in a state of total spin $\vec{S}$, each occupying a different manifold.

We now present a minimal model that captures the combined effect of spin-orbit interaction, exchange energy and the Zeeman effect of an external magnetic field on the two manifolds, restricting our attention to the case $\phi=\pi$.
To do so it is convenient to consider the Andreev states $\{\ket{\downarrow_{a}}, \ket{\uparrow_{a}}, \ket{\downarrow_{b}}, \ket{\uparrow_{b}}\}$ belonging to the $a$ or $b$ manifold and with spin up or down with respect to the $z$-axis, running parallel to the nanowire.
Denoting with $\gamma^\dagger_{i\sigma}$ the operator which creates a quasiparticle with spin $\sigma$ in the $i=a,b$ manifold, the model Hamiltonian is:
\begin{eqnarray}\label{eq:model}
\nonumber
H = & & \sum_{i,\sigma} \left( E_i + \sigma g^*_i B_{z} \right) 
\gamma^\dagger_{i\sigma}\gamma_{i\sigma} - J/2 \vec{S}^2\\
& & + \sum_{\sigma} i \bar{\sigma} \alpha_\perp    \gamma^\dagger_{a\sigma}\gamma_{b\sigma} +
i\alpha_\parallel   \gamma^\dagger_{a\bar{\sigma}}\gamma_{b\sigma} + \textrm{h.c.}\,.
\end{eqnarray}
Here, $E_i$ is the energy of the Andreev manifold in the absence of spin-orbit interaction; \(B_{z}\) is the parallel magnetic field and \(g^*_i\) is a effective $g$-factor which can depend on the manifold; $\vec{S}=\frac{1}{2}\sum_{i,\sigma,\sigma'}\gamma^\dagger_{i\sigma}(\vec{\sigma})_{\sigma,\sigma'}\gamma_{i\sigma'}$ is the total spin, where $\vec{\sigma}$ is the vector of Pauli matrices; and finally, $i\alpha_{||}$ and $i\alpha_{\perp}$ are the matrix elements of the spin-orbit interaction described with a 2D Rashba model, respectively in the direction parallel and perpendicular to the nanowire.
More details about each term are given in Ref.~\cite{wesdorp_supplementary_2022})

Within this minimal model, it is straightforward to find the single-particle and two-particle energy levels, which determine the transitions measured in spectroscopy.
In particular, the simultaneous occupation of the junction by two quasiparticles results in six possible states.
These are two singlet same-manifold states \ketSa~$=\ket{\uparrow_{a}\downarrow_{a}}$ and \ketSb~$=\ket{\uparrow_{b}\downarrow_{b}}$ as well as four states corresponding to a mixed occupation of the two manifolds.
For the latter, it is natural to pick the basis of simultaneous eigenstates of $\vec{S}^2$ and $S_\mathrm{z}$. These are the singlet \ketSab~$=(\ket{\uparrow_{a}\downarrow_{b}} - \ket{\downarrow_{a}\uparrow_{b}})/\sqrt{2}$ and the triplet states \ketTzero~$=(\ket{\uparrow_{a}\downarrow_{b}} +  \ket{\downarrow_{a}\uparrow_{b}})/\sqrt{2}$, \ketTplus~$=\ket{\uparrow_{a}\uparrow_{b}}$, and \ketTmin~$=\ket{\downarrow_{a}\downarrow_{b}}$.
Note that without exchange interaction, a more natural basis of mixed states would be $\{\ket{T_-}, \ket{\uparrow_{a}\downarrow_{b}}, \ket{\downarrow_{a}\uparrow_{b}}, \ket{T_+}\}$.
Also, note that spin-orbit interaction breaks spin-rotation symmetry by hybridizing spin and spatial degrees of freedom.
Therefore, in its presence, spin is in general not a good quantum number, and the singlet and triplet states hybridize.
Nevertheless, for many parameter regimes the eigenstates of Eq.~\eqref{eq:model} are well approximated by the singlet or triplet states, with expectation values of the spin close to zero and one.
With this in mind, in the rest of the manuscript we will for simplicity keep referring to singlet, doublet and triplet states, except in cases where spin-orbit effects change this simple picture appreciably.

In microwave spectroscopy, we only have access to transitions between many-body states of the same fermion parity.
In~\cref{fig:f2}(c) we label the possible transitions in both even and odd parity sectors.
In the even parity sector, we only consider transition from the ground state of the junction $\ket{0}$,  with no quasiparticle excitations.
There are therefore six possible transitions (blue arrows), which we will denote by their final state.
The lowest energy transition is the singlet pair transition \Sa\ from $\ket{0}$ to $\ket{\uparrow_{a}\downarrow_{a}}$.
The four transitions that involve breaking a Cooper pair and splitting over the two different manifolds $a$ and $b$ will be globally denoted as
$\{M_{ab}\}$ [blue arrows on the right side of \cref{fig:f2}(c)].
Note that these four transitions are degenerate in the absence of spin-orbit interaction and exchange interaction.

In the odd parity sector, we denote the lowest doublet intra-manifold transitions as $D_a: \ket{\uparrow_{a}}\leftrightarrow\ket{\downarrow_{a}}$. This is a direct spin-flip of a quasiparticle occupying the lowest Andreev manifold [left yellow arrow in~\cref{fig:f2}(c)].
Furthermore, we denote the set of four inter-manifold transitions of a single quasiparticle from $\{\ket{\uparrow_{a}}, \ket{\downarrow_{a}}\}$ to $\{\ket{\uparrow_{b}}, \ket{\downarrow_{b}}\}$ as $\{I_{ab}\}$ [set of yellow arrows in~\cref{fig:f2}(c)].

In ~\cref{fig:f2}(d) we sketch the resulting modifications to the two-particle spectrum as predicted by the model of Eq.~\eqref{eq:model}.
The exchange interaction lowers the energy of the triplet states and, in doing so, partially lifts the degeneracy between the singlet transition and triplet transitions~[\cref{fig:f2}(d) - left panel].
The role of spin-orbit interaction is different: it breaks the spin-rotation symmetry and lifts the degeneracy of single-particle states away from the time-reversal invariant points $\varphi=0, \pi$.
The combination of spin-orbit interaction and exchange interaction can completely lift the degeneracy of the triplet states even at $\varphi=0, \pi$~\cite{matute-canadas_signatures_2022}. In the minimal model, this occurs partially, by hybridizing \ketTzero\ and \ketSa.
We will denote the transitions to the hybridized states \ketTzerotilde\ and \ketSatilde\ by \Tzerotilde\ and \Satilde\   respectively~[\cref{fig:f2}(d) - middle panel].  
The remaining degeneracy within the manifold of two-particle states, that of the triplet states $\ket{T_\pm}$, is lifted by the external magnetic field via the Zeeman effect~[\cref{fig:f2}(d) - right panel].  

\begin{figure*}
\includegraphics{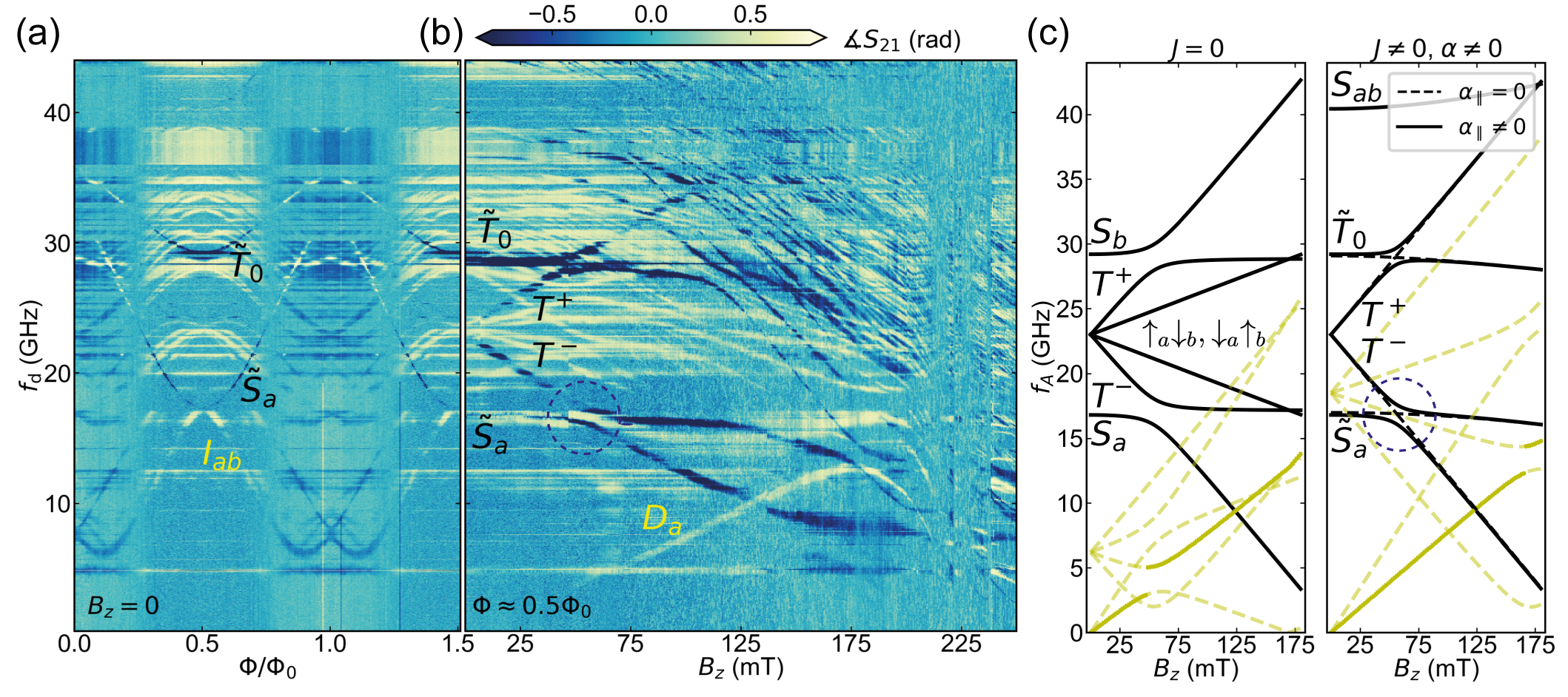}
\caption{\label{fig:f3}
Spectroscopy of singlet, doublet and triplet Andreev states versus flux and magnetic field.
\textbf{(a)} Measured spectrum at zero field versus external flux $\Phi$ at $V_\mathrm{g}=\SI{625}{\milli\volt}$. 
\textbf{(b)} Evolution of Andreev transition spectrum in an aligned parallel magnetic field starting from the right side of (a) ($\Phi\approx0.5)$. At low fields, the transitions disperse linearly with \Bz\ and we can distinguish transitions to singlet, doublet and triplet Andreev states as indicated. At higher fields, a plethora of Andreev transitions move downwards until the low-energy spectrum becomes very crowded and visibility low. Note that $\Phi\approx\Phi_0/2$ holds only for \Bz$ < \SI{160}{\milli\tesla}$ after which \Bz\ no longer is a linear function of phase.
\textbf{(c)} Fit to minimal theoretical model at $\varphi=\pi$ including $\alpha$ and exchange interaction $J$ (see Supplement~\cite{wesdorp_supplementary_2022}). Black lines indicate even-parity transitions with types indicated. Also indicated are the inter-band \Iab\ (yellow dashed lines), the lowest doublet spin-flip transition \Da\ (yellow full line) and \Db\ (yellow dashed line starting from the same point) which is not visible in the data. Left plot indicates a best fit when assuming $J=0$ and the top even transition excites a second sub-band \Sb. When $J=0$, two more transitions (\Sab,\Tzero) should appear that are not in the data and \Iab\ should appear at low frequency. Right plot is best fit with $J=\SI{17}{\giga\hertz},  \alpha_{\perp}=\SI{4.8}{\giga\hertz}$. Here the even-parity transitions and \Iab\ match the data (see Supplement for lines on top of the data~\cite{wesdorp_supplementary_2022}).
Note that $\alpha\neq0$ hybridizes the singlet \Sa\ and triplet \Tzero\ denoted as  \Satilde , \Tzerotilde\  respectively.
The avoided crossing between \Tmin\ and \Satilde\ only occurs in presence of a spin-orbit component $\alpha_{||}$ along the wire. }
\end{figure*}

\section{Andreev spectroscopy: singlet, doublet and triplet transitions}

With the theory developed, we now continue with the measurement results.
We first measure the junction spectrum at zero magnetic field versus applied flux $\Phi$~[\cref{fig:f3}(a)].
The gate is set to \Vg~$=\SI{625}{mV}$, where we have a few Andreev transitions present and the spectrum is dominated by the lowest two manifolds (see Supplement for additional gate dependence~\cite{wesdorp_supplementary_2022}).
Due to the presence of quasiparticle poisoning, the junction fluctuates between the even parity ground state $\ket{0}$ with no Andreev level occupied, and, when a quasiparticle has entered the junction due to a poisoning event, one of the odd parity doublet states $\ket{\uparrow_{a}}$, $\ket{\downarrow_{a}}$.
In a related work performed on the same junction, we measured typical poisoning times of $\approx\SI{0.5}{\milli\second}$~\cite{wesdorp_dynamical_2021}, much smaller than the integration time per point $\sim\SI{100}{\milli\second}$: thus, the measured spectra result to be an average of those resulting from initial states with and without a quasiparticle.
Odd and even parity transitions can be distinguished by their opposite sign in the dispersive frequency shift induced on the resonator~\cite{metzger_circuit-qed_2021-1}. For instance, the phase response near $\Phi=\Phi_0/2$ is negative (blue) for even parity and positive (yellow) for odd parity.

We first establish that we detect the same types of transitions as in recent experimental works~\cite{tosi_spin-orbit_2019,hays_continuous_2020, wesdorp_dynamical_2021}.
These are the even-parity transition with parabolic dispersion around \SI{16}{\giga\hertz} at $\Phi=\Phi_0/2$, and the transitions starting from the poisoned doublet state, \Iab, with the characteristic ``spider-like'' shape due to the spin-orbit splitting of the Andreev levels in manifolds $a,b$.
Note that the lowest bundle is associated with $\{I_{ab}\}$ and higher bundles likely correspond to transitions from manifold $a$ to higher manifolds $c,d,\dots$ present in the junction at higher energies.
We investigate the splitting of the $\{I_\mathrm{ac}\}$ transitions due to \Bx\ and \Bz\ in the Supplement~\cite{wesdorp_supplementary_2022}.
A symmetric splitting due to \Bz\ and asymmetric splitting due to \Bx\ was used in Ref.~\cite{tosi_spin-orbit_2019} to infer that the direction of the effective magnetic field generated by spin-orbit interaction was in-plane and perpendicular to their full-shell nanowire.
Here, we do not observe such a clear differentiation between symmetric and asymmetric splitting. This leads us to suspect that the effective spin-orbit field is not parallel to \Bx, which is consistent with recent findings indicating that, in partial-shell wires, the spin-orbit direction depends strongly on the direction of the local electric field in the wire, which in turns depends on the position and number of Al facets and gate geometry~\cite{bommer_spin-orbit_2019, de_moor_electric_2018}. 

Furthermore, we see a second even-parity transition dispersing in a similar way as the first, but at higher frequency, with a minimum around \SI{30}{\giga\hertz}.
The identification of the final states in the even transitions visible at zero field is resolved later in this Section on the basis of the magnetic field dependence.
The horizontal bands visible in~\cref{fig:f3}(a,b), mostly at higher frequencies, are attributed to resonances in the drive line and connected circuit, resulting in a frequency-dependent driving strength. 
 
Next, we measure a parallel field dependence of the Andreev spectrum, while keeping the gate fixed~[\cref{fig:f3}](b), in order to investigate the spin texture of the excited states.
By aligning the magnetic field, we keep the phase drop over the junction fixed at $\varphi\approx\pi$~(see Supplement for the alignment procedure~\cite{wesdorp_supplementary_2022}) and polarize the spins with $B_{z}$ (see ~\cref{fig:f1}(d)).
A rich spectrum emerges, with several notable features in both the even and odd transitions.

We start by describing the even-parity spectrum observed in~\cref{fig:f3}(b).
Based on the phase-response at $\Phi\approx\Phi_0/2$ in ~\cref{fig:f3}(a), we can distinguish even-parity transitions as dark-blue spectral lines.
The even transitions observed at \SI{16}{\giga\hertz} and 
\SI{31}{\giga\hertz} at \Bz~=~0 remain approximately constant at low fields, as expected from a transition to a final state with a small spin polarization, thus essentially insensitive to the Zeeman effect. We also observe two even-parity transitions that disperse linearly in field in opposite direction starting at approximately \SI{24}{\giga\hertz}.
We thus infer that the final states reached by these transitions are sensitive to the Zeeman effect and must therefore have some degree of spin polarization along the field direction.
Notably, they also display an avoided crossing with the non-dispersing transitions at \Bz~$\approx\SI{50}{\milli\tesla}$. 

In order to label these transitions correctly, we first attempt to fit the main features of the spectrum to our model of Eq.~\eqref{eq:model} without assuming electron-electron interactions.
For this, we assume the even transition at \SI{16}{\giga\hertz} in~\cref{fig:f3}(a,b) is \Sa\  while the one at \SI{31}{\giga\hertz} is due to a second Andreev manifold, i.e. the pair transition \Sb.
We then perform a best fit to the extracted transition frequencies at \Bz~=~0, while imposing a constraint that $J=0$.
While such a fit is possible, this choice of parameters also predicts the presence of two additional spectral lines corresponding to the mixed final states without exchange interaction $\sim\ket{\uparrow_{a}\downarrow_{b}}, \ket{\downarrow_{a}\uparrow_{b}}$. These states disperse with the difference of the effective $g$-factors of the two manifolds and are not observed in the field-dependent data. 
We have investigated, using a standard non-interacting tight-binding model for the nanowire Josephson junction, whether the absence of these transitions could be explained on the basis of a selection rule, i.e. vanishing matrix elements~\cite{wesdorp_supplementary_2022}.
We have indeed found cases where transitions to $\ket{\uparrow_{a}\downarrow_{b}}$ and $ \ket{\downarrow_{a}\uparrow_{b}}$ have vanishingly small matrix elements at $\varphi=\pi$.
However, even in these cases, the non-interacting model predicts them to be typically more visible than \Tplus\ and \Tmin\ at phase differences away from $\varphi=\pi$.
The latter fact can be understood on the basis that, unlike \ketTplus\ and \ketTmin, the final states $\ket{\uparrow_{a}\downarrow_{b}}$ and $\ket{\downarrow_{a}\uparrow_{b}}$ do not require a spin-flip and thus should be more easily observable at small magnetic fields.
Overall, this picture is inconsistent with additional measurements of the phase-dependence of these states at finite magnetic field (see Supplement~\cite{wesdorp_supplementary_2022}), where we did not observe the additional transitions.

Having thus disfavored a scenario based on the absence of interactions between Andreev states, we proceed by analyzing the consequence of setting $J\not=0$ in Eq.~\eqref{eq:model}.
Only in presence of both a finite spin-orbit interaction $\alpha_\perp, \alpha_{||} \neq 0$ and $J\not=0$ we can reproduce the spectrum seen in the data~[\cref{fig:f3}(c) - right panel]. 
From the fit of the data positions at zero field, we find $J=\SI{17}{\giga\hertz}$ and $\alpha_\perp=\SI{4.8}{\giga\hertz}$ (see Supplement for lines on top of the data~\cite{wesdorp_supplementary_2022}). The extracted exchange is comparable to estimated values of the effective charging energy of the normal region in a similar device (\({\sim}0.1\Delta\)) \cite{matute-canadas_signatures_2022}, and thus singlet-doublet ground state phase transitions are not expected. This is different to the situation reported in \cite{fatemi_microwave_2021-3} (\({\sim}\Delta\)), or when a quantum dot is gate-defined in the junction, such as in Ref.~\cite{bargerbos_singlet_doublet_2022} where the interaction is estimated to be \({\sim}10\Delta\).
With these parameters, the two even transitions that do not disperse in field in Fig.~\ref{fig:f3} are identified with the hybridized states \Satilde\ and \Tzerotilde, motivating the ordering of transitions displayed in~\cref{fig:f2}(d).
Note that the fit simultaneously takes into account and matches the position of the odd-parity interband transitions \Iab\ (yellow dashed lines) at zero field. 
On the other hand, the effective $g$-factors of the Andreev manifolds are not varied in the fit, but fixed to values extracted separately, as discussed in the next Section.
From the fit, together with the wire diameter, we can estimate a lower bound on the Rashba spin-orbit strength $\alpha_\mathrm{R}$ of $\alpha_\mathrm{R}\geq\SI{2}{\milli\electronvolt\nano\meter}$ (see Supplement~\cite{wesdorp_supplementary_2022}).
This is on the lower side of typical values of 5-\SI{40}{\milli\electronvolt\nano\meter} found in literature for InAs nanowires ~\cite{liang_strong_2012, albrecht_exponential_2016, van_woerkom_microwave_2017-1, tosi_spin-orbit_2019}.
Finally, the avoided crossings between \Tmin\ and \Satilde, circled in~\cref{fig:f3} and between \Tplus\ and \Tzerotilde\ are only reproduced by the model if we include a finite parallel spin-orbit component $\alpha_\parallel$, set to \SI{1}{\giga\hertz} for visibility.
The extracted size of the \Tmin, \Satilde\ crossing from the data, approximated by half the frequency difference of the transitions in the center of the crossing, is $\approx$ \SI{0.5}{\giga\hertz}.
Overall, the observation of the triplet transitions \Tmin, \Tzerotilde, \Tplus, in finite magnetic field, together with the fact that they have a strong phase-dispersion (see Supplement~\cite{wesdorp_supplementary_2022}), implies that part of the supercurrent flowing in the junction is carried by spin-polarized triplet pairs.  
From the slope of the transition \Tmin\ versus phase at \Bz~$=\SI{95}{\milli\tesla}$, we can estimate a change in current of approximately $\SI{2.3}{\nano\ampere}$ with respect to the supercurrent flowing when the junction is in the ground state (see Supplement~\cite{wesdorp_supplementary_2022}). This is a measure of the supercurrent carried by the spin-polarized pair. 

At higher fields we observe a strong downward trend of the transition frequencies.
We suspect that this is dominated by the orbital effect of the magnetic field in the nanowire, since the \SI{6}{\nano\meter} aluminum shell has a much higher critical field exceeding \SI{1}{\tesla}~\cite{chang_hard_2015}.
In the Supplement, we investigate the presence of a revival of the Andreev spectrum in fields up to 1T~\cite{wesdorp_supplementary_2022}, motivated by observations of a plasma mode revival on similar nanowires in a transmon geometry~\cite{kringhoj_magnetic-field-compatible_2021-1, uilhoorn_quasiparticle_2021} and supercurrent revival~\cite{zuo_supercurrent_2017-1} due to interference effects, but we do not find it.
The presence of a revival would open up the path towards detection of signatures in the microwave response of a topological phase transition in presence of multiple Andreev manifolds~\cite{vayrynen_microwave_2015}, manifesting as the fractional Josephson effect~\cite{fu_josephson_2009, lutchyn_majorana_2010}.

\section{Directly driven Andreev spin-flip}

So far we have mostly considered the even-parity part of the spectrum of~\cref{fig:f3}(b). However, when the junction initially is in one of the doublet states due to QP-poisoning, we can distinguish a linearly upwards dispersing transition \Da~(\cref{fig:f3}(b) yellow line) at finite field.
We attribute this to a directly driven spin-flip between the spin-up $\ket{\uparrow_{a}}$ and spin-down $\ket{\downarrow_{a}}$ levels of the lowest Andreev manifold~[\cref{fig:f3} (c)].

From the slope of \Da\ we can extract an effective $g$-factor $g^*_a=5.3$ of the lowest manifold.
The triplet transitions \Tplus\ and \Tmin\ should disperse in field with the half sum of the effective $g$-factors of the two manifolds: $g^* = \pm 1/2(g_a^*+g_b^*) = 7.8$.
Thus, we infer that the higher doublet has a higher effective $g$-factor of $g_b^* = 10.2$.
These values are used for the fit to the theory model presented in~[\cref{fig:f3}(c)] and are consistent with hybridized states where the $g$-factor should be between $g_\mathrm{Al}\approx2$ and $g_\mathrm{InAs}\approx15$.
On the other hand, \Tzerotilde\ and \Sab\ disperse weakly in field. We attribute this to a competition between the exchange and  the difference in Zeeman energy of each manifold. 
By solving the model without spin-orbit interaction, the eigenenergies of \ketTzero\ and \ketSab\ result in 
$E_a+E_b-J/2\pm \sqrt{(J/2)^2 + (\mu_\mathrm{B} B_{z})^2(g^*_a - g^*_b)^2}$ which is linear in \Bz\ when \Bz\ $\gg J$ and quadratic in \Bz\ when \Bz\ $\ll J$.
Thus for large fields their dispersion converges to that of the non-interacting states $\ket{\uparrow_{a}\downarrow_{b}},\ket{\downarrow_{a}\uparrow_{b}}$.
A possible cause of the large difference in the $g$-factors of the Andreev manifolds is that, the Fermi velocity of the first sub-band is higher than that of the second sub-band, due to the larger distance from the band bottom.
Therefore the effective spin-orbit strength is higher for the first sub-band, reducing $g_a$ more than $g_b$.

We note that we do not observe the intra-doublet transition \Db\ of the higher doublet, which would have a larger slope due to the higher $g$-factor.
This can be explained by the fact that the initial state of this transition is too short-lived: any quasiparticle occupying the higher manifold quickly decays into the lowest manifold. A comparison of the measured parity lifetimes $\sim\SI{ 0.5}{\milli\second}$ that we recently reported for this device~\cite{wesdorp_dynamical_2021}, to measured lifetimes ~$\sim \SI{4}{\micro\second}$ of an excited quasiparticle in the higher manifold in InAs/Al nanowires~\cite{hays_continuous_2020}, supports this.

\begin{figure}
\includegraphics{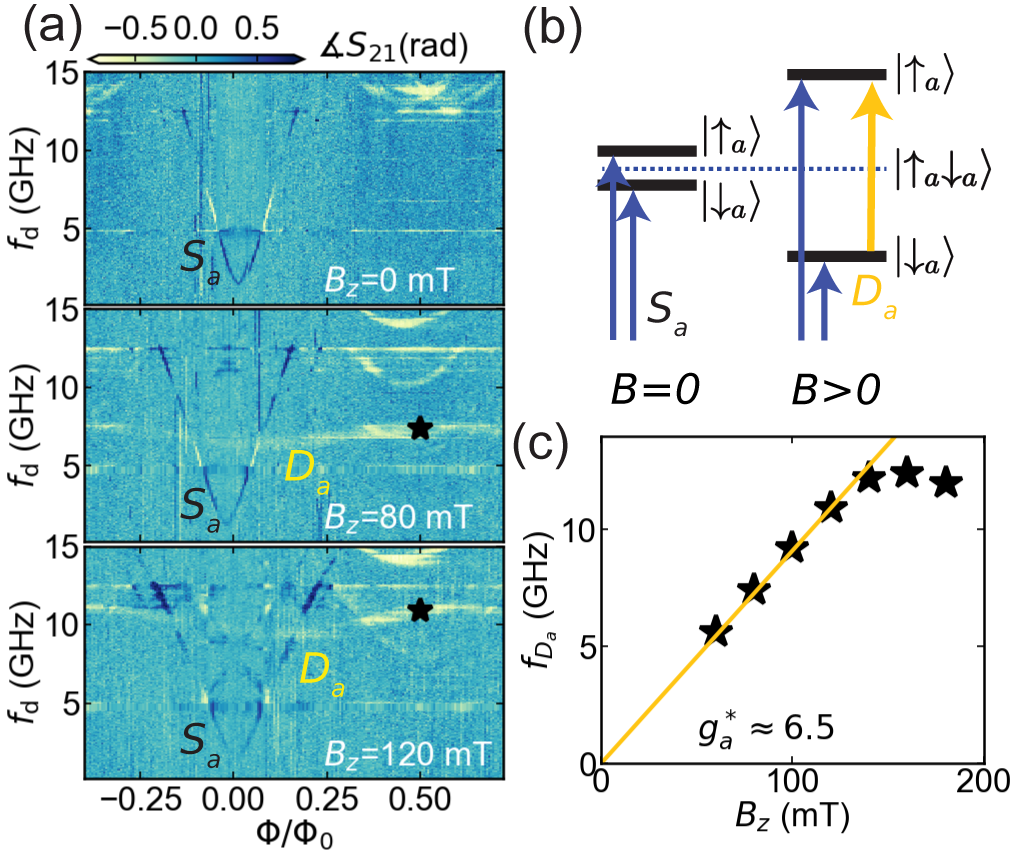}
\caption{\label{fig:f4} Phase dependence of the singlet \Sa\ and directly driven spin-flip doublet \Da\ transition in the lowest Andreev manifold at finite magnetic field.   \textbf{(a)} Measured low energy transition spectra at $V_\mathrm{g}=\SI{628}{\milli\volt}$ for increasing magnetic fields where the transparency of the lowest Andreev state is at a local maximum. The spin-flip doublet transition is visible at \Bz$>\SI{50}{\milli\tesla}$.
\textbf{(b)} Schematic of the two transitions at zero and finite field. A magnetic field induces a finite matrix element to allow observation of the \Da\ : $\ket{\uparrow_{a}}\leftrightarrow\ket{\downarrow_{a}}$ transition in the spectrum.  
\textbf{(c)} Extracted doublet transition frequency versus phase at $\Phi=\Phi_0/2$ indicated by stars in (a) . The transition evolves linearly versus field until spin-orbit interaction causes the lowest Andreev level to interact with higher levels that come down with \Bz\ bringing down the transition frequency. }
\end{figure}

So far, we have exclusively inferred the observation of a direct Andreev spin-flip transition \Da\ from data at $\Phi\approx \Phi_0/2$.
In order to provide additional evidence supporting this observation, we now explore the phase dispersion of \Da\ [\cref{fig:f4}].
To facilitate this, we exploit the gate-tunability of the nanowire Josephson junction to move to a nearby gate setting \Vg~=~\SI{628}{mV} where the lowest manifold has a high transparency and thus \Da\ is energetically separated from the rest of the spectrum (see Supplement for the gate dependence at \Bz~=~\SI{0}{\tesla}~\cite{wesdorp_supplementary_2022}).
In~\cref{fig:f4}(a) we show the evolution of the phase dispersion when increasing \Bz.
At \Bz~=~0, we only see the singlet transition \Sa. 
When we increase \Bz, we observe both \Sa\ and the odd-parity spin-flip doublet \Da\, as indicated in the diagram of~\cref{fig:f4}(b). 
As the dispersive shift in presence of resonator crossings in general can switch sign~\cite{metzger_circuit-qed_2021-1}, which would change their color in~\cref{fig:f3}, we have confirmed the odd-parity nature by performing parity-selective spectroscopy~\cite{wesdorp_dynamical_2021} at \Bz~=~\SI{100}{\milli\tesla} in the Supplement~\cite{wesdorp_supplementary_2022}.
As expected, the phase dispersion of \Sa\ stays constant at small fields since it is spin-singlet or hybridized with \Tzero, while \Da\ moves up in frequency linearly~[\cref{fig:f4}(c)] with $g_a^*\approx 6.5$. Note that $g_a^*$ differs from the previous gate-setting (see Supplement~\cite{wesdorp_supplementary_2022}).

The lack of \Da\ at zero field can be explained by two possible causes. At \Bz~=~0, the steady state population of $\ket{\uparrow_{a}}$ and $\ket{\downarrow_{a}}$ could be nearly equal due to the near-degeneracy in energy, reducing signal when driving \Da. 
Additionally, the matrix element to drive \Da\ is expected to vanish at zero field~\cite{van_heck_zeeman_2017}(see Supplement~\cite{wesdorp_supplementary_2022}), which is why recent works on coherent manipulation of an Andreev spin qubit~\cite{hays_coherent_2021} were forced to utilize Raman transitions to be able to achieve population transfer. Additionally, recent observations of \Da\ at zero field~\cite{metzger_circuit-qed_2021-1} indeed observed a vanishing of the transition around $\varphi=\pi$.
A finite magnetic field in combination with spin-orbit coupling increases the matrix element, thus facilitating direct driving of this transition at $B>0$ [thicker yellow line in~\cref{fig:f4}(b)]. The field also favors the occupation of $\ket{\uparrow_{a}}$, possibly increasing the population difference and therefore the strength of the signal.

Although \Sa\ has a large dispersion, \Da\ only has a small phase dispersion of ($\approx \SI{2}{\giga \hertz}$).
This is consistent with expectations, since the dispersion is only caused by the effective spin-orbit splitting of the Andreev levels.
Finally, note that the minimum of \Da\ is not aligned with \Sa.
Using tight-binding simulations of a similar scenario (see Supplement~\cite{wesdorp_supplementary_2022}), we found that a possible explanation could be due a component of $B_\mathrm{SO}$ parallel to \Bz\, consistent with the earlier mentioned field-dependence of the interband odd-parity transitions.
The observation of the spin-flip transition in a magnetic field opens up the path towards directly driven superconducting spin qubits~\cite{chtchelkatchev_andreev_2003, padurariu_theoretical_2010} and allows tuning the qubit frequency over a wide range of frequency depending on the field strength.

\begin{figure*}
\includegraphics{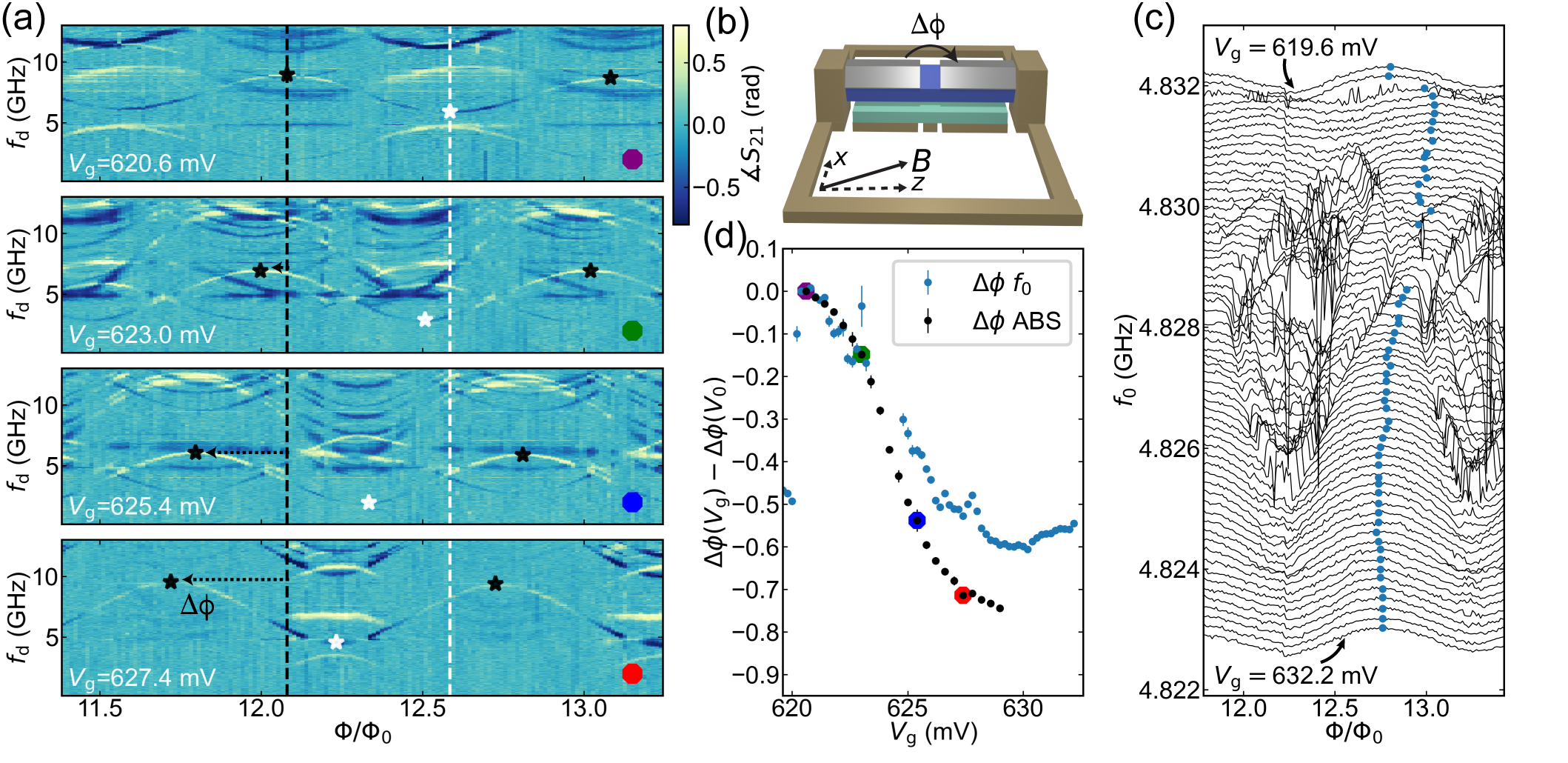}
\caption{\label{fig:f5} Gate-dependent anomalous Josephson effect of both individual Andreev transitions and aggregate supercurrent in presence of finite magnetic field: \Bz~=~\SI{220}{\milli\tesla} and \Bx~$=\SI{45}{\milli\tesla}$. 
\textbf{(a)}
In the two-tone spectra, we follow a transition shifting to the left with increasing gate voltage. We suspect this is an odd-parity transition (see Supplement for the identification and comparison of even and odd-parity phase shifts~\cite{wesdorp_supplementary_2022}). Dashed vertical lines indicate the positions of the maximum (black) and minimum (white) of the reference gate voltage.
\textbf{(b)} Diagram of the SQUID loop with B indicating the direction of the field. 
\textbf{(c)}
SQUID oscillations in the resonator frequency in the same field and gate settings as the spectra in (a). They undergo a leftward phase shift with an increasing gate voltage. The blue dots indicate the positions of the maxima for each \Vg. The distortion of the lines in the middle region are caused by avoided crossings between Andreev state transitions frequencies and the resonator when the transparancy of the junction is high.
\textbf{(d)} Phase shift extracted from the two-tone excitation spectra (colored markers indicate the corresponding panel in (a)) with respect to a reference gate voltage $V_{\mathrm{0}}=\SI{620.6}{\milli\volt}$. Additionally the phase shift $\Delta\phi $ $f_0$ extracted from resonator SQUID oscillations (c) are shown.
}
\end{figure*}

\section{Gate-dependent anomalous Josephson effect at finite fields}

In~\cref{fig:f3}(b) we have shown the field evolution at fixed phase difference.
The entire phase dispersion is also of interest, because of the possible presence of the anomalous Josephson effect (AJE)~\cite{krive_chiral_2004, reynoso_spin-orbit-induced_2012, yokoyama_anomalous_2014, yokoyama_josephson_2013-1, bergeret_theory_2014, campagnano_spinorbit_2015}.
To investigate its occurrence, we measured finite-field spectra at different gate voltages, several of which are shown in~\cref{fig:f5}(a) (see Supplement for all data~\cite{wesdorp_supplementary_2022}).
We track the minima and maxima of this transition, indicated with the white and black stars respectively. 
As we increase the gate voltage, the transition starts to shift horizontally to the left, demonstrating the phase shift in the spectrum. 
Both even and odd-parity transitions (offset with a nearly $\pi$-phase) exhibit a shift. 

In~\cref{fig:f5}(d) we show the extracted shift of the maxima for all measured spectra versus gate, resulting in a continuously gate-tunable relative shift up to $\Delta\phi=0.72\pi$ at $V_\mathrm{g}=\SI{629}{\milli\volt}$ with respect to the reference phase at $V_\mathrm{0}=\SI{620.6}{\milli\volt}$. 
For these measurements, the magnetic field is set to
\Bz~$=\SI{220}{\milli\tesla}$ and we add a perpendicular component \Bx~$=\SI{45}{\milli\tesla}$ (equivalent to $\Phi=12.3\Phi_0$)~[\cref{fig:f5}(b)]. 
The perpendicular component is added to reduce flux-jumps due to zero-field crossings.
Additionally we expected the AJE to be stronger in presence of a perpendicular field component~\cite{yokoyama_josephson_2013-1, szombati_josephson_2016}. 
The choice of field was limited in the \Bx\ direction by the maximum output of the current source.
Beyond \Bz~$=\SI{220}{\milli\tesla}$, we lost visibility in the two-tone spectra.

In essence, the AJE occurs because coupling between different Andreev levels pushes their minima away from $\varphi=0$~\cite{yokoyama_anomalous_2014,yokoyama_josephson_2013-1}.
The minimum of the ground state energy, which is a sum over all the Andreev energies, then also shifts away from $\varphi=0$ and the junction will assume a phase difference that minimizes the ground state energy at $\varphi=\varphi_0$, or, if a phase difference is imposed externally in a loop geometry, a finite current will flow at zero external flux through the loop.

For the AJE to occur, breaking of time-reversal symmetry is a necessary but not sufficient condition. Additional spatial or spin-rotation symmetries need to be broken depending on the setup~\cite{liu_relation_2010, rasmussen_effects_2016,assouline_spin-orbit_2019}.  
A Zeeman field breaks time-reversal symmetry and spatial symmetries can be broken by spin-orbit interaction in presence of a non-symmetric potential~\cite{campagnano_spinorbit_2015} or multiple sub-bands~\cite{yokoyama_anomalous_2014}.
In our system we are clearly in a regime with multiple occupied sub-bands~[\cref{fig:f3}] evident by the many Andreev transitions visible at higher fields. Due to the asymmetry of the gates with respect to the junction (see device images in~\cite{wesdorp_supplementary_2021}) we would not expect a symmetric potential. 
Thus we expect to see the AJE.
In recent experiments demonstrating the AJE, measurements of the DC supercurrent~\cite{szombati_josephson_2016,assouline_spin-orbit_2019, mayer_gate_2020,strambini_josephson_2020}, or of the ground state Josephson energy~\cite{pita-vidal_gate-tunable_2020-1}, were used to probe the anomalous phase shift caused by the summed contributions of all Andreev levels.
In~\cref{fig:f5}(a) we add to this by showing the underlying microscopic origin of the anomalous supercurrent: the phase shifts of Andreev transitions, which imply shifts of the individual Andreev levels, and which we can measure directly in magnetic fields strong enough to produce this effect.  

To compare with supercurrent measurements, we also measure the SQUID oscillations in the resonator in the same gate-regime~[\cref{fig:f5}(c)]. 
The blue datapoints in~\cref{fig:f5}(d) correspond to the maxima in the single tone (ST) resonator traces in~\cref{fig:f5} (c).
The frequency shift of the resonator \fnull\ originates from the dispersive coupling with the junction in the ground state, and so it is a measure of the phase shift of the ground state current-phase relation of the junction.
Since the total $\varphi_0$ results from contributions of different channels, which may lead to cancellation if these channels have different phase shifts (see Supplement~\cite{wesdorp_supplementary_2022} for a larger frequency range than~\cref{fig:f5}(a) illustrating the different shifts per Andreev state), it is not surprising that the phase shift in \fnull\ is smaller than the phase shift of the individual lowest Andreev states in~\cref{fig:f5}(d).

A gate-induced phase shift can have other explanations different than the AJE, and we now discuss measures we took to rule those out.
When sweeping the large vector magnet, flux can be trapped or de-trapped on-chip or drift over time, which can cause unwanted phase shifts.
To rule out flux drift, we measured the 3D map in the spectra of~\cref{fig:f5}(a) by sweeping \fd\ for each \Vg\ before stepping flux. 
This ensures that the change in phase is caused by \Vg.
Alternatively, a change in total supercurrent can change the phase-drop over the junction when the shunt-inductance is large due to a non-linear relation between $\Phi$ and $\varphi$. 
In the Supplement we estimate that we should be in a linear regime for the given \Ls\ and typical $I_\mathrm{c}$~\cite{wesdorp_supplementary_2022}.
Also, we measured data at lower field strengths and we saw no anomalous shift at $B=0$~\cite{wesdorp_supplementary_2022}. 
To exclude a trivial origin of the observed phase shift by a gate-induced change in effective loop size, we have also kept track of the difference between two maxima in the two-tone spectra as an estimate of the total period.
We saw no clear correlation with the phase shift. Fluctuations of the period were around $\Delta \Phi = 6$~m$\Phi_0$.
This can at most account for a phase shift of $\Delta\varphi=0.14 \pi$, much smaller than what we report here.
To further investigate the cause of the gate-dependence of the AJE we performed a series of parallel field sweeps versus gate in the Supplement~\cite{wesdorp_supplementary_2022}.
Here by inspecting the \Da\ transition where it was visible, we extracted $g^*_a$ as a function of \Vg\, which is correlated with the size of the phase shift. This would indicate the AJE scales with the effective Zeeman energy of the lowest Andreev manifold.
We have also performed measurements of the AJE with the field vector reversed, both for the SQUID oscillations and the spectroscopy. Here we observe a reversal of the phase shift as expected (see Supplement~\cite{wesdorp_supplementary_2022}).
We thus conclude that the observed phase shift~[\cref{fig:f5}] is indeed due the AJE and not due to the alternative causes mentioned above. 

\section{Conclusions}

In this work, we have performed microwave spectroscopy of Andreev bound states in a nanowire Josephson junction in a magnetic field,
using a field-compatible superconducting resonator. 
By aligning the magnetic field parallel to the nanowire, we have investigated the field dependence of the many-body spectrum at fixed phase difference over the junction ($\varphi=\pi$), for both even and odd fermion parity. 
In the even parity sector, we distinguished singlet and triplet-like Andreev states, hybridized by spin-orbit interaction and split by exchange interaction.
In the odd parity, at finite field, we observed the direct doublet spin-flip transition in the lowest Andreev manifold.
At fields larger than \Bz=\SI{170}{\milli\tesla}, we found a strong gate-tunable anomalous Josephson effect in the many-body spectrum, currently of interest due to its envisioned application in spintronics~\cite{linder_superconducting_2015}.
Our findings confirm that both spin-orbit interaction and electron-electron interactions are important to understand Andreev spectra in InAs/Al Josephson junctions.

The observed hybridization of triplet and singlet Andreev transitions is consistent with predictions that in a finite magnetic field, the induced superconducting pairing in the semiconducting nanowire is a mixture of singlet and triplet components~\cite{lutchyn_majorana_2010, oreg_helical_2010}.
However, our measurements probe states localized at the Josephson junction, which depend on both local and bulk properties, and we cannot exclude that spin-orbit interaction is only active in the junction, but not in the leads. 
Thus, our measurements should be complemented with methods that can single out the bulk properties of the nanowire~\cite{rosdahl_andreev_2018, phan_detecting_2022, splitthoff_gate-tunable_2022}.  

An open question remains to explore the density dependence of the proximitized leads and that of the Andreev spectrum.
Signatures of topological phase diagrams could be observable in microwave spectroscopy~\cite{vayrynen_microwave_2015, peng_signatures_2016, murthy_energy_2020} due to the onset of the fractional Josephson effect, but it seems crucial to extend available theory to understand the effect of interactions and $g$-factor renormalization.
To prevent the closing of the spectral gap due to orbital interference, it would be interesting to perform these measurements in devices with a lower density of states.
This could be aided by another choice of material, e.g. InSb, which has a lower effective mass and band-offset with Al~\cite{winkler_unified_2019-1} compared to InAs.

Spectroscopy of Andreev states using superconducting circuitry allow the combination of spectroscopic measurements with high time-resolution, allowing e.g. parity-selective spectroscopy as we have shown recently at zero field~\cite{wesdorp_dynamical_2021}.
In future, when combined with on-chip flux control and parametric amplification readily available in the superconducting circuit community, this combination should allow for fast measurements of the phase-periodicity of individual Andreev levels in timescales of \SI{}{\giga\hertz} to \SI{}{\mega\hertz}. 
This type of measurement could provide a more controlled way towards the detection of the fractional Josephson effect, not hindered by the presence of QP-poisoning~\cite{lutchyn_majorana_2010} or Landau-Zener effects~\cite{laroche_observation_2019}.

Additionally, the observation of the spin-flip transition as well as a singlet-triplet avoided crossing can provide new ways to manipulate Andreev (spin) qubits~\cite{janvier_coherent_2015, hays_direct_2018, hays_continuous_2020, hays_coherent_2021}, that exploit an external field.
In particular, the direct spin-flip transition activated by the magnetic field makes it possible to circumvent the need to use used Raman techiques~\cite{hays_coherent_2021} involving a second bound state in order to manipulate the spin of an Andreev state~\cite{park_andreev_2017}.
Furthermore, the singlet-triplet avoided crossing is particularly interesting as it opens up the possibility to manipulate Andreev pairs in analogy with singlet-triplet qubits in semiconducting quantum dots~\cite{burkard_semiconductor_2021, padurariu_spin_2012}.

\begin{acknowledgments}
We would like to thank Ruben Grigoryan for the PCB and enclosure design, Guanzhong Wang, Tom Dvir and Cyril Metzger for useful discussions and Valla Fatemi for valuable comments on the manuscript. 
This work is part of the research project ‘Scalable circuits of Majorana qubits with topological protection’ (i39, SCMQ) with project number 14SCMQ02, which is (partly) financed by the Dutch Research Council (NWO). It has further been supported by the Microsoft Quantum initiative. FMC and ALY acknowledge funding
by FET-Open contract AndQC, by the Spanish AEI through Grant No.~PID2020-117671GB-I00 and through the ``Mar\'{\i}a de Maeztu'' Programme for Units of Excellence in R\&D (CEX2018-000805-M), and by the Spanish Ministry of
Universities (FPU20/01871).

\textbf{Data availability}
Raw data and scripts for the presented figures will be made available online.

\textbf{Author contributions}
JJW, BvH, GdL, DvW, SR contributed to the device design.
JJW, AV, LJS, MPV contributed
to sample fabrication and inspection. JJW,
AV contributed to the data acquisition and analysis
with input from GdL, BvH, AB, MPV, LG. JJW, TL, BvH contributed to the tight-binding simulations without interactions and FMC, ALY to  the minimal model with interactions. JJW, FMC,
AV, BvH wrote the manuscript with comments and input from
all other co-authors. Nanowires were grown by
PK. Project was supervised by GdL, BvH.

\end{acknowledgments}

\bibliography{bib/main_mac.bib}

\begin{thebibliography}{92}%
\makeatletter
\providecommand \@ifxundefined [1]{%
 \@ifx{#1\undefined}
}%
\providecommand \@ifnum [1]{%
 \ifnum #1\expandafter \@firstoftwo
 \else \expandafter \@secondoftwo
 \fi
}%
\providecommand \@ifx [1]{%
 \ifx #1\expandafter \@firstoftwo
 \else \expandafter \@secondoftwo
 \fi
}%
\providecommand \natexlab [1]{#1}%
\providecommand \enquote  [1]{``#1''}%
\providecommand \bibnamefont  [1]{#1}%
\providecommand \bibfnamefont [1]{#1}%
\providecommand \citenamefont [1]{#1}%
\providecommand \href@noop [0]{\@secondoftwo}%
\providecommand \href [0]{\begingroup \@sanitize@url \@href}%
\providecommand \@href[1]{\@@startlink{#1}\@@href}%
\providecommand \@@href[1]{\endgroup#1\@@endlink}%
\providecommand \@sanitize@url [0]{\catcode `\\12\catcode `\$12\catcode
  `\&12\catcode `\#12\catcode `\^12\catcode `\_12\catcode `\%12\relax}%
\providecommand \@@startlink[1]{}%
\providecommand \@@endlink[0]{}%
\providecommand \url  [0]{\begingroup\@sanitize@url \@url }%
\providecommand \@url [1]{\endgroup\@href {#1}{\urlprefix }}%
\providecommand \urlprefix  [0]{URL }%
\providecommand \Eprint [0]{\href }%
\providecommand \doibase [0]{https://doi.org/}%
\providecommand \selectlanguage [0]{\@gobble}%
\providecommand \bibinfo  [0]{\@secondoftwo}%
\providecommand \bibfield  [0]{\@secondoftwo}%
\providecommand \translation [1]{[#1]}%
\providecommand \BibitemOpen [0]{}%
\providecommand \bibitemStop [0]{}%
\providecommand \bibitemNoStop [0]{.\EOS\space}%
\providecommand \EOS [0]{\spacefactor3000\relax}%
\providecommand \BibitemShut  [1]{\csname bibitem#1\endcsname}%
\let\auto@bib@innerbib\@empty
\bibitem [{\citenamefont {Kulik}(1969)}]{kulik_macroscopic_1969}%
  \BibitemOpen
  \bibfield  {author} {\bibinfo {author} {\bibfnamefont {I.}~\bibnamefont
  {Kulik}},\ }\bibfield  {title} {\bibinfo {title} {Macroscopic quantization
  and the proximity effect in {{S-N-S}} junctions},\ }\href@noop {} {\bibfield
  {journal} {\bibinfo  {journal} {Soviet Journal of Experimental and
  Theoretical Physics}\ }\textbf {\bibinfo {volume} {30}},\ \bibinfo {pages}
  {944} (\bibinfo {year} {1969})}\BibitemShut {NoStop}%
\bibitem [{\citenamefont {Beenakker}(1991)}]{beenakker_universal_1991}%
  \BibitemOpen
  \bibfield  {author} {\bibinfo {author} {\bibfnamefont {C.~W.~J.}\
  \bibnamefont {Beenakker}},\ }\bibfield  {title} {\bibinfo {title} {Universal
  limit of critical-current fluctuations in mesoscopic {{Josephson}}
  junctions},\ }\href {https://doi.org/10.1103/PhysRevLett.67.3836} {\bibfield
  {journal} {\bibinfo  {journal} {Physical Review Letters}\ }\textbf {\bibinfo
  {volume} {67}},\ \bibinfo {pages} {3836} (\bibinfo {year}
  {1991})}\BibitemShut {NoStop}%
\bibitem [{\citenamefont {Klapwijk}(2004)}]{klapwijk_proximity_2004}%
  \BibitemOpen
  \bibfield  {author} {\bibinfo {author} {\bibfnamefont {T.~M.}\ \bibnamefont
  {Klapwijk}},\ }\bibfield  {title} {\bibinfo {title} {Proximity {{Effect
  From}} an {{Andreev Perspective}}},\ }\href
  {https://doi.org/10.1007/s10948-004-0773-0} {\bibfield  {journal} {\bibinfo
  {journal} {Journal of Superconductivity}\ }\textbf {\bibinfo {volume} {17}},\
  \bibinfo {pages} {593} (\bibinfo {year} {2004})}\BibitemShut {NoStop}%
\bibitem [{\citenamefont {Bretheau}\ \emph
  {et~al.}(2013{\natexlab{a}})\citenamefont {Bretheau}, \citenamefont {Girit},
  \citenamefont {Pothier}, \citenamefont {Esteve},\ and\ \citenamefont
  {Urbina}}]{bretheau_exciting_2013}%
  \BibitemOpen
  \bibfield  {author} {\bibinfo {author} {\bibfnamefont {L.}~\bibnamefont
  {Bretheau}}, \bibinfo {author} {\bibfnamefont {{\c C}.~{\"O}.}\ \bibnamefont
  {Girit}}, \bibinfo {author} {\bibfnamefont {H.}~\bibnamefont {Pothier}},
  \bibinfo {author} {\bibfnamefont {D.}~\bibnamefont {Esteve}},\ and\ \bibinfo
  {author} {\bibfnamefont {C.}~\bibnamefont {Urbina}},\ }\bibfield  {title}
  {\bibinfo {title} {Exciting {{Andreev}} pairs in a superconducting atomic
  contact},\ }\href {https://doi.org/10.1038/nature12315} {\bibfield  {journal}
  {\bibinfo  {journal} {Nature}\ }\textbf {\bibinfo {volume} {499}},\ \bibinfo
  {pages} {312} (\bibinfo {year} {2013}{\natexlab{a}})}\BibitemShut {NoStop}%
\bibitem [{\citenamefont {Janvier}\ \emph {et~al.}(2015)\citenamefont
  {Janvier}, \citenamefont {Tosi}, \citenamefont {Bretheau}, \citenamefont
  {Girit}, \citenamefont {Stern}, \citenamefont {Bertet}, \citenamefont
  {Joyez}, \citenamefont {Vion}, \citenamefont {Esteve}, \citenamefont
  {Goffman}, \citenamefont {Pothier},\ and\ \citenamefont
  {Urbina}}]{janvier_coherent_2015}%
  \BibitemOpen
  \bibfield  {author} {\bibinfo {author} {\bibfnamefont {C.}~\bibnamefont
  {Janvier}}, \bibinfo {author} {\bibfnamefont {L.}~\bibnamefont {Tosi}},
  \bibinfo {author} {\bibfnamefont {L.}~\bibnamefont {Bretheau}}, \bibinfo
  {author} {\bibfnamefont {{\c C}.~{\"O}.}\ \bibnamefont {Girit}}, \bibinfo
  {author} {\bibfnamefont {M.}~\bibnamefont {Stern}}, \bibinfo {author}
  {\bibfnamefont {P.}~\bibnamefont {Bertet}}, \bibinfo {author} {\bibfnamefont
  {P.}~\bibnamefont {Joyez}}, \bibinfo {author} {\bibfnamefont
  {D.}~\bibnamefont {Vion}}, \bibinfo {author} {\bibfnamefont {D.}~\bibnamefont
  {Esteve}}, \bibinfo {author} {\bibfnamefont {M.~F.}\ \bibnamefont {Goffman}},
  \bibinfo {author} {\bibfnamefont {H.}~\bibnamefont {Pothier}},\ and\ \bibinfo
  {author} {\bibfnamefont {C.}~\bibnamefont {Urbina}},\ }\bibfield  {title}
  {\bibinfo {title} {Coherent manipulation of {{Andreev}} states in
  superconducting atomic contacts},\ }\href
  {https://doi.org/10.1126/science.aab2179} {\bibfield  {journal} {\bibinfo
  {journal} {Science}\ }\textbf {\bibinfo {volume} {349}},\ \bibinfo {pages}
  {1199} (\bibinfo {year} {2015})}\BibitemShut {NoStop}%
\bibitem [{\citenamefont {Bretheau}\ \emph
  {et~al.}(2013{\natexlab{b}})\citenamefont {Bretheau}, \citenamefont {Girit},
  \citenamefont {Urbina}, \citenamefont {Esteve},\ and\ \citenamefont
  {Pothier}}]{bretheau_supercurrent_2013-1}%
  \BibitemOpen
  \bibfield  {author} {\bibinfo {author} {\bibfnamefont {L.}~\bibnamefont
  {Bretheau}}, \bibinfo {author} {\bibfnamefont {{\c C}.~{\"O}.}\ \bibnamefont
  {Girit}}, \bibinfo {author} {\bibfnamefont {C.}~\bibnamefont {Urbina}},
  \bibinfo {author} {\bibfnamefont {D.}~\bibnamefont {Esteve}},\ and\ \bibinfo
  {author} {\bibfnamefont {H.}~\bibnamefont {Pothier}},\ }\bibfield  {title}
  {\bibinfo {title} {Supercurrent {{Spectroscopy}} of {{Andreev States}}},\
  }\href {https://doi.org/10.1103/PhysRevX.3.041034} {\bibfield  {journal}
  {\bibinfo  {journal} {Physical Review X}\ }\textbf {\bibinfo {volume} {3}},\
  \bibinfo {pages} {041034} (\bibinfo {year} {2013}{\natexlab{b}})}\BibitemShut
  {NoStop}%
\bibitem [{\citenamefont {{van Woerkom}}\ \emph {et~al.}(2017)\citenamefont
  {{van Woerkom}}, \citenamefont {Proutski}, \citenamefont {{van Heck}},
  \citenamefont {Bouman}, \citenamefont {V{\"a}yrynen}, \citenamefont
  {Glazman}, \citenamefont {Krogstrup}, \citenamefont {Nyg{\aa}rd},
  \citenamefont {Kouwenhoven},\ and\ \citenamefont
  {Geresdi}}]{van_woerkom_microwave_2017-1}%
  \BibitemOpen
  \bibfield  {author} {\bibinfo {author} {\bibfnamefont {D.~J.}\ \bibnamefont
  {{van Woerkom}}}, \bibinfo {author} {\bibfnamefont {A.}~\bibnamefont
  {Proutski}}, \bibinfo {author} {\bibfnamefont {B.}~\bibnamefont {{van
  Heck}}}, \bibinfo {author} {\bibfnamefont {D.}~\bibnamefont {Bouman}},
  \bibinfo {author} {\bibfnamefont {J.~I.}\ \bibnamefont {V{\"a}yrynen}},
  \bibinfo {author} {\bibfnamefont {L.~I.}\ \bibnamefont {Glazman}}, \bibinfo
  {author} {\bibfnamefont {P.}~\bibnamefont {Krogstrup}}, \bibinfo {author}
  {\bibfnamefont {J.}~\bibnamefont {Nyg{\aa}rd}}, \bibinfo {author}
  {\bibfnamefont {L.~P.}\ \bibnamefont {Kouwenhoven}},\ and\ \bibinfo {author}
  {\bibfnamefont {A.}~\bibnamefont {Geresdi}},\ }\bibfield  {title} {\bibinfo
  {title} {Microwave spectroscopy of spinful {{Andreev}} bound states in
  ballistic semiconductor {{Josephson}} junctions},\ }\href
  {https://doi.org/10.1038/nphys4150} {\bibfield  {journal} {\bibinfo
  {journal} {Nature Physics}\ }\textbf {\bibinfo {volume} {13}},\ \bibinfo
  {pages} {876} (\bibinfo {year} {2017})}\BibitemShut {NoStop}%
\bibitem [{\citenamefont {Hays}\ \emph {et~al.}(2018)\citenamefont {Hays},
  \citenamefont {{de Lange}}, \citenamefont {Serniak}, \citenamefont {{van
  Woerkom}}, \citenamefont {Bouman}, \citenamefont {Krogstrup}, \citenamefont
  {Nyg{\aa}rd}, \citenamefont {Geresdi},\ and\ \citenamefont
  {Devoret}}]{hays_direct_2018}%
  \BibitemOpen
  \bibfield  {author} {\bibinfo {author} {\bibfnamefont {M.}~\bibnamefont
  {Hays}}, \bibinfo {author} {\bibfnamefont {G.}~\bibnamefont {{de Lange}}},
  \bibinfo {author} {\bibfnamefont {K.}~\bibnamefont {Serniak}}, \bibinfo
  {author} {\bibfnamefont {D.~J.}\ \bibnamefont {{van Woerkom}}}, \bibinfo
  {author} {\bibfnamefont {D.}~\bibnamefont {Bouman}}, \bibinfo {author}
  {\bibfnamefont {P.}~\bibnamefont {Krogstrup}}, \bibinfo {author}
  {\bibfnamefont {J.}~\bibnamefont {Nyg{\aa}rd}}, \bibinfo {author}
  {\bibfnamefont {A.}~\bibnamefont {Geresdi}},\ and\ \bibinfo {author}
  {\bibfnamefont {M.~H.}\ \bibnamefont {Devoret}},\ }\bibfield  {title}
  {\bibinfo {title} {Direct microwave measurement of {{Andreev-bound-state}}
  dynamics in a proximitized semiconducting nanowire},\ }\href
  {https://doi.org/10.1103/PhysRevLett.121.047001} {\bibfield  {journal}
  {\bibinfo  {journal} {Physical Review Letters}\ }\textbf {\bibinfo {volume}
  {121}},\ \bibinfo {pages} {047001} (\bibinfo {year} {2018})},\ \Eprint
  {https://arxiv.org/abs/1711.01645} {arXiv:1711.01645} \BibitemShut {NoStop}%
\bibitem [{\citenamefont {Tosi}\ \emph {et~al.}(2019)\citenamefont {Tosi},
  \citenamefont {Metzger}, \citenamefont {Goffman}, \citenamefont {Urbina},
  \citenamefont {Pothier}, \citenamefont {Park}, \citenamefont {Yeyati},
  \citenamefont {Nyg{\aa}rd},\ and\ \citenamefont
  {Krogstrup}}]{tosi_spin-orbit_2019}%
  \BibitemOpen
  \bibfield  {author} {\bibinfo {author} {\bibfnamefont {L.}~\bibnamefont
  {Tosi}}, \bibinfo {author} {\bibfnamefont {C.}~\bibnamefont {Metzger}},
  \bibinfo {author} {\bibfnamefont {M.~F.}\ \bibnamefont {Goffman}}, \bibinfo
  {author} {\bibfnamefont {C.}~\bibnamefont {Urbina}}, \bibinfo {author}
  {\bibfnamefont {H.}~\bibnamefont {Pothier}}, \bibinfo {author} {\bibfnamefont
  {S.}~\bibnamefont {Park}}, \bibinfo {author} {\bibfnamefont {A.~L.}\
  \bibnamefont {Yeyati}}, \bibinfo {author} {\bibfnamefont {J.}~\bibnamefont
  {Nyg{\aa}rd}},\ and\ \bibinfo {author} {\bibfnamefont {P.}~\bibnamefont
  {Krogstrup}},\ }\bibfield  {title} {\bibinfo {title} {Spin-orbit splitting of
  {{Andreev}} states revealed by microwave spectroscopy},\ }\href
  {https://doi.org/10.1103/PhysRevX.9.011010} {\bibfield  {journal} {\bibinfo
  {journal} {Physical Review X}\ }\textbf {\bibinfo {volume} {9}},\ \bibinfo
  {pages} {011010} (\bibinfo {year} {2019})},\ \Eprint
  {https://arxiv.org/abs/1810.02591} {arXiv:1810.02591} \BibitemShut {NoStop}%
\bibitem [{\citenamefont {Hays}\ \emph {et~al.}(2020)\citenamefont {Hays},
  \citenamefont {Fatemi}, \citenamefont {Serniak}, \citenamefont {Bouman},
  \citenamefont {Diamond}, \citenamefont {{de Lange}}, \citenamefont
  {Krogstrup}, \citenamefont {Nyg{\aa}rd}, \citenamefont {Geresdi},\ and\
  \citenamefont {Devoret}}]{hays_continuous_2020}%
  \BibitemOpen
  \bibfield  {author} {\bibinfo {author} {\bibfnamefont {M.}~\bibnamefont
  {Hays}}, \bibinfo {author} {\bibfnamefont {V.}~\bibnamefont {Fatemi}},
  \bibinfo {author} {\bibfnamefont {K.}~\bibnamefont {Serniak}}, \bibinfo
  {author} {\bibfnamefont {D.}~\bibnamefont {Bouman}}, \bibinfo {author}
  {\bibfnamefont {S.}~\bibnamefont {Diamond}}, \bibinfo {author} {\bibfnamefont
  {G.}~\bibnamefont {{de Lange}}}, \bibinfo {author} {\bibfnamefont
  {P.}~\bibnamefont {Krogstrup}}, \bibinfo {author} {\bibfnamefont
  {J.}~\bibnamefont {Nyg{\aa}rd}}, \bibinfo {author} {\bibfnamefont
  {A.}~\bibnamefont {Geresdi}},\ and\ \bibinfo {author} {\bibfnamefont {M.~H.}\
  \bibnamefont {Devoret}},\ }\bibfield  {title} {\bibinfo {title} {Continuous
  monitoring of a trapped, superconducting spin},\ }\href
  {https://doi.org/10.1038/s41567-020-0952-3} {\bibfield  {journal} {\bibinfo
  {journal} {Nature Physics}\ }\textbf {\bibinfo {volume} {16}},\ \bibinfo
  {pages} {1103} (\bibinfo {year} {2020})},\ \Eprint
  {https://arxiv.org/abs/1908.02800} {arXiv:1908.02800} \BibitemShut {NoStop}%
\bibitem [{\citenamefont {Hays}\ \emph {et~al.}(2021)\citenamefont {Hays},
  \citenamefont {Fatemi}, \citenamefont {Bouman}, \citenamefont {Cerrillo},
  \citenamefont {Diamond}, \citenamefont {Serniak}, \citenamefont {Connolly},
  \citenamefont {Krogstrup}, \citenamefont {Nyg{\aa}rd}, \citenamefont
  {Levy~Yeyati}, \citenamefont {Geresdi},\ and\ \citenamefont
  {Devoret}}]{hays_coherent_2021}%
  \BibitemOpen
  \bibfield  {author} {\bibinfo {author} {\bibfnamefont {M.}~\bibnamefont
  {Hays}}, \bibinfo {author} {\bibfnamefont {V.}~\bibnamefont {Fatemi}},
  \bibinfo {author} {\bibfnamefont {D.}~\bibnamefont {Bouman}}, \bibinfo
  {author} {\bibfnamefont {J.}~\bibnamefont {Cerrillo}}, \bibinfo {author}
  {\bibfnamefont {S.}~\bibnamefont {Diamond}}, \bibinfo {author} {\bibfnamefont
  {K.}~\bibnamefont {Serniak}}, \bibinfo {author} {\bibfnamefont
  {T.}~\bibnamefont {Connolly}}, \bibinfo {author} {\bibfnamefont
  {P.}~\bibnamefont {Krogstrup}}, \bibinfo {author} {\bibfnamefont
  {J.}~\bibnamefont {Nyg{\aa}rd}}, \bibinfo {author} {\bibfnamefont
  {A.}~\bibnamefont {Levy~Yeyati}}, \bibinfo {author} {\bibfnamefont
  {A.}~\bibnamefont {Geresdi}},\ and\ \bibinfo {author} {\bibfnamefont {M.~H.}\
  \bibnamefont {Devoret}},\ }\bibfield  {title} {\bibinfo {title} {Coherent
  manipulation of an {{Andreev}} spin qubit},\ }\href
  {https://doi.org/10.1126/science.abf0345} {\bibfield  {journal} {\bibinfo
  {journal} {Science}\ }\textbf {\bibinfo {volume} {373}},\ \bibinfo {pages}
  {430} (\bibinfo {year} {2021})}\BibitemShut {NoStop}%
\bibitem [{\citenamefont {Metzger}\ \emph {et~al.}(2021)\citenamefont
  {Metzger}, \citenamefont {Park}, \citenamefont {Tosi}, \citenamefont
  {Janvier}, \citenamefont {Reynoso}, \citenamefont {Goffman}, \citenamefont
  {Urbina}, \citenamefont {Yeyati},\ and\ \citenamefont
  {Pothier}}]{metzger_circuit-qed_2021-1}%
  \BibitemOpen
  \bibfield  {author} {\bibinfo {author} {\bibfnamefont {C.}~\bibnamefont
  {Metzger}}, \bibinfo {author} {\bibfnamefont {S.}~\bibnamefont {Park}},
  \bibinfo {author} {\bibfnamefont {L.}~\bibnamefont {Tosi}}, \bibinfo {author}
  {\bibfnamefont {C.}~\bibnamefont {Janvier}}, \bibinfo {author} {\bibfnamefont
  {A.~A.}\ \bibnamefont {Reynoso}}, \bibinfo {author} {\bibfnamefont {M.~F.}\
  \bibnamefont {Goffman}}, \bibinfo {author} {\bibfnamefont {C.}~\bibnamefont
  {Urbina}}, \bibinfo {author} {\bibfnamefont {A.~L.}\ \bibnamefont {Yeyati}},\
  and\ \bibinfo {author} {\bibfnamefont {H.}~\bibnamefont {Pothier}},\
  }\bibfield  {title} {\bibinfo {title} {Circuit-{{QED}} with phase-biased
  {{Josephson}} weak links},\ }\href
  {https://doi.org/10.1103/PhysRevResearch.3.013036} {\bibfield  {journal}
  {\bibinfo  {journal} {Physical Review Research}\ }\textbf {\bibinfo {volume}
  {3}},\ \bibinfo {pages} {013036} (\bibinfo {year} {2021})},\ \Eprint
  {https://arxiv.org/abs/2010.00430} {arXiv:2010.00430} \BibitemShut {NoStop}%
\bibitem [{\citenamefont {Wesdorp}\ \emph
  {et~al.}(2021{\natexlab{a}})\citenamefont {Wesdorp}, \citenamefont
  {Gr{\"u}nhaupt}, \citenamefont {Vaartjes}, \citenamefont {{Pita-Vidal}},
  \citenamefont {Bargerbos}, \citenamefont {Splitthoff}, \citenamefont
  {Krogstrup}, \citenamefont {{van Heck}},\ and\ \citenamefont {{de
  Lange}}}]{wesdorp_dynamical_2021}%
  \BibitemOpen
  \bibfield  {author} {\bibinfo {author} {\bibfnamefont {J.~J.}\ \bibnamefont
  {Wesdorp}}, \bibinfo {author} {\bibfnamefont {L.}~\bibnamefont
  {Gr{\"u}nhaupt}}, \bibinfo {author} {\bibfnamefont {A.}~\bibnamefont
  {Vaartjes}}, \bibinfo {author} {\bibfnamefont {M.}~\bibnamefont
  {{Pita-Vidal}}}, \bibinfo {author} {\bibfnamefont {A.}~\bibnamefont
  {Bargerbos}}, \bibinfo {author} {\bibfnamefont {L.~J.}\ \bibnamefont
  {Splitthoff}}, \bibinfo {author} {\bibfnamefont {P.}~\bibnamefont
  {Krogstrup}}, \bibinfo {author} {\bibfnamefont {B.}~\bibnamefont {{van
  Heck}}},\ and\ \bibinfo {author} {\bibfnamefont {G.}~\bibnamefont {{de
  Lange}}},\ }\href@noop {} {\bibinfo {title} {Dynamical polarization of the
  fermion parity in a nanowire {{Josephson}} junction}} (\bibinfo {year}
  {2021}{\natexlab{a}}),\ \Eprint {https://arxiv.org/abs/2112.01936}
  {arXiv:2112.01936} \BibitemShut {NoStop}%
\bibitem [{\citenamefont {Fatemi}\ \emph {et~al.}(2021)\citenamefont {Fatemi},
  \citenamefont {Kurilovich}, \citenamefont {Hays}, \citenamefont {Bouman},
  \citenamefont {Connolly}, \citenamefont {Diamond}, \citenamefont {Frattini},
  \citenamefont {Kurilovich}, \citenamefont {Krogstrup}, \citenamefont
  {Nygard}, \citenamefont {Geresdi}, \citenamefont {Glazman},\ and\
  \citenamefont {Devoret}}]{fatemi_microwave_2021-3}%
  \BibitemOpen
  \bibfield  {author} {\bibinfo {author} {\bibfnamefont {V.}~\bibnamefont
  {Fatemi}}, \bibinfo {author} {\bibfnamefont {P.~D.}\ \bibnamefont
  {Kurilovich}}, \bibinfo {author} {\bibfnamefont {M.}~\bibnamefont {Hays}},
  \bibinfo {author} {\bibfnamefont {D.}~\bibnamefont {Bouman}}, \bibinfo
  {author} {\bibfnamefont {T.}~\bibnamefont {Connolly}}, \bibinfo {author}
  {\bibfnamefont {S.}~\bibnamefont {Diamond}}, \bibinfo {author} {\bibfnamefont
  {N.~E.}\ \bibnamefont {Frattini}}, \bibinfo {author} {\bibfnamefont {V.~D.}\
  \bibnamefont {Kurilovich}}, \bibinfo {author} {\bibfnamefont
  {P.}~\bibnamefont {Krogstrup}}, \bibinfo {author} {\bibfnamefont
  {J.}~\bibnamefont {Nygard}}, \bibinfo {author} {\bibfnamefont
  {A.}~\bibnamefont {Geresdi}}, \bibinfo {author} {\bibfnamefont {L.~I.}\
  \bibnamefont {Glazman}},\ and\ \bibinfo {author} {\bibfnamefont {M.~H.}\
  \bibnamefont {Devoret}},\ }\bibfield  {title} {\bibinfo {title} {Microwave
  susceptibility observation of interacting many-body {{Andreev}} states},\
  }\href@noop {} {\bibfield  {journal} {\bibinfo  {journal} {arXiv:2112.05624
  [cond-mat]}\ } (\bibinfo {year} {2021})},\ \Eprint
  {https://arxiv.org/abs/2112.05624} {arXiv:2112.05624 [cond-mat]} \BibitemShut
  {NoStop}%
\bibitem [{\citenamefont {{Matute-Ca{\~n}adas}}\ \emph
  {et~al.}(2022)\citenamefont {{Matute-Ca{\~n}adas}}, \citenamefont {Metzger},
  \citenamefont {Park}, \citenamefont {Tosi}, \citenamefont {Krogstrup},
  \citenamefont {Nyg{\aa}rd}, \citenamefont {Goffman}, \citenamefont {Urbina},
  \citenamefont {Pothier},\ and\ \citenamefont
  {Yeyati}}]{matute-canadas_signatures_2022}%
  \BibitemOpen
  \bibfield  {author} {\bibinfo {author} {\bibfnamefont {F.~J.}\ \bibnamefont
  {{Matute-Ca{\~n}adas}}}, \bibinfo {author} {\bibfnamefont {C.}~\bibnamefont
  {Metzger}}, \bibinfo {author} {\bibfnamefont {S.}~\bibnamefont {Park}},
  \bibinfo {author} {\bibfnamefont {L.}~\bibnamefont {Tosi}}, \bibinfo {author}
  {\bibfnamefont {P.}~\bibnamefont {Krogstrup}}, \bibinfo {author}
  {\bibfnamefont {J.}~\bibnamefont {Nyg{\aa}rd}}, \bibinfo {author}
  {\bibfnamefont {M.~F.}\ \bibnamefont {Goffman}}, \bibinfo {author}
  {\bibfnamefont {C.}~\bibnamefont {Urbina}}, \bibinfo {author} {\bibfnamefont
  {H.}~\bibnamefont {Pothier}},\ and\ \bibinfo {author} {\bibfnamefont {A.~L.}\
  \bibnamefont {Yeyati}},\ }\bibfield  {title} {\bibinfo {title} {Signatures of
  {{Interactions}} in the {{Andreev Spectrum}} of {{Nanowire Josephson
  Junctions}}},\ }\href {https://doi.org/10.1103/PhysRevLett.128.197702}
  {\bibfield  {journal} {\bibinfo  {journal} {Physical Review Letters}\
  }\textbf {\bibinfo {volume} {128}},\ \bibinfo {pages} {197702} (\bibinfo
  {year} {2022})}\BibitemShut {NoStop}%
\bibitem [{\citenamefont {Gor'kov}\ and\ \citenamefont
  {Rashba}(2001)}]{gorkov_superconducting_2001}%
  \BibitemOpen
  \bibfield  {author} {\bibinfo {author} {\bibfnamefont {L.~P.}\ \bibnamefont
  {Gor'kov}}\ and\ \bibinfo {author} {\bibfnamefont {E.~I.}\ \bibnamefont
  {Rashba}},\ }\bibfield  {title} {\bibinfo {title} {Superconducting {{2D
  System}} with {{Lifted Spin Degeneracy}}: {{Mixed Singlet-Triplet State}}},\
  }\href {https://doi.org/10.1103/PhysRevLett.87.037004} {\bibfield  {journal}
  {\bibinfo  {journal} {Physical Review Letters}\ }\textbf {\bibinfo {volume}
  {87}},\ \bibinfo {pages} {037004} (\bibinfo {year} {2001})}\BibitemShut
  {NoStop}%
\bibitem [{\citenamefont {Reeg}\ and\ \citenamefont
  {Maslov}(2015)}]{reeg_proximity-induced_2015}%
  \BibitemOpen
  \bibfield  {author} {\bibinfo {author} {\bibfnamefont {C.~R.}\ \bibnamefont
  {Reeg}}\ and\ \bibinfo {author} {\bibfnamefont {D.~L.}\ \bibnamefont
  {Maslov}},\ }\bibfield  {title} {\bibinfo {title} {Proximity-induced triplet
  superconductivity in {{Rashba}} materials},\ }\href
  {https://doi.org/10.1103/PhysRevB.92.134512} {\bibfield  {journal} {\bibinfo
  {journal} {Physical Review B}\ }\textbf {\bibinfo {volume} {92}},\ \bibinfo
  {pages} {134512} (\bibinfo {year} {2015})}\BibitemShut {NoStop}%
\bibitem [{\citenamefont {Lutchyn}\ \emph {et~al.}(2010)\citenamefont
  {Lutchyn}, \citenamefont {Sau},\ and\ \citenamefont
  {Sarma}}]{lutchyn_majorana_2010}%
  \BibitemOpen
  \bibfield  {author} {\bibinfo {author} {\bibfnamefont {R.~M.}\ \bibnamefont
  {Lutchyn}}, \bibinfo {author} {\bibfnamefont {J.~D.}\ \bibnamefont {Sau}},\
  and\ \bibinfo {author} {\bibfnamefont {S.~D.}\ \bibnamefont {Sarma}},\
  }\bibfield  {title} {\bibinfo {title} {Majorana {{Fermions}} and a
  {{Topological Phase Transition}} in {{Semiconductor-Superconductor
  Heterostructures}}},\ }\href {https://doi.org/10.1103/PhysRevLett.105.077001}
  {\bibfield  {journal} {\bibinfo  {journal} {Physical Review Letters}\
  }\textbf {\bibinfo {volume} {105}},\ \bibinfo {pages} {077001} (\bibinfo
  {year} {2010})},\ \Eprint {https://arxiv.org/abs/1002.4033} {arXiv:1002.4033}
  \BibitemShut {NoStop}%
\bibitem [{\citenamefont {Oreg}\ \emph {et~al.}(2010)\citenamefont {Oreg},
  \citenamefont {Refael},\ and\ \citenamefont {{von
  Oppen}}}]{oreg_helical_2010}%
  \BibitemOpen
  \bibfield  {author} {\bibinfo {author} {\bibfnamefont {Y.}~\bibnamefont
  {Oreg}}, \bibinfo {author} {\bibfnamefont {G.}~\bibnamefont {Refael}},\ and\
  \bibinfo {author} {\bibfnamefont {F.}~\bibnamefont {{von Oppen}}},\
  }\bibfield  {title} {\bibinfo {title} {Helical liquids and {{Majorana}} bound
  states in quantum wires},\ }\href
  {https://doi.org/10.1103/PhysRevLett.105.177002} {\bibfield  {journal}
  {\bibinfo  {journal} {Physical Review Letters}\ }\textbf {\bibinfo {volume}
  {105}},\ \bibinfo {pages} {177002} (\bibinfo {year} {2010})},\ \Eprint
  {https://arxiv.org/abs/1003.1145} {arXiv:1003.1145 [cond-mat]} \BibitemShut
  {NoStop}%
\bibitem [{\citenamefont {Alicea}(2010)}]{alicea_majorana_2010}%
  \BibitemOpen
  \bibfield  {author} {\bibinfo {author} {\bibfnamefont {J.}~\bibnamefont
  {Alicea}},\ }\bibfield  {title} {\bibinfo {title} {Majorana fermions in a
  tunable semiconductor device},\ }\href
  {https://doi.org/10.1103/PhysRevB.81.125318} {\bibfield  {journal} {\bibinfo
  {journal} {Physical Review B}\ }\textbf {\bibinfo {volume} {81}},\ \bibinfo
  {pages} {125318} (\bibinfo {year} {2010})}\BibitemShut {NoStop}%
\bibitem [{\citenamefont {Potter}\ and\ \citenamefont
  {Lee}(2011)}]{potter_majorana_2011}%
  \BibitemOpen
  \bibfield  {author} {\bibinfo {author} {\bibfnamefont {A.~C.}\ \bibnamefont
  {Potter}}\ and\ \bibinfo {author} {\bibfnamefont {P.~A.}\ \bibnamefont
  {Lee}},\ }\bibfield  {title} {\bibinfo {title} {Majorana end states in
  multiband microstructures with {{Rashba}} spin-orbit coupling},\ }\href
  {https://doi.org/10.1103/PhysRevB.83.094525} {\bibfield  {journal} {\bibinfo
  {journal} {Physical Review B}\ }\textbf {\bibinfo {volume} {83}},\ \bibinfo
  {pages} {094525} (\bibinfo {year} {2011})}\BibitemShut {NoStop}%
\bibitem [{\citenamefont {Read}\ and\ \citenamefont
  {Green}(2000)}]{read_paired_2000-1}%
  \BibitemOpen
  \bibfield  {author} {\bibinfo {author} {\bibfnamefont {N.}~\bibnamefont
  {Read}}\ and\ \bibinfo {author} {\bibfnamefont {D.}~\bibnamefont {Green}},\
  }\bibfield  {title} {\bibinfo {title} {Paired states of fermions in two
  dimensions with breaking of parity and time-reversal symmetries, and the
  fractional quantum {{Hall}} effect},\ }\href
  {https://doi.org/10.1103/PhysRevB.61.10267} {\bibfield  {journal} {\bibinfo
  {journal} {Physical Review B}\ }\textbf {\bibinfo {volume} {61}},\ \bibinfo
  {pages} {10267} (\bibinfo {year} {2000})},\ \Eprint
  {https://arxiv.org/abs/cond-mat/9906453} {arXiv:cond-mat/9906453}
  \BibitemShut {NoStop}%
\bibitem [{\citenamefont {Ivanov}(2001)}]{ivanov_non-abelian_2001}%
  \BibitemOpen
  \bibfield  {author} {\bibinfo {author} {\bibfnamefont {D.~A.}\ \bibnamefont
  {Ivanov}},\ }\bibfield  {title} {\bibinfo {title} {Non-{{Abelian Statistics}}
  of {{Half-Quantum Vortices}} in p -{{Wave Superconductors}}},\ }\href
  {https://doi.org/10.1103/PhysRevLett.86.268} {\bibfield  {journal} {\bibinfo
  {journal} {Physical Review Letters}\ }\textbf {\bibinfo {volume} {86}},\
  \bibinfo {pages} {268} (\bibinfo {year} {2001})}\BibitemShut {NoStop}%
\bibitem [{\citenamefont {Kitaev}(2001)}]{kitaev_unpaired_2001}%
  \BibitemOpen
  \bibfield  {author} {\bibinfo {author} {\bibfnamefont {A.}~\bibnamefont
  {Kitaev}},\ }\bibfield  {title} {\bibinfo {title} {Unpaired {{Majorana}}
  fermions in quantum wires},\ }\href
  {https://doi.org/10.1070/1063-7869/44/10S/S29} {\bibfield  {journal}
  {\bibinfo  {journal} {Physics-Uspekhi}\ }\textbf {\bibinfo {volume} {44}},\
  \bibinfo {pages} {131} (\bibinfo {year} {2001})},\ \Eprint
  {https://arxiv.org/abs/cond-mat/0010440} {arXiv:cond-mat/0010440}
  \BibitemShut {NoStop}%
\bibitem [{\citenamefont {Krive}\ \emph {et~al.}(2004)\citenamefont {Krive},
  \citenamefont {Gorelik}, \citenamefont {Shekhter},\ and\ \citenamefont
  {Jonson}}]{krive_chiral_2004}%
  \BibitemOpen
  \bibfield  {author} {\bibinfo {author} {\bibfnamefont {I.~V.}\ \bibnamefont
  {Krive}}, \bibinfo {author} {\bibfnamefont {L.~Y.}\ \bibnamefont {Gorelik}},
  \bibinfo {author} {\bibfnamefont {R.~I.}\ \bibnamefont {Shekhter}},\ and\
  \bibinfo {author} {\bibfnamefont {M.}~\bibnamefont {Jonson}},\ }\bibfield
  {title} {\bibinfo {title} {Chiral symmetry breaking and the {{Josephson}}
  current in a ballistic superconductor\textendash quantum wire\textendash
  superconductor junction},\ }\href {https://doi.org/10.1063/1.1739160}
  {\bibfield  {journal} {\bibinfo  {journal} {Low Temperature Physics}\
  }\textbf {\bibinfo {volume} {30}},\ \bibinfo {pages} {398} (\bibinfo {year}
  {2004})}\BibitemShut {NoStop}%
\bibitem [{\citenamefont {Buzdin}(2005)}]{buzdin_proximity_2005}%
  \BibitemOpen
  \bibfield  {author} {\bibinfo {author} {\bibfnamefont {A.~I.}\ \bibnamefont
  {Buzdin}},\ }\bibfield  {title} {\bibinfo {title} {Proximity effects in
  superconductor-ferromagnet heterostructures},\ }\href
  {https://doi.org/10.1103/RevModPhys.77.935} {\bibfield  {journal} {\bibinfo
  {journal} {Reviews of Modern Physics}\ }\textbf {\bibinfo {volume} {77}},\
  \bibinfo {pages} {935} (\bibinfo {year} {2005})}\BibitemShut {NoStop}%
\bibitem [{\citenamefont {Reynoso}\ \emph {et~al.}(2008)\citenamefont
  {Reynoso}, \citenamefont {Usaj}, \citenamefont {Balseiro}, \citenamefont
  {Feinberg},\ and\ \citenamefont {Avignon}}]{reynoso_anomalous_2008}%
  \BibitemOpen
  \bibfield  {author} {\bibinfo {author} {\bibfnamefont {A.~A.}\ \bibnamefont
  {Reynoso}}, \bibinfo {author} {\bibfnamefont {G.}~\bibnamefont {Usaj}},
  \bibinfo {author} {\bibfnamefont {C.~A.}\ \bibnamefont {Balseiro}}, \bibinfo
  {author} {\bibfnamefont {D.}~\bibnamefont {Feinberg}},\ and\ \bibinfo
  {author} {\bibfnamefont {M.}~\bibnamefont {Avignon}},\ }\bibfield  {title}
  {\bibinfo {title} {Anomalous {{Josephson Current}} in {{Junctions}} with
  {{Spin Polarizing Quantum Point Contacts}}},\ }\href
  {https://doi.org/10.1103/PhysRevLett.101.107001} {\bibfield  {journal}
  {\bibinfo  {journal} {Physical Review Letters}\ }\textbf {\bibinfo {volume}
  {101}},\ \bibinfo {pages} {107001} (\bibinfo {year} {2008})}\BibitemShut
  {NoStop}%
\bibitem [{\citenamefont {Yokoyama}\ \emph {et~al.}(2014)\citenamefont
  {Yokoyama}, \citenamefont {Eto},\ and\ \citenamefont
  {Nazarov}}]{yokoyama_anomalous_2014}%
  \BibitemOpen
  \bibfield  {author} {\bibinfo {author} {\bibfnamefont {T.}~\bibnamefont
  {Yokoyama}}, \bibinfo {author} {\bibfnamefont {M.}~\bibnamefont {Eto}},\ and\
  \bibinfo {author} {\bibfnamefont {Y.~V.}\ \bibnamefont {Nazarov}},\
  }\bibfield  {title} {\bibinfo {title} {Anomalous {{Josephson}} effect induced
  by spin-orbit interaction and {{Zeeman}} effect in semiconductor nanowires},\
  }\href {https://doi.org/10.1103/PhysRevB.89.195407} {\bibfield  {journal}
  {\bibinfo  {journal} {Physical Review B}\ }\textbf {\bibinfo {volume} {89}},\
  \bibinfo {pages} {195407} (\bibinfo {year} {2014})},\ \Eprint
  {https://arxiv.org/abs/1402.0305} {arXiv:1402.0305} \BibitemShut {NoStop}%
\bibitem [{\citenamefont {Konschelle}\ \emph {et~al.}(2015)\citenamefont
  {Konschelle}, \citenamefont {Tokatly},\ and\ \citenamefont
  {Bergeret}}]{konschelle_theory_2015}%
  \BibitemOpen
  \bibfield  {author} {\bibinfo {author} {\bibfnamefont {F.}~\bibnamefont
  {Konschelle}}, \bibinfo {author} {\bibfnamefont {I.~V.}\ \bibnamefont
  {Tokatly}},\ and\ \bibinfo {author} {\bibfnamefont {F.~S.}\ \bibnamefont
  {Bergeret}},\ }\bibfield  {title} {\bibinfo {title} {Theory of the
  spin-galvanic effect and the anomalous phase shift {$\varphi_0$} in
  superconductors and {{Josephson}} junctions with intrinsic spin-orbit
  coupling},\ }\href {https://doi.org/10.1103/PhysRevB.92.125443} {\bibfield
  {journal} {\bibinfo  {journal} {Physical Review B}\ }\textbf {\bibinfo
  {volume} {92}},\ \bibinfo {pages} {125443} (\bibinfo {year}
  {2015})}\BibitemShut {NoStop}%
\bibitem [{\citenamefont {Khaire}\ \emph {et~al.}(2010)\citenamefont {Khaire},
  \citenamefont {Khasawneh}, \citenamefont {Pratt},\ and\ \citenamefont
  {Birge}}]{khaire_observation_2010}%
  \BibitemOpen
  \bibfield  {author} {\bibinfo {author} {\bibfnamefont {T.~S.}\ \bibnamefont
  {Khaire}}, \bibinfo {author} {\bibfnamefont {M.~A.}\ \bibnamefont
  {Khasawneh}}, \bibinfo {author} {\bibfnamefont {W.~P.}\ \bibnamefont
  {Pratt}},\ and\ \bibinfo {author} {\bibfnamefont {N.~O.}\ \bibnamefont
  {Birge}},\ }\bibfield  {title} {\bibinfo {title} {Observation of
  {{Spin-Triplet Superconductivity}} in {{Co-Based Josephson Junctions}}},\
  }\href {https://doi.org/10.1103/PhysRevLett.104.137002} {\bibfield  {journal}
  {\bibinfo  {journal} {Physical Review Letters}\ }\textbf {\bibinfo {volume}
  {104}},\ \bibinfo {pages} {137002} (\bibinfo {year} {2010})}\BibitemShut
  {NoStop}%
\bibitem [{\citenamefont {Robinson}\ \emph {et~al.}(2010)\citenamefont
  {Robinson}, \citenamefont {Witt},\ and\ \citenamefont
  {Blamire}}]{robinson_controlled_2010}%
  \BibitemOpen
  \bibfield  {author} {\bibinfo {author} {\bibfnamefont {J.~W.~A.}\
  \bibnamefont {Robinson}}, \bibinfo {author} {\bibfnamefont {J.~D.~S.}\
  \bibnamefont {Witt}},\ and\ \bibinfo {author} {\bibfnamefont {M.~G.}\
  \bibnamefont {Blamire}},\ }\bibfield  {title} {\bibinfo {title} {Controlled
  {{Injection}} of {{Spin-Triplet Supercurrents}} into a {{Strong
  Ferromagnet}}},\ }\href {https://doi.org/10.1126/science.1189246} {\bibfield
  {journal} {\bibinfo  {journal} {Science}\ }\textbf {\bibinfo {volume}
  {329}},\ \bibinfo {pages} {59} (\bibinfo {year} {2010})}\BibitemShut
  {NoStop}%
\bibitem [{\citenamefont {Sprungmann}\ \emph {et~al.}(2010)\citenamefont
  {Sprungmann}, \citenamefont {Westerholt}, \citenamefont {Zabel},
  \citenamefont {Weides},\ and\ \citenamefont
  {Kohlstedt}}]{sprungmann_evidence_2010}%
  \BibitemOpen
  \bibfield  {author} {\bibinfo {author} {\bibfnamefont {D.}~\bibnamefont
  {Sprungmann}}, \bibinfo {author} {\bibfnamefont {K.}~\bibnamefont
  {Westerholt}}, \bibinfo {author} {\bibfnamefont {H.}~\bibnamefont {Zabel}},
  \bibinfo {author} {\bibfnamefont {M.}~\bibnamefont {Weides}},\ and\ \bibinfo
  {author} {\bibfnamefont {H.}~\bibnamefont {Kohlstedt}},\ }\href@noop {}
  {\bibinfo {title} {Evidence for triplet superconductivity in {{Josephson}}
  junctions with ferromagnetic {{Cu}}$_{2}${{MnAl-Heusler}} barriers}}
  (\bibinfo {year} {2010}),\ \Eprint {https://arxiv.org/abs/1003.2082}
  {arXiv:1003.2082 [cond-mat]} \BibitemShut {NoStop}%
\bibitem [{\citenamefont {Linder}\ and\ \citenamefont
  {Robinson}(2015{\natexlab{a}})}]{linder_superconducting_2015-1}%
  \BibitemOpen
  \bibfield  {author} {\bibinfo {author} {\bibfnamefont {J.}~\bibnamefont
  {Linder}}\ and\ \bibinfo {author} {\bibfnamefont {J.~W.~A.}\ \bibnamefont
  {Robinson}},\ }\bibfield  {title} {\bibinfo {title} {Superconducting
  spintronics},\ }\href {https://doi.org/10.1038/nphys3242} {\bibfield
  {journal} {\bibinfo  {journal} {Nature Physics}\ }\textbf {\bibinfo {volume}
  {11}},\ \bibinfo {pages} {307} (\bibinfo {year}
  {2015}{\natexlab{a}})}\BibitemShut {NoStop}%
\bibitem [{\citenamefont {Jeon}\ \emph {et~al.}(2020)\citenamefont {Jeon},
  \citenamefont {Montiel}, \citenamefont {Komori}, \citenamefont {Ciccarelli},
  \citenamefont {Haigh}, \citenamefont {Kurebayashi}, \citenamefont {Cohen},
  \citenamefont {Chan}, \citenamefont {Stenning}, \citenamefont {Lee},
  \citenamefont {Eschrig}, \citenamefont {Blamire},\ and\ \citenamefont
  {Robinson}}]{jeon_tunable_2020}%
  \BibitemOpen
  \bibfield  {author} {\bibinfo {author} {\bibfnamefont {K.-R.}\ \bibnamefont
  {Jeon}}, \bibinfo {author} {\bibfnamefont {X.}~\bibnamefont {Montiel}},
  \bibinfo {author} {\bibfnamefont {S.}~\bibnamefont {Komori}}, \bibinfo
  {author} {\bibfnamefont {C.}~\bibnamefont {Ciccarelli}}, \bibinfo {author}
  {\bibfnamefont {J.}~\bibnamefont {Haigh}}, \bibinfo {author} {\bibfnamefont
  {H.}~\bibnamefont {Kurebayashi}}, \bibinfo {author} {\bibfnamefont {L.~F.}\
  \bibnamefont {Cohen}}, \bibinfo {author} {\bibfnamefont {A.~K.}\ \bibnamefont
  {Chan}}, \bibinfo {author} {\bibfnamefont {K.~D.}\ \bibnamefont {Stenning}},
  \bibinfo {author} {\bibfnamefont {C.-M.}\ \bibnamefont {Lee}}, \bibinfo
  {author} {\bibfnamefont {M.}~\bibnamefont {Eschrig}}, \bibinfo {author}
  {\bibfnamefont {M.~G.}\ \bibnamefont {Blamire}},\ and\ \bibinfo {author}
  {\bibfnamefont {J.~W.~A.}\ \bibnamefont {Robinson}},\ }\bibfield  {title}
  {\bibinfo {title} {Tunable {{Pure Spin Supercurrents}} and the
  {{Demonstration}} of {{Their Gateability}} in a {{Spin-Wave Device}}},\
  }\href {https://doi.org/10.1103/PhysRevX.10.031020} {\bibfield  {journal}
  {\bibinfo  {journal} {Physical Review X}\ }\textbf {\bibinfo {volume} {10}},\
  \bibinfo {pages} {031020} (\bibinfo {year} {2020})}\BibitemShut {NoStop}%
\bibitem [{\citenamefont {Cai}\ \emph {et~al.}(2021)\citenamefont {Cai},
  \citenamefont {Yao}, \citenamefont {Lv}, \citenamefont {Ma}, \citenamefont
  {Xing}, \citenamefont {Li}, \citenamefont {Ji}, \citenamefont {Zhou},
  \citenamefont {Shen}, \citenamefont {Jia}, \citenamefont {Xie}, \citenamefont
  {{\v Z}uti{\'c}}, \citenamefont {Sun},\ and\ \citenamefont
  {Han}}]{cai_evidence_2021}%
  \BibitemOpen
  \bibfield  {author} {\bibinfo {author} {\bibfnamefont {R.}~\bibnamefont
  {Cai}}, \bibinfo {author} {\bibfnamefont {Y.}~\bibnamefont {Yao}}, \bibinfo
  {author} {\bibfnamefont {P.}~\bibnamefont {Lv}}, \bibinfo {author}
  {\bibfnamefont {Y.}~\bibnamefont {Ma}}, \bibinfo {author} {\bibfnamefont
  {W.}~\bibnamefont {Xing}}, \bibinfo {author} {\bibfnamefont {B.}~\bibnamefont
  {Li}}, \bibinfo {author} {\bibfnamefont {Y.}~\bibnamefont {Ji}}, \bibinfo
  {author} {\bibfnamefont {H.}~\bibnamefont {Zhou}}, \bibinfo {author}
  {\bibfnamefont {C.}~\bibnamefont {Shen}}, \bibinfo {author} {\bibfnamefont
  {S.}~\bibnamefont {Jia}}, \bibinfo {author} {\bibfnamefont {X.~C.}\
  \bibnamefont {Xie}}, \bibinfo {author} {\bibfnamefont {I.}~\bibnamefont {{\v
  Z}uti{\'c}}}, \bibinfo {author} {\bibfnamefont {Q.-F.}\ \bibnamefont {Sun}},\
  and\ \bibinfo {author} {\bibfnamefont {W.}~\bibnamefont {Han}},\ }\bibfield
  {title} {\bibinfo {title} {Evidence for anisotropic spin-triplet {{Andreev}}
  reflection at the {{2D}} van der {{Waals}} ferromagnet/superconductor
  interface},\ }\href {https://doi.org/10.1038/s41467-021-27041-w} {\bibfield
  {journal} {\bibinfo  {journal} {Nature Communications}\ }\textbf {\bibinfo
  {volume} {12}},\ \bibinfo {pages} {6725} (\bibinfo {year}
  {2021})}\BibitemShut {NoStop}%
\bibitem [{\citenamefont {Yang}\ \emph {et~al.}(2021)\citenamefont {Yang},
  \citenamefont {Ciccarelli},\ and\ \citenamefont
  {Robinson}}]{yang_boosting_2021}%
  \BibitemOpen
  \bibfield  {author} {\bibinfo {author} {\bibfnamefont {G.}~\bibnamefont
  {Yang}}, \bibinfo {author} {\bibfnamefont {C.}~\bibnamefont {Ciccarelli}},\
  and\ \bibinfo {author} {\bibfnamefont {J.~W.~A.}\ \bibnamefont {Robinson}},\
  }\bibfield  {title} {\bibinfo {title} {Boosting spintronics with
  superconductivity},\ }\href {https://doi.org/10.1063/5.0048904} {\bibfield
  {journal} {\bibinfo  {journal} {APL Materials}\ }\textbf {\bibinfo {volume}
  {9}},\ \bibinfo {pages} {050703} (\bibinfo {year} {2021})}\BibitemShut
  {NoStop}%
\bibitem [{\citenamefont {Ahmad}\ \emph {et~al.}(2022)\citenamefont {Ahmad},
  \citenamefont {Minutillo}, \citenamefont {Capecelatro}, \citenamefont {Pal},
  \citenamefont {Caruso}, \citenamefont {Passarelli}, \citenamefont {Blamire},
  \citenamefont {Tafuri}, \citenamefont {Lucignano},\ and\ \citenamefont
  {Massarotti}}]{ahmad_coexistence_2022}%
  \BibitemOpen
  \bibfield  {author} {\bibinfo {author} {\bibfnamefont {H.~G.}\ \bibnamefont
  {Ahmad}}, \bibinfo {author} {\bibfnamefont {M.}~\bibnamefont {Minutillo}},
  \bibinfo {author} {\bibfnamefont {R.}~\bibnamefont {Capecelatro}}, \bibinfo
  {author} {\bibfnamefont {A.}~\bibnamefont {Pal}}, \bibinfo {author}
  {\bibfnamefont {R.}~\bibnamefont {Caruso}}, \bibinfo {author} {\bibfnamefont
  {G.}~\bibnamefont {Passarelli}}, \bibinfo {author} {\bibfnamefont {M.~G.}\
  \bibnamefont {Blamire}}, \bibinfo {author} {\bibfnamefont {F.}~\bibnamefont
  {Tafuri}}, \bibinfo {author} {\bibfnamefont {P.}~\bibnamefont {Lucignano}},\
  and\ \bibinfo {author} {\bibfnamefont {D.}~\bibnamefont {Massarotti}},\
  }\bibfield  {title} {\bibinfo {title} {Coexistence and tuning of spin-singlet
  and triplet transport in spin-filter {{Josephson}} junctions},\ }\href
  {https://doi.org/10.1038/s42005-021-00783-1} {\bibfield  {journal} {\bibinfo
  {journal} {Communications Physics}\ }\textbf {\bibinfo {volume} {5}},\
  \bibinfo {pages} {2} (\bibinfo {year} {2022})}\BibitemShut {NoStop}%
\bibitem [{\citenamefont {Szombati}\ \emph {et~al.}(2016)\citenamefont
  {Szombati}, \citenamefont {{Nadj-Perge}}, \citenamefont {Car}, \citenamefont
  {Plissard}, \citenamefont {Bakkers},\ and\ \citenamefont
  {Kouwenhoven}}]{szombati_josephson_2016}%
  \BibitemOpen
  \bibfield  {author} {\bibinfo {author} {\bibfnamefont {D.~B.}\ \bibnamefont
  {Szombati}}, \bibinfo {author} {\bibfnamefont {S.}~\bibnamefont
  {{Nadj-Perge}}}, \bibinfo {author} {\bibfnamefont {D.}~\bibnamefont {Car}},
  \bibinfo {author} {\bibfnamefont {S.~R.}\ \bibnamefont {Plissard}}, \bibinfo
  {author} {\bibfnamefont {E.~P. A.~M.}\ \bibnamefont {Bakkers}},\ and\
  \bibinfo {author} {\bibfnamefont {L.~P.}\ \bibnamefont {Kouwenhoven}},\
  }\bibfield  {title} {\bibinfo {title} {Josephson {$\Phi_0$}-junction in
  nanowire quantum dots},\ }\href {https://doi.org/10.1038/nphys3742}
  {\bibfield  {journal} {\bibinfo  {journal} {Nature Physics}\ }\textbf
  {\bibinfo {volume} {12}},\ \bibinfo {pages} {568} (\bibinfo {year}
  {2016})}\BibitemShut {NoStop}%
\bibitem [{\citenamefont {Strambini}\ \emph {et~al.}(2020)\citenamefont
  {Strambini}, \citenamefont {Iorio}, \citenamefont {Durante}, \citenamefont
  {Citro}, \citenamefont {{Sanz-Fern{\'a}ndez}}, \citenamefont {Guarcello},
  \citenamefont {Tokatly}, \citenamefont {Braggio}, \citenamefont {Rocci},
  \citenamefont {Ligato}, \citenamefont {Zannier}, \citenamefont {Sorba},
  \citenamefont {Bergeret},\ and\ \citenamefont
  {Giazotto}}]{strambini_josephson_2020}%
  \BibitemOpen
  \bibfield  {author} {\bibinfo {author} {\bibfnamefont {E.}~\bibnamefont
  {Strambini}}, \bibinfo {author} {\bibfnamefont {A.}~\bibnamefont {Iorio}},
  \bibinfo {author} {\bibfnamefont {O.}~\bibnamefont {Durante}}, \bibinfo
  {author} {\bibfnamefont {R.}~\bibnamefont {Citro}}, \bibinfo {author}
  {\bibfnamefont {C.}~\bibnamefont {{Sanz-Fern{\'a}ndez}}}, \bibinfo {author}
  {\bibfnamefont {C.}~\bibnamefont {Guarcello}}, \bibinfo {author}
  {\bibfnamefont {I.~V.}\ \bibnamefont {Tokatly}}, \bibinfo {author}
  {\bibfnamefont {A.}~\bibnamefont {Braggio}}, \bibinfo {author} {\bibfnamefont
  {M.}~\bibnamefont {Rocci}}, \bibinfo {author} {\bibfnamefont
  {N.}~\bibnamefont {Ligato}}, \bibinfo {author} {\bibfnamefont
  {V.}~\bibnamefont {Zannier}}, \bibinfo {author} {\bibfnamefont
  {L.}~\bibnamefont {Sorba}}, \bibinfo {author} {\bibfnamefont {F.~S.}\
  \bibnamefont {Bergeret}},\ and\ \bibinfo {author} {\bibfnamefont
  {F.}~\bibnamefont {Giazotto}},\ }\bibfield  {title} {\bibinfo {title} {A
  {{Josephson}} quantum phase battery},\ }\href@noop {} {\bibfield  {journal}
  {\bibinfo  {journal} {arXiv:2001.03393 [cond-mat, physics:quant-ph]}\ }
  (\bibinfo {year} {2020})},\ \Eprint {https://arxiv.org/abs/2001.03393}
  {arXiv:2001.03393 [cond-mat, physics:quant-ph]} \BibitemShut {NoStop}%
\bibitem [{\citenamefont {Mayer}\ \emph {et~al.}(2020)\citenamefont {Mayer},
  \citenamefont {Dartiailh}, \citenamefont {Yuan}, \citenamefont
  {Wickramasinghe}, \citenamefont {Rossi},\ and\ \citenamefont
  {Shabani}}]{mayer_gate_2020}%
  \BibitemOpen
  \bibfield  {author} {\bibinfo {author} {\bibfnamefont {W.}~\bibnamefont
  {Mayer}}, \bibinfo {author} {\bibfnamefont {M.~C.}\ \bibnamefont
  {Dartiailh}}, \bibinfo {author} {\bibfnamefont {J.}~\bibnamefont {Yuan}},
  \bibinfo {author} {\bibfnamefont {K.~S.}\ \bibnamefont {Wickramasinghe}},
  \bibinfo {author} {\bibfnamefont {E.}~\bibnamefont {Rossi}},\ and\ \bibinfo
  {author} {\bibfnamefont {J.}~\bibnamefont {Shabani}},\ }\bibfield  {title}
  {\bibinfo {title} {Gate controlled anomalous phase shift in {{Al}}/{{InAs
  Josephson}} junctions},\ }\href {https://doi.org/10.1038/s41467-019-14094-1}
  {\bibfield  {journal} {\bibinfo  {journal} {Nature Communications}\ }\textbf
  {\bibinfo {volume} {11}},\ \bibinfo {pages} {212} (\bibinfo {year}
  {2020})}\BibitemShut {NoStop}%
\bibitem [{\citenamefont {Phan}\ \emph {et~al.}(2022)\citenamefont {Phan},
  \citenamefont {Senior}, \citenamefont {Ghazaryan}, \citenamefont
  {Hatefipour}, \citenamefont {Strickland}, \citenamefont {Shabani},
  \citenamefont {Serbyn},\ and\ \citenamefont
  {Higginbotham}}]{phan_detecting_2022}%
  \BibitemOpen
  \bibfield  {author} {\bibinfo {author} {\bibfnamefont {D.}~\bibnamefont
  {Phan}}, \bibinfo {author} {\bibfnamefont {J.}~\bibnamefont {Senior}},
  \bibinfo {author} {\bibfnamefont {A.}~\bibnamefont {Ghazaryan}}, \bibinfo
  {author} {\bibfnamefont {M.}~\bibnamefont {Hatefipour}}, \bibinfo {author}
  {\bibfnamefont {W.~M.}\ \bibnamefont {Strickland}}, \bibinfo {author}
  {\bibfnamefont {J.}~\bibnamefont {Shabani}}, \bibinfo {author} {\bibfnamefont
  {M.}~\bibnamefont {Serbyn}},\ and\ \bibinfo {author} {\bibfnamefont {A.~P.}\
  \bibnamefont {Higginbotham}},\ }\bibfield  {title} {\bibinfo {title}
  {Detecting induced \$p \textbackslash pm ip\$ pairing at the {{Al-InAs}}
  interface with a quantum microwave circuit},\ }\href
  {https://doi.org/10.1103/PhysRevLett.128.107701} {\bibfield  {journal}
  {\bibinfo  {journal} {Physical Review Letters}\ }\textbf {\bibinfo {volume}
  {128}},\ \bibinfo {pages} {107701} (\bibinfo {year} {2022})},\ \Eprint
  {https://arxiv.org/abs/2107.03695} {arXiv:2107.03695 [cond-mat,
  physics:quant-ph]} \BibitemShut {NoStop}%
\bibitem [{\citenamefont {Wang}\ \emph {et~al.}(2022)\citenamefont {Wang},
  \citenamefont {Dvir}, \citenamefont {Mazur}, \citenamefont {Liu},\ and\
  \citenamefont {{van Loo}}}]{wang_singlet_2022}%
  \BibitemOpen
  \bibfield  {author} {\bibinfo {author} {\bibfnamefont {G.}~\bibnamefont
  {Wang}}, \bibinfo {author} {\bibfnamefont {T.}~\bibnamefont {Dvir}}, \bibinfo
  {author} {\bibfnamefont {G.~P.}\ \bibnamefont {Mazur}}, \bibinfo {author}
  {\bibfnamefont {C.-X.}\ \bibnamefont {Liu}},\ and\ \bibinfo {author}
  {\bibfnamefont {N.}~\bibnamefont {{van Loo}}},\ }\href@noop {} {\bibinfo
  {title} {Singlet and triplet {{Cooper}} pair splitting in
  superconducting-semiconducting hybrid nanowires}} (\bibinfo {year}
  {2022})\BibitemShut {NoStop}%
\bibitem [{\citenamefont {Glazman}\ and\ \citenamefont
  {Catelani}(2021)}]{glazman_bogoliubov_2021}%
  \BibitemOpen
  \bibfield  {author} {\bibinfo {author} {\bibfnamefont {L.~I.}\ \bibnamefont
  {Glazman}}\ and\ \bibinfo {author} {\bibfnamefont {G.}~\bibnamefont
  {Catelani}},\ }\bibfield  {title} {\bibinfo {title} {Bogoliubov
  {{Quasiparticles}} in {{Superconducting Qubits}}},\ }\href
  {https://doi.org/10.21468/SciPostPhysLectNotes.31} {\bibfield  {journal}
  {\bibinfo  {journal} {SciPost Physics Lecture Notes}\ ,\ \bibinfo {pages}
  {31}} (\bibinfo {year} {2021})},\ \Eprint {https://arxiv.org/abs/2003.04366}
  {arXiv:2003.04366} \BibitemShut {NoStop}%
\bibitem [{\citenamefont {Wesdorp}(2022)}]{wesdorp_supplementary_2022}%
  \BibitemOpen
  \bibfield  {author} {\bibinfo {author} {\bibfnamefont {J.}~\bibnamefont
  {Wesdorp}, \bibfnamefont {J}},\ }\href@noop {} {\bibinfo {title}
  {Supplementary information to: {{Microwave}} spectroscopy of {{Andreev}}
  bound states in a magnetic field}} (\bibinfo {year} {2022})\BibitemShut
  {NoStop}%
\bibitem [{\citenamefont {{Pita-Vidal}}\ \emph {et~al.}(2020)\citenamefont
  {{Pita-Vidal}}, \citenamefont {Bargerbos}, \citenamefont {Yang},
  \citenamefont {{van Woerkom}}, \citenamefont {Pfaff}, \citenamefont {Haider},
  \citenamefont {Krogstrup}, \citenamefont {Kouwenhoven}, \citenamefont {{de
  Lange}},\ and\ \citenamefont {Kou}}]{pita-vidal_gate-tunable_2020-1}%
  \BibitemOpen
  \bibfield  {author} {\bibinfo {author} {\bibfnamefont {M.}~\bibnamefont
  {{Pita-Vidal}}}, \bibinfo {author} {\bibfnamefont {A.}~\bibnamefont
  {Bargerbos}}, \bibinfo {author} {\bibfnamefont {C.-K.}\ \bibnamefont {Yang}},
  \bibinfo {author} {\bibfnamefont {D.~J.}\ \bibnamefont {{van Woerkom}}},
  \bibinfo {author} {\bibfnamefont {W.}~\bibnamefont {Pfaff}}, \bibinfo
  {author} {\bibfnamefont {N.}~\bibnamefont {Haider}}, \bibinfo {author}
  {\bibfnamefont {P.}~\bibnamefont {Krogstrup}}, \bibinfo {author}
  {\bibfnamefont {L.~P.}\ \bibnamefont {Kouwenhoven}}, \bibinfo {author}
  {\bibfnamefont {G.}~\bibnamefont {{de Lange}}},\ and\ \bibinfo {author}
  {\bibfnamefont {A.}~\bibnamefont {Kou}},\ }\bibfield  {title} {\bibinfo
  {title} {A gate-tunable, field-compatible fluxonium},\ }\href
  {https://doi.org/10.1103/PhysRevApplied.14.064038} {\bibfield  {journal}
  {\bibinfo  {journal} {Physical Review Applied}\ }\textbf {\bibinfo {volume}
  {14}},\ \bibinfo {pages} {064038} (\bibinfo {year} {2020})},\ \Eprint
  {https://arxiv.org/abs/1910.07978} {arXiv:1910.07978} \BibitemShut {NoStop}%
\bibitem [{\citenamefont {Annunziata}\ \emph {et~al.}(2010)\citenamefont
  {Annunziata}, \citenamefont {Santavicca}, \citenamefont {Frunzio},
  \citenamefont {Catelani}, \citenamefont {Rooks}, \citenamefont {Frydman},\
  and\ \citenamefont {Prober}}]{annunziata_tunable_2010}%
  \BibitemOpen
  \bibfield  {author} {\bibinfo {author} {\bibfnamefont {A.~J.}\ \bibnamefont
  {Annunziata}}, \bibinfo {author} {\bibfnamefont {D.~F.}\ \bibnamefont
  {Santavicca}}, \bibinfo {author} {\bibfnamefont {L.}~\bibnamefont {Frunzio}},
  \bibinfo {author} {\bibfnamefont {G.}~\bibnamefont {Catelani}}, \bibinfo
  {author} {\bibfnamefont {M.~J.}\ \bibnamefont {Rooks}}, \bibinfo {author}
  {\bibfnamefont {A.}~\bibnamefont {Frydman}},\ and\ \bibinfo {author}
  {\bibfnamefont {D.~E.}\ \bibnamefont {Prober}},\ }\bibfield  {title}
  {\bibinfo {title} {Tunable superconducting nanoinductors},\ }\href
  {https://doi.org/10.1088/0957-4484/21/44/445202} {\bibfield  {journal}
  {\bibinfo  {journal} {Nanotechnology}\ }\textbf {\bibinfo {volume} {21}},\
  \bibinfo {pages} {445202} (\bibinfo {year} {2010})}\BibitemShut {NoStop}%
\bibitem [{\citenamefont {Kroll}\ \emph {et~al.}(2019)\citenamefont {Kroll},
  \citenamefont {Borsoi}, \citenamefont {{van der Enden}}, \citenamefont
  {Uilhoorn}, \citenamefont {{de Jong}}, \citenamefont {{Quintero-P{\'e}rez}},
  \citenamefont {{van Woerkom}}, \citenamefont {Bruno}, \citenamefont
  {Plissard}, \citenamefont {Car}, \citenamefont {Bakkers}, \citenamefont
  {Cassidy},\ and\ \citenamefont {Kouwenhoven}}]{kroll_magnetic_2019-2}%
  \BibitemOpen
  \bibfield  {author} {\bibinfo {author} {\bibfnamefont {J.~G.}\ \bibnamefont
  {Kroll}}, \bibinfo {author} {\bibfnamefont {F.}~\bibnamefont {Borsoi}},
  \bibinfo {author} {\bibfnamefont {K.~L.}\ \bibnamefont {{van der Enden}}},
  \bibinfo {author} {\bibfnamefont {W.}~\bibnamefont {Uilhoorn}}, \bibinfo
  {author} {\bibfnamefont {D.}~\bibnamefont {{de Jong}}}, \bibinfo {author}
  {\bibfnamefont {M.}~\bibnamefont {{Quintero-P{\'e}rez}}}, \bibinfo {author}
  {\bibfnamefont {D.~J.}\ \bibnamefont {{van Woerkom}}}, \bibinfo {author}
  {\bibfnamefont {A.}~\bibnamefont {Bruno}}, \bibinfo {author} {\bibfnamefont
  {S.~R.}\ \bibnamefont {Plissard}}, \bibinfo {author} {\bibfnamefont
  {D.}~\bibnamefont {Car}}, \bibinfo {author} {\bibfnamefont {E.~P. A.~M.}\
  \bibnamefont {Bakkers}}, \bibinfo {author} {\bibfnamefont {M.~C.}\
  \bibnamefont {Cassidy}},\ and\ \bibinfo {author} {\bibfnamefont {L.~P.}\
  \bibnamefont {Kouwenhoven}},\ }\bibfield  {title} {\bibinfo {title} {Magnetic
  field resilient superconducting coplanar waveguide resonators for hybrid
  {{cQED}} experiments},\ }\href
  {https://doi.org/10.1103/PhysRevApplied.11.064053} {\bibfield  {journal}
  {\bibinfo  {journal} {Physical Review Applied}\ }\textbf {\bibinfo {volume}
  {11}},\ \bibinfo {pages} {064053} (\bibinfo {year} {2019})},\ \Eprint
  {https://arxiv.org/abs/1809.03932} {arXiv:1809.03932} \BibitemShut {NoStop}%
\bibitem [{Note1()}]{Note1}%
  \BibitemOpen
  \bibinfo {note} {We found that holes closer spaced to the edges reduced the
  flux jumps significantly, compared to a $\SI {1}{\micro \meter }$ spacing
  used in Ref.~\cite {kroll_magnetic_2019-2}}\BibitemShut {NoStop}%
\bibitem [{\citenamefont {Chang}\ \emph {et~al.}(2015)\citenamefont {Chang},
  \citenamefont {Albrecht}, \citenamefont {Jespersen}, \citenamefont
  {Kuemmeth}, \citenamefont {Krogstrup}, \citenamefont {Nyg{\aa}rd},\ and\
  \citenamefont {Marcus}}]{chang_hard_2015}%
  \BibitemOpen
  \bibfield  {author} {\bibinfo {author} {\bibfnamefont {W.}~\bibnamefont
  {Chang}}, \bibinfo {author} {\bibfnamefont {S.~M.}\ \bibnamefont {Albrecht}},
  \bibinfo {author} {\bibfnamefont {T.~S.}\ \bibnamefont {Jespersen}}, \bibinfo
  {author} {\bibfnamefont {F.}~\bibnamefont {Kuemmeth}}, \bibinfo {author}
  {\bibfnamefont {P.}~\bibnamefont {Krogstrup}}, \bibinfo {author}
  {\bibfnamefont {J.}~\bibnamefont {Nyg{\aa}rd}},\ and\ \bibinfo {author}
  {\bibfnamefont {C.~M.}\ \bibnamefont {Marcus}},\ }\bibfield  {title}
  {\bibinfo {title} {Hard gap in epitaxial semiconductor-superconductor
  nanowires},\ }\href {https://doi.org/10.1038/nnano.2014.306} {\bibfield
  {journal} {\bibinfo  {journal} {Nature Nanotechnology}\ }\textbf {\bibinfo
  {volume} {10}},\ \bibinfo {pages} {232} (\bibinfo {year} {2015})},\ \Eprint
  {https://arxiv.org/abs/1411.6255} {arXiv:1411.6255} \BibitemShut {NoStop}%
\bibitem [{\citenamefont {Benfenati}\ \emph {et~al.}(2020)\citenamefont
  {Benfenati}, \citenamefont {Maiani}, \citenamefont {Rybakov},\ and\
  \citenamefont {Babaev}}]{benfenati_vortex_2020}%
  \BibitemOpen
  \bibfield  {author} {\bibinfo {author} {\bibfnamefont {A.}~\bibnamefont
  {Benfenati}}, \bibinfo {author} {\bibfnamefont {A.}~\bibnamefont {Maiani}},
  \bibinfo {author} {\bibfnamefont {F.~N.}\ \bibnamefont {Rybakov}},\ and\
  \bibinfo {author} {\bibfnamefont {E.}~\bibnamefont {Babaev}},\ }\bibfield
  {title} {\bibinfo {title} {Vortex nucleation barrier in superconductors
  beyond the {{Bean-Livingston}} approximation: {{A}} numerical approach for
  the sphaleron problem in a gauge theory},\ }\href
  {https://doi.org/10.1103/PhysRevB.101.220505} {\bibfield  {journal} {\bibinfo
   {journal} {Physical Review B}\ }\textbf {\bibinfo {volume} {101}},\ \bibinfo
  {pages} {220505} (\bibinfo {year} {2020})},\ \Eprint
  {https://arxiv.org/abs/1911.09513} {arXiv:1911.09513 [cond-mat]} \BibitemShut
  {NoStop}%
\bibitem [{\citenamefont {Tinkham}(2015)}]{tinkham_introduction_2015}%
  \BibitemOpen
  \bibfield  {author} {\bibinfo {author} {\bibfnamefont {M.}~\bibnamefont
  {Tinkham}},\ }\href@noop {} {\emph {\bibinfo {title} {Introduction to
  Superconductivity}}},\ \bibinfo {edition} {2nd}\ ed.,\ Dover Books on
  Physics\ (\bibinfo  {publisher} {{Dover Publ}},\ \bibinfo {address}
  {{Mineola, NY}},\ \bibinfo {year} {2015})\BibitemShut {NoStop}%
\bibitem [{\citenamefont {Zazunov}\ \emph {et~al.}(2003)\citenamefont
  {Zazunov}, \citenamefont {Shumeiko}, \citenamefont {Bratus'}, \citenamefont
  {Lantz},\ and\ \citenamefont {Wendin}}]{zazunov_andreev_2003}%
  \BibitemOpen
  \bibfield  {author} {\bibinfo {author} {\bibfnamefont {A.}~\bibnamefont
  {Zazunov}}, \bibinfo {author} {\bibfnamefont {V.~S.}\ \bibnamefont
  {Shumeiko}}, \bibinfo {author} {\bibfnamefont {E.~N.}\ \bibnamefont
  {Bratus'}}, \bibinfo {author} {\bibfnamefont {J.}~\bibnamefont {Lantz}},\
  and\ \bibinfo {author} {\bibfnamefont {G.}~\bibnamefont {Wendin}},\
  }\bibfield  {title} {\bibinfo {title} {Andreev {{Level Qubit}}},\ }\href
  {https://doi.org/10.1103/PhysRevLett.90.087003} {\bibfield  {journal}
  {\bibinfo  {journal} {Physical Review Letters}\ }\textbf {\bibinfo {volume}
  {90}},\ \bibinfo {pages} {087003} (\bibinfo {year} {2003})},\ \Eprint
  {https://arxiv.org/abs/cond-mat/0206342} {arXiv:cond-mat/0206342}
  \BibitemShut {NoStop}%
\bibitem [{\citenamefont {Bretheau}(2013)}]{bretheau_localized_2013}%
  \BibitemOpen
  \bibfield  {author} {\bibinfo {author} {\bibfnamefont {L.}~\bibnamefont
  {Bretheau}},\ }\emph {\bibinfo {title} {Localized {{Excitations}} in
  {{Superconducting Atomic Contacts}}: {{Probing}} the {{Andreev}} Doublet}},\
  \href@noop {} {Ph.D. thesis} (\bibinfo {year} {2013})\BibitemShut {NoStop}%
\bibitem [{\citenamefont {Park}\ \emph {et~al.}(2020)\citenamefont {Park},
  \citenamefont {Metzger}, \citenamefont {Tosi}, \citenamefont {Goffman},
  \citenamefont {Urbina}, \citenamefont {Pothier},\ and\ \citenamefont
  {Yeyati}}]{park_adiabatic_2020-1}%
  \BibitemOpen
  \bibfield  {author} {\bibinfo {author} {\bibfnamefont {S.}~\bibnamefont
  {Park}}, \bibinfo {author} {\bibfnamefont {C.}~\bibnamefont {Metzger}},
  \bibinfo {author} {\bibfnamefont {L.}~\bibnamefont {Tosi}}, \bibinfo {author}
  {\bibfnamefont {M.~F.}\ \bibnamefont {Goffman}}, \bibinfo {author}
  {\bibfnamefont {C.}~\bibnamefont {Urbina}}, \bibinfo {author} {\bibfnamefont
  {H.}~\bibnamefont {Pothier}},\ and\ \bibinfo {author} {\bibfnamefont {A.~L.}\
  \bibnamefont {Yeyati}},\ }\bibfield  {title} {\bibinfo {title} {From
  adiabatic to dispersive readout of quantum circuits},\ }\href
  {https://doi.org/10.1103/PhysRevLett.125.077701} {\bibfield  {journal}
  {\bibinfo  {journal} {Physical Review Letters}\ }\textbf {\bibinfo {volume}
  {125}},\ \bibinfo {pages} {077701} (\bibinfo {year} {2020})},\ \Eprint
  {https://arxiv.org/abs/2007.05030} {arXiv:2007.05030} \BibitemShut {NoStop}%
\bibitem [{\citenamefont {Doh}(2005)}]{doh_tunable_2005}%
  \BibitemOpen
  \bibfield  {author} {\bibinfo {author} {\bibfnamefont {Y.-J.}\ \bibnamefont
  {Doh}},\ }\bibfield  {title} {\bibinfo {title} {Tunable {{Supercurrent
  Through Semiconductor Nanowires}}},\ }\href
  {https://doi.org/10.1126/science.1113523} {\bibfield  {journal} {\bibinfo
  {journal} {Science}\ }\textbf {\bibinfo {volume} {309}},\ \bibinfo {pages}
  {272} (\bibinfo {year} {2005})}\BibitemShut {NoStop}%
\bibitem [{\citenamefont {Goffman}\ \emph {et~al.}(2017)\citenamefont
  {Goffman}, \citenamefont {Urbina}, \citenamefont {Pothier}, \citenamefont
  {Nyg{\aa}rd}, \citenamefont {Marcus},\ and\ \citenamefont
  {Krogstrup}}]{goffman_conduction_2017}%
  \BibitemOpen
  \bibfield  {author} {\bibinfo {author} {\bibfnamefont {M.~F.}\ \bibnamefont
  {Goffman}}, \bibinfo {author} {\bibfnamefont {C.}~\bibnamefont {Urbina}},
  \bibinfo {author} {\bibfnamefont {H.}~\bibnamefont {Pothier}}, \bibinfo
  {author} {\bibfnamefont {J.}~\bibnamefont {Nyg{\aa}rd}}, \bibinfo {author}
  {\bibfnamefont {C.~M.}\ \bibnamefont {Marcus}},\ and\ \bibinfo {author}
  {\bibfnamefont {P.}~\bibnamefont {Krogstrup}},\ }\bibfield  {title} {\bibinfo
  {title} {Conduction channels of an {{InAs-Al}} nanowire {{Josephson}} weak
  link},\ }\href {https://doi.org/10.1088/1367-2630/aa7641} {\bibfield
  {journal} {\bibinfo  {journal} {New Journal of Physics}\ }\textbf {\bibinfo
  {volume} {19}},\ \bibinfo {pages} {092002} (\bibinfo {year}
  {2017})}\BibitemShut {NoStop}%
\bibitem [{Note2()}]{Note2}%
  \BibitemOpen
  \bibinfo {note} {Note that the labels can refer either to manifolds that
  originate from the same transverse sub-band, due to finite-length effects, or
  to orbitals from different transverse sub-bands}\BibitemShut {NoStop}%
\bibitem [{\citenamefont {Park}\ and\ \citenamefont
  {Yeyati}(2017)}]{park_andreev_2017}%
  \BibitemOpen
  \bibfield  {author} {\bibinfo {author} {\bibfnamefont {S.}~\bibnamefont
  {Park}}\ and\ \bibinfo {author} {\bibfnamefont {A.~L.}\ \bibnamefont
  {Yeyati}},\ }\bibfield  {title} {\bibinfo {title} {Andreev spin qubits in
  multichannel {{Rashba}} nanowires},\ }\href
  {https://doi.org/10.1103/PhysRevB.96.125416} {\bibfield  {journal} {\bibinfo
  {journal} {Physical Review B}\ }\textbf {\bibinfo {volume} {96}},\ \bibinfo
  {pages} {125416} (\bibinfo {year} {2017})},\ \Eprint
  {https://arxiv.org/abs/1707.04273} {arXiv:1707.04273} \BibitemShut {NoStop}%
\bibitem [{\citenamefont {Governale}\ and\ \citenamefont
  {Z{\"u}licke}(2002)}]{governale_spin_2002}%
  \BibitemOpen
  \bibfield  {author} {\bibinfo {author} {\bibfnamefont {M.}~\bibnamefont
  {Governale}}\ and\ \bibinfo {author} {\bibfnamefont {U.}~\bibnamefont
  {Z{\"u}licke}},\ }\bibfield  {title} {\bibinfo {title} {Spin accumulation in
  quantum wires with strong {{Rashba}} spin-orbit coupling},\ }\href
  {https://doi.org/10.1103/PhysRevB.66.073311} {\bibfield  {journal} {\bibinfo
  {journal} {Physical Review B}\ }\textbf {\bibinfo {volume} {66}},\ \bibinfo
  {pages} {073311} (\bibinfo {year} {2002})}\BibitemShut {NoStop}%
\bibitem [{\citenamefont {Chtchelkatchev}\ and\ \citenamefont
  {Nazarov}(2003)}]{chtchelkatchev_andreev_2003}%
  \BibitemOpen
  \bibfield  {author} {\bibinfo {author} {\bibfnamefont {N.~M.}\ \bibnamefont
  {Chtchelkatchev}}\ and\ \bibinfo {author} {\bibfnamefont {Y.~V.}\
  \bibnamefont {Nazarov}},\ }\bibfield  {title} {\bibinfo {title} {Andreev
  {{Quantum Dots}} for {{Spin Manipulation}}},\ }\href
  {https://doi.org/10.1103/PhysRevLett.90.226806} {\bibfield  {journal}
  {\bibinfo  {journal} {Physical Review Letters}\ }\textbf {\bibinfo {volume}
  {90}},\ \bibinfo {pages} {226806} (\bibinfo {year} {2003})}\BibitemShut
  {NoStop}%
\bibitem [{\citenamefont {B\'eri}\ \emph {et~al.}(2008)\citenamefont {B\'eri},
  \citenamefont {Bardarson},\ and\ \citenamefont {Beenakker}}]{beri2008}%
  \BibitemOpen
  \bibfield  {author} {\bibinfo {author} {\bibfnamefont {B.}~\bibnamefont
  {B\'eri}}, \bibinfo {author} {\bibfnamefont {J.~H.}\ \bibnamefont
  {Bardarson}},\ and\ \bibinfo {author} {\bibfnamefont {C.~W.~J.}\ \bibnamefont
  {Beenakker}},\ }\bibfield  {title} {\bibinfo {title} {Splitting of andreev
  levels in a josephson junction by spin-orbit coupling},\ }\href
  {https://doi.org/10.1103/PhysRevB.77.045311} {\bibfield  {journal} {\bibinfo
  {journal} {Phys. Rev. B}\ }\textbf {\bibinfo {volume} {77}},\ \bibinfo
  {pages} {045311} (\bibinfo {year} {2008})}\BibitemShut {NoStop}%
\bibitem [{\citenamefont {Yokoyama}\ \emph {et~al.}(2013)\citenamefont
  {Yokoyama}, \citenamefont {Eto},\ and\ \citenamefont
  {Nazarov}}]{yokoyama_josephson_2013-1}%
  \BibitemOpen
  \bibfield  {author} {\bibinfo {author} {\bibfnamefont {T.}~\bibnamefont
  {Yokoyama}}, \bibinfo {author} {\bibfnamefont {M.}~\bibnamefont {Eto}},\ and\
  \bibinfo {author} {\bibfnamefont {Y.~V.}\ \bibnamefont {Nazarov}},\
  }\bibfield  {title} {\bibinfo {title} {Josephson {{Current}} through
  {{Semiconductor Nanowire}} with {{Spin-Orbit Interaction}} in {{Magnetic
  Field}}},\ }\href {https://doi.org/10.7566/JPSJ.82.054703} {\bibfield
  {journal} {\bibinfo  {journal} {Journal of the Physical Society of Japan}\
  }\textbf {\bibinfo {volume} {82}},\ \bibinfo {pages} {054703} (\bibinfo
  {year} {2013})},\ \Eprint {https://arxiv.org/abs/1212.5390} {arXiv:1212.5390
  [cond-mat]} \BibitemShut {NoStop}%
\bibitem [{\citenamefont {Konschelle}\ \emph {et~al.}(2016)\citenamefont
  {Konschelle}, \citenamefont {Bergeret},\ and\ \citenamefont
  {Tokatly}}]{konschelle_semiclassical_2016}%
  \BibitemOpen
  \bibfield  {author} {\bibinfo {author} {\bibfnamefont {F.}~\bibnamefont
  {Konschelle}}, \bibinfo {author} {\bibfnamefont {F.~S.}\ \bibnamefont
  {Bergeret}},\ and\ \bibinfo {author} {\bibfnamefont {I.~V.}\ \bibnamefont
  {Tokatly}},\ }\bibfield  {title} {\bibinfo {title} {Semiclassical
  {{Quantization}} of {{Spinning Quasiparticles}} in {{Ballistic Josephson
  Junctions}}},\ }\href {https://doi.org/10.1103/PhysRevLett.116.237002}
  {\bibfield  {journal} {\bibinfo  {journal} {Physical Review Letters}\
  }\textbf {\bibinfo {volume} {116}},\ \bibinfo {pages} {237002} (\bibinfo
  {year} {2016})}\BibitemShut {NoStop}%
\bibitem [{\citenamefont {Kurland}\ \emph {et~al.}(2000)\citenamefont
  {Kurland}, \citenamefont {Aleiner},\ and\ \citenamefont
  {Altshuler}}]{kurland_mesoscopic_2000}%
  \BibitemOpen
  \bibfield  {author} {\bibinfo {author} {\bibfnamefont {I.~L.}\ \bibnamefont
  {Kurland}}, \bibinfo {author} {\bibfnamefont {I.~L.}\ \bibnamefont
  {Aleiner}},\ and\ \bibinfo {author} {\bibfnamefont {B.~L.}\ \bibnamefont
  {Altshuler}},\ }\bibfield  {title} {\bibinfo {title} {Mesoscopic
  magnetization fluctuations for metallic grains close to the {{Stoner}}
  instability},\ }\href {https://doi.org/10.1103/PhysRevB.62.14886} {\bibfield
  {journal} {\bibinfo  {journal} {Physical Review B}\ }\textbf {\bibinfo
  {volume} {62}},\ \bibinfo {pages} {14886} (\bibinfo {year}
  {2000})}\BibitemShut {NoStop}%
\bibitem [{\citenamefont {Bommer}\ \emph {et~al.}(2019)\citenamefont {Bommer},
  \citenamefont {Zhang}, \citenamefont {G{\"u}l}, \citenamefont {Nijholt},
  \citenamefont {Wimmer}, \citenamefont {Rybakov}, \citenamefont {Garaud},
  \citenamefont {Rodic}, \citenamefont {Babaev}, \citenamefont {Troyer},
  \citenamefont {Car}, \citenamefont {Plissard}, \citenamefont {Bakkers},
  \citenamefont {Watanabe}, \citenamefont {Taniguchi},\ and\ \citenamefont
  {Kouwenhoven}}]{bommer_spin-orbit_2019}%
  \BibitemOpen
  \bibfield  {author} {\bibinfo {author} {\bibfnamefont {J.~D.~S.}\
  \bibnamefont {Bommer}}, \bibinfo {author} {\bibfnamefont {H.}~\bibnamefont
  {Zhang}}, \bibinfo {author} {\bibfnamefont {{\"O}.}~\bibnamefont {G{\"u}l}},
  \bibinfo {author} {\bibfnamefont {B.}~\bibnamefont {Nijholt}}, \bibinfo
  {author} {\bibfnamefont {M.}~\bibnamefont {Wimmer}}, \bibinfo {author}
  {\bibfnamefont {F.~N.}\ \bibnamefont {Rybakov}}, \bibinfo {author}
  {\bibfnamefont {J.}~\bibnamefont {Garaud}}, \bibinfo {author} {\bibfnamefont
  {D.}~\bibnamefont {Rodic}}, \bibinfo {author} {\bibfnamefont
  {E.}~\bibnamefont {Babaev}}, \bibinfo {author} {\bibfnamefont
  {M.}~\bibnamefont {Troyer}}, \bibinfo {author} {\bibfnamefont
  {D.}~\bibnamefont {Car}}, \bibinfo {author} {\bibfnamefont {S.~R.}\
  \bibnamefont {Plissard}}, \bibinfo {author} {\bibfnamefont {E.~P. A.~M.}\
  \bibnamefont {Bakkers}}, \bibinfo {author} {\bibfnamefont {K.}~\bibnamefont
  {Watanabe}}, \bibinfo {author} {\bibfnamefont {T.}~\bibnamefont
  {Taniguchi}},\ and\ \bibinfo {author} {\bibfnamefont {L.~P.}\ \bibnamefont
  {Kouwenhoven}},\ }\bibfield  {title} {\bibinfo {title} {Spin-{{Orbit
  Protection}} of {{Induced Superconductivity}} in {{Majorana Nanowires}}},\
  }\href {https://doi.org/10.1103/PhysRevLett.122.187702} {\bibfield  {journal}
  {\bibinfo  {journal} {Physical Review Letters}\ }\textbf {\bibinfo {volume}
  {122}},\ \bibinfo {pages} {187702} (\bibinfo {year} {2019})}\BibitemShut
  {NoStop}%
\bibitem [{\citenamefont {{de Moor}}\ \emph {et~al.}(2018)\citenamefont {{de
  Moor}}, \citenamefont {Bommer}, \citenamefont {Xu}, \citenamefont {Winkler},
  \citenamefont {Antipov}, \citenamefont {Bargerbos}, \citenamefont {Wang},
  \citenamefont {van Loo}, \citenamefont {{Op het Veld}}, \citenamefont
  {Gazibegovic}, \citenamefont {Car}, \citenamefont {Logan}, \citenamefont
  {Pendharkar}, \citenamefont {Lee}, \citenamefont {M~Bakkers}, \citenamefont
  {Palmstr{\o}m}, \citenamefont {Lutchyn}, \citenamefont {Kouwenhoven},\ and\
  \citenamefont {Zhang}}]{de_moor_electric_2018}%
  \BibitemOpen
  \bibfield  {author} {\bibinfo {author} {\bibfnamefont {M.~W.~A.}\
  \bibnamefont {{de Moor}}}, \bibinfo {author} {\bibfnamefont {J.~D.~S.}\
  \bibnamefont {Bommer}}, \bibinfo {author} {\bibfnamefont {D.}~\bibnamefont
  {Xu}}, \bibinfo {author} {\bibfnamefont {G.~W.}\ \bibnamefont {Winkler}},
  \bibinfo {author} {\bibfnamefont {A.~E.}\ \bibnamefont {Antipov}}, \bibinfo
  {author} {\bibfnamefont {A.}~\bibnamefont {Bargerbos}}, \bibinfo {author}
  {\bibfnamefont {G.}~\bibnamefont {Wang}}, \bibinfo {author} {\bibfnamefont
  {N.}~\bibnamefont {van Loo}}, \bibinfo {author} {\bibfnamefont {R.~L.~M.}\
  \bibnamefont {{Op het Veld}}}, \bibinfo {author} {\bibfnamefont
  {S.}~\bibnamefont {Gazibegovic}}, \bibinfo {author} {\bibfnamefont
  {D.}~\bibnamefont {Car}}, \bibinfo {author} {\bibfnamefont {J.~A.}\
  \bibnamefont {Logan}}, \bibinfo {author} {\bibfnamefont {M.}~\bibnamefont
  {Pendharkar}}, \bibinfo {author} {\bibfnamefont {J.~S.}\ \bibnamefont {Lee}},
  \bibinfo {author} {\bibfnamefont {E.~P.~A.}\ \bibnamefont {M~Bakkers}},
  \bibinfo {author} {\bibfnamefont {C.~J.}\ \bibnamefont {Palmstr{\o}m}},
  \bibinfo {author} {\bibfnamefont {R.~M.}\ \bibnamefont {Lutchyn}}, \bibinfo
  {author} {\bibfnamefont {L.~P.}\ \bibnamefont {Kouwenhoven}},\ and\ \bibinfo
  {author} {\bibfnamefont {H.}~\bibnamefont {Zhang}},\ }\bibfield  {title}
  {\bibinfo {title} {Electric field tunable superconductor-semiconductor
  coupling in {{Majorana}} nanowires},\ }\href
  {https://doi.org/10.1088/1367-2630/aae61d} {\bibfield  {journal} {\bibinfo
  {journal} {New Journal of Physics}\ }\textbf {\bibinfo {volume} {20}},\
  \bibinfo {pages} {103049} (\bibinfo {year} {2018})}\BibitemShut {NoStop}%
\bibitem [{\citenamefont {Bargerbos}\ \emph {et~al.}(2022)\citenamefont
  {Bargerbos}, \citenamefont {Pita-Vidal}, \citenamefont
  {\ifmmode~\check{Z}\else \v{Z}\fi{}itko}, \citenamefont {\'Avila},
  \citenamefont {Splitthoff}, \citenamefont {Gr\"unhaupt}, \citenamefont
  {Wesdorp}, \citenamefont {Andersen}, \citenamefont {Liu}, \citenamefont
  {Kouwenhoven}, \citenamefont {Aguado}, \citenamefont {Kou},\ and\
  \citenamefont {van Heck}}]{bargerbos_singlet_doublet_2022}%
  \BibitemOpen
  \bibfield  {author} {\bibinfo {author} {\bibfnamefont {A.}~\bibnamefont
  {Bargerbos}}, \bibinfo {author} {\bibfnamefont {M.}~\bibnamefont
  {Pita-Vidal}}, \bibinfo {author} {\bibfnamefont {R.}~\bibnamefont
  {\ifmmode~\check{Z}\else \v{Z}\fi{}itko}}, \bibinfo {author} {\bibfnamefont
  {J.}~\bibnamefont {\'Avila}}, \bibinfo {author} {\bibfnamefont {L.~J.}\
  \bibnamefont {Splitthoff}}, \bibinfo {author} {\bibfnamefont
  {L.}~\bibnamefont {Gr\"unhaupt}}, \bibinfo {author} {\bibfnamefont {J.~J.}\
  \bibnamefont {Wesdorp}}, \bibinfo {author} {\bibfnamefont {C.~K.}\
  \bibnamefont {Andersen}}, \bibinfo {author} {\bibfnamefont {Y.}~\bibnamefont
  {Liu}}, \bibinfo {author} {\bibfnamefont {L.~P.}\ \bibnamefont
  {Kouwenhoven}}, \bibinfo {author} {\bibfnamefont {R.}~\bibnamefont {Aguado}},
  \bibinfo {author} {\bibfnamefont {A.}~\bibnamefont {Kou}},\ and\ \bibinfo
  {author} {\bibfnamefont {B.}~\bibnamefont {van Heck}},\ }\bibfield  {title}
  {\bibinfo {title} {Singlet-doublet transitions of a quantum dot josephson
  junction detected in a transmon circuit},\ }\href
  {https://doi.org/10.1103/PRXQuantum.3.030311} {\bibfield  {journal} {\bibinfo
   {journal} {PRX Quantum}\ }\textbf {\bibinfo {volume} {3}},\ \bibinfo {pages}
  {030311} (\bibinfo {year} {2022})}\BibitemShut {NoStop}%
\bibitem [{\citenamefont {Liang}\ and\ \citenamefont
  {Gao}(2012)}]{liang_strong_2012}%
  \BibitemOpen
  \bibfield  {author} {\bibinfo {author} {\bibfnamefont {D.}~\bibnamefont
  {Liang}}\ and\ \bibinfo {author} {\bibfnamefont {X.~P.}\ \bibnamefont
  {Gao}},\ }\bibfield  {title} {\bibinfo {title} {Strong {{Tuning}} of {{Rashba
  Spin}}\textendash{{Orbit Interaction}} in {{Single InAs Nanowires}}},\ }\href
  {https://doi.org/10.1021/nl301325h} {\bibfield  {journal} {\bibinfo
  {journal} {Nano Letters}\ }\textbf {\bibinfo {volume} {12}},\ \bibinfo
  {pages} {3263} (\bibinfo {year} {2012})}\BibitemShut {NoStop}%
\bibitem [{\citenamefont {Albrecht}\ \emph {et~al.}(2016)\citenamefont
  {Albrecht}, \citenamefont {Higginbotham}, \citenamefont {Madsen},
  \citenamefont {Kuemmeth}, \citenamefont {Jespersen}, \citenamefont
  {Nyg{\aa}rd}, \citenamefont {Krogstrup},\ and\ \citenamefont
  {Marcus}}]{albrecht_exponential_2016}%
  \BibitemOpen
  \bibfield  {author} {\bibinfo {author} {\bibfnamefont {S.~M.}\ \bibnamefont
  {Albrecht}}, \bibinfo {author} {\bibfnamefont {A.~P.}\ \bibnamefont
  {Higginbotham}}, \bibinfo {author} {\bibfnamefont {M.}~\bibnamefont
  {Madsen}}, \bibinfo {author} {\bibfnamefont {F.}~\bibnamefont {Kuemmeth}},
  \bibinfo {author} {\bibfnamefont {T.~S.}\ \bibnamefont {Jespersen}}, \bibinfo
  {author} {\bibfnamefont {J.}~\bibnamefont {Nyg{\aa}rd}}, \bibinfo {author}
  {\bibfnamefont {P.}~\bibnamefont {Krogstrup}},\ and\ \bibinfo {author}
  {\bibfnamefont {C.~M.}\ \bibnamefont {Marcus}},\ }\bibfield  {title}
  {\bibinfo {title} {Exponential protection of zero modes in {{Majorana}}
  islands},\ }\href {https://doi.org/10.1038/nature17162} {\bibfield  {journal}
  {\bibinfo  {journal} {Nature}\ }\textbf {\bibinfo {volume} {531}},\ \bibinfo
  {pages} {206} (\bibinfo {year} {2016})}\BibitemShut {NoStop}%
\bibitem [{\citenamefont {Kringh{\o}j}\ \emph {et~al.}(2021)\citenamefont
  {Kringh{\o}j}, \citenamefont {Larsen}, \citenamefont {Erlandsson},
  \citenamefont {Uilhoorn}, \citenamefont {Kroll}, \citenamefont {Hesselberg},
  \citenamefont {McNeil}, \citenamefont {Krogstrup}, \citenamefont {Casparis},
  \citenamefont {Marcus},\ and\ \citenamefont
  {Petersson}}]{kringhoj_magnetic-field-compatible_2021-1}%
  \BibitemOpen
  \bibfield  {author} {\bibinfo {author} {\bibfnamefont {A.}~\bibnamefont
  {Kringh{\o}j}}, \bibinfo {author} {\bibfnamefont {T.~W.}\ \bibnamefont
  {Larsen}}, \bibinfo {author} {\bibfnamefont {O.}~\bibnamefont {Erlandsson}},
  \bibinfo {author} {\bibfnamefont {W.}~\bibnamefont {Uilhoorn}}, \bibinfo
  {author} {\bibfnamefont {J.}~\bibnamefont {Kroll}}, \bibinfo {author}
  {\bibfnamefont {M.}~\bibnamefont {Hesselberg}}, \bibinfo {author}
  {\bibfnamefont {R.}~\bibnamefont {McNeil}}, \bibinfo {author} {\bibfnamefont
  {P.}~\bibnamefont {Krogstrup}}, \bibinfo {author} {\bibfnamefont
  {L.}~\bibnamefont {Casparis}}, \bibinfo {author} {\bibfnamefont
  {C.}~\bibnamefont {Marcus}},\ and\ \bibinfo {author} {\bibfnamefont
  {K.}~\bibnamefont {Petersson}},\ }\bibfield  {title} {\bibinfo {title}
  {Magnetic-{{Field-Compatible Superconducting Transmon Qubit}}},\ }\href
  {https://doi.org/10.1103/PhysRevApplied.15.054001} {\bibfield  {journal}
  {\bibinfo  {journal} {Physical Review Applied}\ }\textbf {\bibinfo {volume}
  {15}},\ \bibinfo {pages} {054001} (\bibinfo {year} {2021})}\BibitemShut
  {NoStop}%
\bibitem [{\citenamefont {Uilhoorn}\ \emph {et~al.}(2021)\citenamefont
  {Uilhoorn}, \citenamefont {Kroll}, \citenamefont {Bargerbos}, \citenamefont
  {Nabi}, \citenamefont {Yang}, \citenamefont {Krogstrup}, \citenamefont
  {Kouwenhoven}, \citenamefont {Kou},\ and\ \citenamefont {{de
  Lange}}}]{uilhoorn_quasiparticle_2021}%
  \BibitemOpen
  \bibfield  {author} {\bibinfo {author} {\bibfnamefont {W.}~\bibnamefont
  {Uilhoorn}}, \bibinfo {author} {\bibfnamefont {J.~G.}\ \bibnamefont {Kroll}},
  \bibinfo {author} {\bibfnamefont {A.}~\bibnamefont {Bargerbos}}, \bibinfo
  {author} {\bibfnamefont {S.~D.}\ \bibnamefont {Nabi}}, \bibinfo {author}
  {\bibfnamefont {C.-K.}\ \bibnamefont {Yang}}, \bibinfo {author}
  {\bibfnamefont {P.}~\bibnamefont {Krogstrup}}, \bibinfo {author}
  {\bibfnamefont {L.~P.}\ \bibnamefont {Kouwenhoven}}, \bibinfo {author}
  {\bibfnamefont {A.}~\bibnamefont {Kou}},\ and\ \bibinfo {author}
  {\bibfnamefont {G.}~\bibnamefont {{de Lange}}},\ }\bibfield  {title}
  {\bibinfo {title} {Quasiparticle trapping by orbital effect in a hybrid
  superconducting-semiconducting circuit},\ }\href@noop {} {\bibfield
  {journal} {\bibinfo  {journal} {arXiv:2105.11038 [cond-mat]}\ } (\bibinfo
  {year} {2021})},\ \Eprint {https://arxiv.org/abs/2105.11038}
  {arXiv:2105.11038 [cond-mat]} \BibitemShut {NoStop}%
\bibitem [{\citenamefont {Zuo}\ \emph {et~al.}(2017)\citenamefont {Zuo},
  \citenamefont {Mourik}, \citenamefont {Szombati}, \citenamefont {Nijholt},
  \citenamefont {{van Woerkom}}, \citenamefont {Geresdi}, \citenamefont {Chen},
  \citenamefont {Ostroukh}, \citenamefont {Akhmerov}, \citenamefont {Plissard},
  \citenamefont {Car}, \citenamefont {Bakkers}, \citenamefont {Pikulin},
  \citenamefont {Kouwenhoven},\ and\ \citenamefont
  {Frolov}}]{zuo_supercurrent_2017-1}%
  \BibitemOpen
  \bibfield  {author} {\bibinfo {author} {\bibfnamefont {K.}~\bibnamefont
  {Zuo}}, \bibinfo {author} {\bibfnamefont {V.}~\bibnamefont {Mourik}},
  \bibinfo {author} {\bibfnamefont {D.~B.}\ \bibnamefont {Szombati}}, \bibinfo
  {author} {\bibfnamefont {B.}~\bibnamefont {Nijholt}}, \bibinfo {author}
  {\bibfnamefont {D.~J.}\ \bibnamefont {{van Woerkom}}}, \bibinfo {author}
  {\bibfnamefont {A.}~\bibnamefont {Geresdi}}, \bibinfo {author} {\bibfnamefont
  {J.}~\bibnamefont {Chen}}, \bibinfo {author} {\bibfnamefont {V.~P.}\
  \bibnamefont {Ostroukh}}, \bibinfo {author} {\bibfnamefont {A.~R.}\
  \bibnamefont {Akhmerov}}, \bibinfo {author} {\bibfnamefont {S.~R.}\
  \bibnamefont {Plissard}}, \bibinfo {author} {\bibfnamefont {D.}~\bibnamefont
  {Car}}, \bibinfo {author} {\bibfnamefont {E.~P. A.~M.}\ \bibnamefont
  {Bakkers}}, \bibinfo {author} {\bibfnamefont {D.~I.}\ \bibnamefont
  {Pikulin}}, \bibinfo {author} {\bibfnamefont {L.~P.}\ \bibnamefont
  {Kouwenhoven}},\ and\ \bibinfo {author} {\bibfnamefont {S.~M.}\ \bibnamefont
  {Frolov}},\ }\bibfield  {title} {\bibinfo {title} {Supercurrent
  {{Interference}} in {{Few-Mode Nanowire Josephson Junctions}}},\ }\href
  {https://doi.org/10.1103/PhysRevLett.119.187704} {\bibfield  {journal}
  {\bibinfo  {journal} {Physical Review Letters}\ }\textbf {\bibinfo {volume}
  {119}},\ \bibinfo {pages} {187704} (\bibinfo {year} {2017})}\BibitemShut
  {NoStop}%
\bibitem [{\citenamefont {V{\"a}yrynen}\ \emph {et~al.}(2015)\citenamefont
  {V{\"a}yrynen}, \citenamefont {Rastelli}, \citenamefont {Belzig},\ and\
  \citenamefont {Glazman}}]{vayrynen_microwave_2015}%
  \BibitemOpen
  \bibfield  {author} {\bibinfo {author} {\bibfnamefont {J.~I.}\ \bibnamefont
  {V{\"a}yrynen}}, \bibinfo {author} {\bibfnamefont {G.}~\bibnamefont
  {Rastelli}}, \bibinfo {author} {\bibfnamefont {W.}~\bibnamefont {Belzig}},\
  and\ \bibinfo {author} {\bibfnamefont {L.~I.}\ \bibnamefont {Glazman}},\
  }\bibfield  {title} {\bibinfo {title} {Microwave signatures of {{Majorana}}
  states in a topological {{Josephson}} junction},\ }\href
  {https://doi.org/10.1103/PhysRevB.92.134508} {\bibfield  {journal} {\bibinfo
  {journal} {Physical Review B}\ }\textbf {\bibinfo {volume} {92}},\ \bibinfo
  {pages} {134508} (\bibinfo {year} {2015})}\BibitemShut {NoStop}%
\bibitem [{\citenamefont {Fu}\ and\ \citenamefont
  {Kane}(2009)}]{fu_josephson_2009}%
  \BibitemOpen
  \bibfield  {author} {\bibinfo {author} {\bibfnamefont {L.}~\bibnamefont
  {Fu}}\ and\ \bibinfo {author} {\bibfnamefont {C.~L.}\ \bibnamefont {Kane}},\
  }\bibfield  {title} {\bibinfo {title} {Josephson current and noise at a
  superconductor/quantum-spin-{{Hall-insulator}}/superconductor junction},\
  }\href {https://doi.org/10.1103/PhysRevB.79.161408} {\bibfield  {journal}
  {\bibinfo  {journal} {Physical Review B}\ }\textbf {\bibinfo {volume} {79}},\
  \bibinfo {pages} {161408} (\bibinfo {year} {2009})}\BibitemShut {NoStop}%
\bibitem [{\citenamefont {{van Heck}}\ \emph {et~al.}(2017)\citenamefont {{van
  Heck}}, \citenamefont {V{\"a}yrynen},\ and\ \citenamefont
  {Glazman}}]{van_heck_zeeman_2017}%
  \BibitemOpen
  \bibfield  {author} {\bibinfo {author} {\bibfnamefont {B.}~\bibnamefont {{van
  Heck}}}, \bibinfo {author} {\bibfnamefont {J.~I.}\ \bibnamefont
  {V{\"a}yrynen}},\ and\ \bibinfo {author} {\bibfnamefont {L.~I.}\ \bibnamefont
  {Glazman}},\ }\bibfield  {title} {\bibinfo {title} {Zeeman and spin-orbit
  effects in the {{Andreev}} spectra of nanowire junctions},\ }\href
  {https://doi.org/10.1103/PhysRevB.96.075404} {\bibfield  {journal} {\bibinfo
  {journal} {Physical Review B}\ }\textbf {\bibinfo {volume} {96}},\ \bibinfo
  {pages} {075404} (\bibinfo {year} {2017})}\BibitemShut {NoStop}%
\bibitem [{\citenamefont {Padurariu}\ and\ \citenamefont
  {Nazarov}(2010)}]{padurariu_theoretical_2010}%
  \BibitemOpen
  \bibfield  {author} {\bibinfo {author} {\bibfnamefont {C.}~\bibnamefont
  {Padurariu}}\ and\ \bibinfo {author} {\bibfnamefont {Y.~V.}\ \bibnamefont
  {Nazarov}},\ }\bibfield  {title} {\bibinfo {title} {Theoretical proposal for
  superconducting spin qubits},\ }\href
  {https://doi.org/10.1103/PhysRevB.81.144519} {\bibfield  {journal} {\bibinfo
  {journal} {Physical Review B}\ }\textbf {\bibinfo {volume} {81}},\ \bibinfo
  {pages} {144519} (\bibinfo {year} {2010})}\BibitemShut {NoStop}%
\bibitem [{\citenamefont {Reynoso}\ \emph {et~al.}(2012)\citenamefont
  {Reynoso}, \citenamefont {Usaj}, \citenamefont {Balseiro}, \citenamefont
  {Feinberg},\ and\ \citenamefont {Avignon}}]{reynoso_spin-orbit-induced_2012}%
  \BibitemOpen
  \bibfield  {author} {\bibinfo {author} {\bibfnamefont {A.~A.}\ \bibnamefont
  {Reynoso}}, \bibinfo {author} {\bibfnamefont {G.}~\bibnamefont {Usaj}},
  \bibinfo {author} {\bibfnamefont {C.~A.}\ \bibnamefont {Balseiro}}, \bibinfo
  {author} {\bibfnamefont {D.}~\bibnamefont {Feinberg}},\ and\ \bibinfo
  {author} {\bibfnamefont {M.}~\bibnamefont {Avignon}},\ }\bibfield  {title}
  {\bibinfo {title} {Spin-orbit-induced chirality of {{Andreev}} states in
  {{Josephson}} junctions},\ }\href
  {https://doi.org/10.1103/PhysRevB.86.214519} {\bibfield  {journal} {\bibinfo
  {journal} {Physical Review B}\ }\textbf {\bibinfo {volume} {86}},\ \bibinfo
  {pages} {214519} (\bibinfo {year} {2012})}\BibitemShut {NoStop}%
\bibitem [{\citenamefont {Bergeret}\ and\ \citenamefont
  {Tokatly}(2014)}]{bergeret_theory_2014}%
  \BibitemOpen
  \bibfield  {author} {\bibinfo {author} {\bibfnamefont {F.~S.}\ \bibnamefont
  {Bergeret}}\ and\ \bibinfo {author} {\bibfnamefont {I.~V.}\ \bibnamefont
  {Tokatly}},\ }\href@noop {} {\bibinfo {title} {Theory of diffusive $ \phi_0$
  {{Josephson}} junctions in the presence of spin-orbit coupling}} (\bibinfo
  {year} {2014}),\ \Eprint {https://arxiv.org/abs/1409.4563} {arXiv:1409.4563
  [cond-mat]} \BibitemShut {NoStop}%
\bibitem [{\citenamefont {Campagnano}\ \emph {et~al.}(2015)\citenamefont
  {Campagnano}, \citenamefont {Lucignano}, \citenamefont {Giuliano},\ and\
  \citenamefont {Tagliacozzo}}]{campagnano_spinorbit_2015}%
  \BibitemOpen
  \bibfield  {author} {\bibinfo {author} {\bibfnamefont {G.}~\bibnamefont
  {Campagnano}}, \bibinfo {author} {\bibfnamefont {P.}~\bibnamefont
  {Lucignano}}, \bibinfo {author} {\bibfnamefont {D.}~\bibnamefont
  {Giuliano}},\ and\ \bibinfo {author} {\bibfnamefont {A.}~\bibnamefont
  {Tagliacozzo}},\ }\bibfield  {title} {\bibinfo {title} {Spin\textendash orbit
  coupling and anomalous {{Josephson}} effect in nanowires},\ }\href
  {https://doi.org/10.1088/0953-8984/27/20/205301} {\bibfield  {journal}
  {\bibinfo  {journal} {Journal of Physics: Condensed Matter}\ }\textbf
  {\bibinfo {volume} {27}},\ \bibinfo {pages} {205301} (\bibinfo {year}
  {2015})}\BibitemShut {NoStop}%
\bibitem [{\citenamefont {Liu}\ and\ \citenamefont
  {Chan}(2010)}]{liu_relation_2010}%
  \BibitemOpen
  \bibfield  {author} {\bibinfo {author} {\bibfnamefont {J.-F.}\ \bibnamefont
  {Liu}}\ and\ \bibinfo {author} {\bibfnamefont {K.~S.}\ \bibnamefont {Chan}},\
  }\bibfield  {title} {\bibinfo {title} {Relation between symmetry breaking and
  the anomalous josephson effect},\ }\href
  {https://doi.org/10.1103/PhysRevB.82.125305} {\bibfield  {journal} {\bibinfo
  {journal} {Phys. Rev. B}\ }\textbf {\bibinfo {volume} {82}},\ \bibinfo
  {pages} {125305} (\bibinfo {year} {2010})}\BibitemShut {NoStop}%
\bibitem [{\citenamefont {Rasmussen}\ \emph {et~al.}(2016)\citenamefont
  {Rasmussen}, \citenamefont {Danon}, \citenamefont {Suominen}, \citenamefont
  {Nichele}, \citenamefont {Kjaergaard},\ and\ \citenamefont
  {Flensberg}}]{rasmussen_effects_2016}%
  \BibitemOpen
  \bibfield  {author} {\bibinfo {author} {\bibfnamefont {A.}~\bibnamefont
  {Rasmussen}}, \bibinfo {author} {\bibfnamefont {J.}~\bibnamefont {Danon}},
  \bibinfo {author} {\bibfnamefont {H.}~\bibnamefont {Suominen}}, \bibinfo
  {author} {\bibfnamefont {F.}~\bibnamefont {Nichele}}, \bibinfo {author}
  {\bibfnamefont {M.}~\bibnamefont {Kjaergaard}},\ and\ \bibinfo {author}
  {\bibfnamefont {K.}~\bibnamefont {Flensberg}},\ }\bibfield  {title} {\bibinfo
  {title} {Effects of spin-orbit coupling and spatial symmetries on the
  {{Josephson}} current in {{SNS}} junctions},\ }\href
  {https://doi.org/10.1103/PhysRevB.93.155406} {\bibfield  {journal} {\bibinfo
  {journal} {Physical Review B}\ }\textbf {\bibinfo {volume} {93}},\ \bibinfo
  {pages} {155406} (\bibinfo {year} {2016})}\BibitemShut {NoStop}%
\bibitem [{\citenamefont {Assouline}\ \emph {et~al.}(2019)\citenamefont
  {Assouline}, \citenamefont {{Feuillet-Palma}}, \citenamefont {Bergeal},
  \citenamefont {Zhang}, \citenamefont {Mottaghizadeh}, \citenamefont
  {Zimmers}, \citenamefont {Lhuillier}, \citenamefont {Eddrie}, \citenamefont
  {Atkinson}, \citenamefont {Aprili},\ and\ \citenamefont
  {Aubin}}]{assouline_spin-orbit_2019}%
  \BibitemOpen
  \bibfield  {author} {\bibinfo {author} {\bibfnamefont {A.}~\bibnamefont
  {Assouline}}, \bibinfo {author} {\bibfnamefont {C.}~\bibnamefont
  {{Feuillet-Palma}}}, \bibinfo {author} {\bibfnamefont {N.}~\bibnamefont
  {Bergeal}}, \bibinfo {author} {\bibfnamefont {T.}~\bibnamefont {Zhang}},
  \bibinfo {author} {\bibfnamefont {A.}~\bibnamefont {Mottaghizadeh}}, \bibinfo
  {author} {\bibfnamefont {A.}~\bibnamefont {Zimmers}}, \bibinfo {author}
  {\bibfnamefont {E.}~\bibnamefont {Lhuillier}}, \bibinfo {author}
  {\bibfnamefont {M.}~\bibnamefont {Eddrie}}, \bibinfo {author} {\bibfnamefont
  {P.}~\bibnamefont {Atkinson}}, \bibinfo {author} {\bibfnamefont
  {M.}~\bibnamefont {Aprili}},\ and\ \bibinfo {author} {\bibfnamefont
  {H.}~\bibnamefont {Aubin}},\ }\bibfield  {title} {\bibinfo {title}
  {Spin-{{Orbit}} induced phase-shift in {{Bi2Se3 Josephson}} junctions},\
  }\href {https://doi.org/10.1038/s41467-018-08022-y} {\bibfield  {journal}
  {\bibinfo  {journal} {Nature Communications}\ }\textbf {\bibinfo {volume}
  {10}},\ \bibinfo {pages} {126} (\bibinfo {year} {2019})}\BibitemShut
  {NoStop}%
\bibitem [{\citenamefont {Wesdorp}\ \emph
  {et~al.}(2021{\natexlab{b}})\citenamefont {Wesdorp}, \citenamefont
  {Grunhaupt}, \citenamefont {Vaartjes}, \citenamefont {{Pita-vidal}},
  \citenamefont {Bargerbos}, \citenamefont {Splitthoff}, \citenamefont
  {Krogstrup}, \citenamefont {Van~Heck},\ and\ \citenamefont
  {De~Lange}}]{wesdorp_supplementary_2021}%
  \BibitemOpen
  \bibfield  {author} {\bibinfo {author} {\bibfnamefont {J.}~\bibnamefont
  {Wesdorp}}, \bibinfo {author} {\bibfnamefont {L.}~\bibnamefont {Grunhaupt}},
  \bibinfo {author} {\bibfnamefont {A.}~\bibnamefont {Vaartjes}}, \bibinfo
  {author} {\bibfnamefont {M.}~\bibnamefont {{Pita-vidal}}}, \bibinfo {author}
  {\bibfnamefont {A.}~\bibnamefont {Bargerbos}}, \bibinfo {author}
  {\bibfnamefont {L.~J.}\ \bibnamefont {Splitthoff}}, \bibinfo {author}
  {\bibfnamefont {P.}~\bibnamefont {Krogstrup}}, \bibinfo {author}
  {\bibfnamefont {B.}~\bibnamefont {Van~Heck}},\ and\ \bibinfo {author}
  {\bibfnamefont {G.}~\bibnamefont {De~Lange}},\ }\href@noop {} {\bibinfo
  {title} {Supplementary information: {{Dynamical}} polarization of the parity
  of andreev bound states in a nanowire {{Josephson}} junction.}} (\bibinfo
  {year} {2021}{\natexlab{b}})\BibitemShut {NoStop}%
\bibitem [{\citenamefont {Linder}\ and\ \citenamefont
  {Robinson}(2015{\natexlab{b}})}]{linder_superconducting_2015}%
  \BibitemOpen
  \bibfield  {author} {\bibinfo {author} {\bibfnamefont {J.}~\bibnamefont
  {Linder}}\ and\ \bibinfo {author} {\bibfnamefont {J.~W.~A.}\ \bibnamefont
  {Robinson}},\ }\bibfield  {title} {\bibinfo {title} {Superconducting
  spintronics},\ }\href {https://doi.org/10.1038/nphys3242} {\bibfield
  {journal} {\bibinfo  {journal} {Nature Physics}\ }\textbf {\bibinfo {volume}
  {11}},\ \bibinfo {pages} {307} (\bibinfo {year}
  {2015}{\natexlab{b}})}\BibitemShut {NoStop}%
\bibitem [{\citenamefont {Rosdahl}\ \emph {et~al.}(2018)\citenamefont
  {Rosdahl}, \citenamefont {Vuik}, \citenamefont {Kjaergaard},\ and\
  \citenamefont {Akhmerov}}]{rosdahl_andreev_2018}%
  \BibitemOpen
  \bibfield  {author} {\bibinfo {author} {\bibfnamefont {T.~{\"O}.}\
  \bibnamefont {Rosdahl}}, \bibinfo {author} {\bibfnamefont {A.}~\bibnamefont
  {Vuik}}, \bibinfo {author} {\bibfnamefont {M.}~\bibnamefont {Kjaergaard}},\
  and\ \bibinfo {author} {\bibfnamefont {A.~R.}\ \bibnamefont {Akhmerov}},\
  }\bibfield  {title} {\bibinfo {title} {Andreev rectifier: A nonlocal
  conductance signature of topological phase transitions},\ }\href
  {https://doi.org/10.1103/PhysRevB.97.045421} {\bibfield  {journal} {\bibinfo
  {journal} {Physical Review B}\ }\textbf {\bibinfo {volume} {97}},\ \bibinfo
  {pages} {045421} (\bibinfo {year} {2018})},\ \Eprint
  {https://arxiv.org/abs/1706.08888} {arXiv:1706.08888 [cond-mat]} \BibitemShut
  {NoStop}%
\bibitem [{\citenamefont {Splitthoff}\ \emph {et~al.}(2022)\citenamefont
  {Splitthoff}, \citenamefont {Bargerbos}, \citenamefont {Gr{\"u}nhaupt},
  \citenamefont {{Pita-Vidal}}, \citenamefont {Wesdorp}, \citenamefont {Liu},
  \citenamefont {Kou}, \citenamefont {Andersen},\ and\ \citenamefont {{van
  Heck}}}]{splitthoff_gate-tunable_2022}%
  \BibitemOpen
  \bibfield  {author} {\bibinfo {author} {\bibfnamefont {L.~J.}\ \bibnamefont
  {Splitthoff}}, \bibinfo {author} {\bibfnamefont {A.}~\bibnamefont
  {Bargerbos}}, \bibinfo {author} {\bibfnamefont {L.}~\bibnamefont
  {Gr{\"u}nhaupt}}, \bibinfo {author} {\bibfnamefont {M.}~\bibnamefont
  {{Pita-Vidal}}}, \bibinfo {author} {\bibfnamefont {J.~J.}\ \bibnamefont
  {Wesdorp}}, \bibinfo {author} {\bibfnamefont {Y.}~\bibnamefont {Liu}},
  \bibinfo {author} {\bibfnamefont {A.}~\bibnamefont {Kou}}, \bibinfo {author}
  {\bibfnamefont {C.~K.}\ \bibnamefont {Andersen}},\ and\ \bibinfo {author}
  {\bibfnamefont {B.}~\bibnamefont {{van Heck}}},\ }\href@noop {} {\bibinfo
  {title} {Gate-tunable kinetic inductance in proximitized nanowires}}
  (\bibinfo {year} {2022}),\ \Eprint {https://arxiv.org/abs/2202.08729}
  {arXiv:2202.08729 [cond-mat]} \BibitemShut {NoStop}%
\bibitem [{\citenamefont {Peng}\ \emph {et~al.}(2016)\citenamefont {Peng},
  \citenamefont {Pientka}, \citenamefont {Berg}, \citenamefont {Oreg},\ and\
  \citenamefont {{von Oppen}}}]{peng_signatures_2016}%
  \BibitemOpen
  \bibfield  {author} {\bibinfo {author} {\bibfnamefont {Y.}~\bibnamefont
  {Peng}}, \bibinfo {author} {\bibfnamefont {F.}~\bibnamefont {Pientka}},
  \bibinfo {author} {\bibfnamefont {E.}~\bibnamefont {Berg}}, \bibinfo {author}
  {\bibfnamefont {Y.}~\bibnamefont {Oreg}},\ and\ \bibinfo {author}
  {\bibfnamefont {F.}~\bibnamefont {{von Oppen}}},\ }\bibfield  {title}
  {\bibinfo {title} {Signatures of topological {{Josephson}} junctions},\
  }\href {https://doi.org/10.1103/PhysRevB.94.085409} {\bibfield  {journal}
  {\bibinfo  {journal} {Physical Review B}\ }\textbf {\bibinfo {volume} {94}},\
  \bibinfo {pages} {085409} (\bibinfo {year} {2016})}\BibitemShut {NoStop}%
\bibitem [{\citenamefont {Murthy}\ \emph {et~al.}(2020)\citenamefont {Murthy},
  \citenamefont {Kurilovich}, \citenamefont {Kurilovich}, \citenamefont {{van
  Heck}}, \citenamefont {Glazman},\ and\ \citenamefont
  {Nayak}}]{murthy_energy_2020}%
  \BibitemOpen
  \bibfield  {author} {\bibinfo {author} {\bibfnamefont {C.}~\bibnamefont
  {Murthy}}, \bibinfo {author} {\bibfnamefont {V.~D.}\ \bibnamefont
  {Kurilovich}}, \bibinfo {author} {\bibfnamefont {P.~D.}\ \bibnamefont
  {Kurilovich}}, \bibinfo {author} {\bibfnamefont {B.}~\bibnamefont {{van
  Heck}}}, \bibinfo {author} {\bibfnamefont {L.~I.}\ \bibnamefont {Glazman}},\
  and\ \bibinfo {author} {\bibfnamefont {C.}~\bibnamefont {Nayak}},\ }\bibfield
   {title} {\bibinfo {title} {Energy spectrum and current-phase relation of a
  nanowire {{Josephson}} junction close to the topological transition},\ }\href
  {https://doi.org/10.1103/PhysRevB.101.224501} {\bibfield  {journal} {\bibinfo
   {journal} {Physical Review B}\ }\textbf {\bibinfo {volume} {101}},\ \bibinfo
  {pages} {224501} (\bibinfo {year} {2020})},\ \Eprint
  {https://arxiv.org/abs/1912.04952} {arXiv:1912.04952 [cond-mat]} \BibitemShut
  {NoStop}%
\bibitem [{\citenamefont {Winkler}\ \emph {et~al.}(2019)\citenamefont
  {Winkler}, \citenamefont {Antipov}, \citenamefont {{van Heck}}, \citenamefont
  {Soluyanov}, \citenamefont {Glazman}, \citenamefont {Wimmer},\ and\
  \citenamefont {Lutchyn}}]{winkler_unified_2019-1}%
  \BibitemOpen
  \bibfield  {author} {\bibinfo {author} {\bibfnamefont {G.~W.}\ \bibnamefont
  {Winkler}}, \bibinfo {author} {\bibfnamefont {A.~E.}\ \bibnamefont
  {Antipov}}, \bibinfo {author} {\bibfnamefont {B.}~\bibnamefont {{van Heck}}},
  \bibinfo {author} {\bibfnamefont {A.~A.}\ \bibnamefont {Soluyanov}}, \bibinfo
  {author} {\bibfnamefont {L.~I.}\ \bibnamefont {Glazman}}, \bibinfo {author}
  {\bibfnamefont {M.}~\bibnamefont {Wimmer}},\ and\ \bibinfo {author}
  {\bibfnamefont {R.~M.}\ \bibnamefont {Lutchyn}},\ }\bibfield  {title}
  {\bibinfo {title} {A unified numerical approach to
  semiconductor-superconductor heterostructures},\ }\href
  {https://doi.org/10.1103/PhysRevB.99.245408} {\bibfield  {journal} {\bibinfo
  {journal} {Physical Review B}\ }\textbf {\bibinfo {volume} {99}},\ \bibinfo
  {pages} {245408} (\bibinfo {year} {2019})},\ \Eprint
  {https://arxiv.org/abs/1810.04180} {arXiv:1810.04180 [cond-mat]} \BibitemShut
  {NoStop}%
\bibitem [{\citenamefont {Laroche}\ \emph {et~al.}(2019)\citenamefont
  {Laroche}, \citenamefont {Bouman}, \citenamefont {{van Woerkom}},
  \citenamefont {Proutski}, \citenamefont {Murthy}, \citenamefont {Pikulin},
  \citenamefont {Nayak}, \citenamefont {{van Gulik}}, \citenamefont
  {Nyg{\aa}rd}, \citenamefont {Krogstrup}, \citenamefont {Kouwenhoven},\ and\
  \citenamefont {Geresdi}}]{laroche_observation_2019}%
  \BibitemOpen
  \bibfield  {author} {\bibinfo {author} {\bibfnamefont {D.}~\bibnamefont
  {Laroche}}, \bibinfo {author} {\bibfnamefont {D.}~\bibnamefont {Bouman}},
  \bibinfo {author} {\bibfnamefont {D.~J.}\ \bibnamefont {{van Woerkom}}},
  \bibinfo {author} {\bibfnamefont {A.}~\bibnamefont {Proutski}}, \bibinfo
  {author} {\bibfnamefont {C.}~\bibnamefont {Murthy}}, \bibinfo {author}
  {\bibfnamefont {D.~I.}\ \bibnamefont {Pikulin}}, \bibinfo {author}
  {\bibfnamefont {C.}~\bibnamefont {Nayak}}, \bibinfo {author} {\bibfnamefont
  {R.~J.~J.}\ \bibnamefont {{van Gulik}}}, \bibinfo {author} {\bibfnamefont
  {J.}~\bibnamefont {Nyg{\aa}rd}}, \bibinfo {author} {\bibfnamefont
  {P.}~\bibnamefont {Krogstrup}}, \bibinfo {author} {\bibfnamefont {L.~P.}\
  \bibnamefont {Kouwenhoven}},\ and\ \bibinfo {author} {\bibfnamefont
  {A.}~\bibnamefont {Geresdi}},\ }\bibfield  {title} {\bibinfo {title}
  {Observation of the 4$\pi$-periodic {{Josephson}} effect in indium arsenide
  nanowires},\ }\href {https://doi.org/10.1038/s41467-018-08161-2} {\bibfield
  {journal} {\bibinfo  {journal} {Nature Communications}\ }\textbf {\bibinfo
  {volume} {10}},\ \bibinfo {pages} {245} (\bibinfo {year} {2019})},\ \Eprint
  {https://arxiv.org/abs/1712.08459} {arXiv:1712.08459} \BibitemShut {NoStop}%
\bibitem [{\citenamefont {Burkard}\ \emph {et~al.}(2021)\citenamefont
  {Burkard}, \citenamefont {Ladd}, \citenamefont {Nichol}, \citenamefont
  {Pan},\ and\ \citenamefont {Petta}}]{burkard_semiconductor_2021}%
  \BibitemOpen
  \bibfield  {author} {\bibinfo {author} {\bibfnamefont {G.}~\bibnamefont
  {Burkard}}, \bibinfo {author} {\bibfnamefont {T.~D.}\ \bibnamefont {Ladd}},
  \bibinfo {author} {\bibfnamefont {J.~M.}\ \bibnamefont {Nichol}}, \bibinfo
  {author} {\bibfnamefont {A.}~\bibnamefont {Pan}},\ and\ \bibinfo {author}
  {\bibfnamefont {J.~R.}\ \bibnamefont {Petta}},\ }\href@noop {} {\bibinfo
  {title} {Semiconductor {{Spin Qubits}}}} (\bibinfo {year} {2021}),\ \Eprint
  {https://arxiv.org/abs/2112.08863} {arXiv:2112.08863 [cond-mat,
  physics:physics, physics:quant-ph]} \BibitemShut {NoStop}%
\bibitem [{\citenamefont {Padurariu}\ and\ \citenamefont
  {Nazarov}(2012)}]{padurariu_spin_2012}%
  \BibitemOpen
  \bibfield  {author} {\bibinfo {author} {\bibfnamefont {C.}~\bibnamefont
  {Padurariu}}\ and\ \bibinfo {author} {\bibfnamefont {Y.~V.}\ \bibnamefont
  {Nazarov}},\ }\bibfield  {title} {\bibinfo {title} {Spin blockade qubit in a
  superconducting junction},\ }\href
  {https://doi.org/10.1209/0295-5075/100/57006} {\bibfield  {journal} {\bibinfo
   {journal} {EPL (Europhysics Letters)}\ }\textbf {\bibinfo {volume} {100}},\
  \bibinfo {pages} {57006} (\bibinfo {year} {2012})}\BibitemShut {NoStop}%
\end{thebibliography}%


\begin{thebibliography}{25}%
\makeatletter
\providecommand \@ifxundefined [1]{%
 \@ifx{#1\undefined}
}%
\providecommand \@ifnum [1]{%
 \ifnum #1\expandafter \@firstoftwo
 \else \expandafter \@secondoftwo
 \fi
}%
\providecommand \@ifx [1]{%
 \ifx #1\expandafter \@firstoftwo
 \else \expandafter \@secondoftwo
 \fi
}%
\providecommand \natexlab [1]{#1}%
\providecommand \enquote  [1]{``#1''}%
\providecommand \bibnamefont  [1]{#1}%
\providecommand \bibfnamefont [1]{#1}%
\providecommand \citenamefont [1]{#1}%
\providecommand \href@noop [0]{\@secondoftwo}%
\providecommand \href [0]{\begingroup \@sanitize@url \@href}%
\providecommand \@href[1]{\@@startlink{#1}\@@href}%
\providecommand \@@href[1]{\endgroup#1\@@endlink}%
\providecommand \@sanitize@url [0]{\catcode `\\12\catcode `\$12\catcode
  `\&12\catcode `\#12\catcode `\^12\catcode `\_12\catcode `\%12\relax}%
\providecommand \@@startlink[1]{}%
\providecommand \@@endlink[0]{}%
\providecommand \url  [0]{\begingroup\@sanitize@url \@url }%
\providecommand \@url [1]{\endgroup\@href {#1}{\urlprefix }}%
\providecommand \urlprefix  [0]{URL }%
\providecommand \Eprint [0]{\href }%
\providecommand \doibase [0]{https://doi.org/}%
\providecommand \selectlanguage [0]{\@gobble}%
\providecommand \bibinfo  [0]{\@secondoftwo}%
\providecommand \bibfield  [0]{\@secondoftwo}%
\providecommand \translation [1]{[#1]}%
\providecommand \BibitemOpen [0]{}%
\providecommand \bibitemStop [0]{}%
\providecommand \bibitemNoStop [0]{.\EOS\space}%
\providecommand \EOS [0]{\spacefactor3000\relax}%
\providecommand \BibitemShut  [1]{\csname bibitem#1\endcsname}%
\let\auto@bib@innerbib\@empty
\bibitem [{\citenamefont {Wesdorp}\ \emph {et~al.}(2021)\citenamefont
  {Wesdorp}, \citenamefont {Gr{\"u}nhaupt}, \citenamefont {Vaartjes},
  \citenamefont {{Pita-Vidal}}, \citenamefont {Bargerbos}, \citenamefont
  {Splitthoff}, \citenamefont {Krogstrup}, \citenamefont {{van Heck}},\ and\
  \citenamefont {{de Lange}}}]{wesdorp_dynamical_2021}%
  \BibitemOpen
  \bibfield  {author} {\bibinfo {author} {\bibfnamefont {J.~J.}\ \bibnamefont
  {Wesdorp}}, \bibinfo {author} {\bibfnamefont {L.}~\bibnamefont
  {Gr{\"u}nhaupt}}, \bibinfo {author} {\bibfnamefont {A.}~\bibnamefont
  {Vaartjes}}, \bibinfo {author} {\bibfnamefont {M.}~\bibnamefont
  {{Pita-Vidal}}}, \bibinfo {author} {\bibfnamefont {A.}~\bibnamefont
  {Bargerbos}}, \bibinfo {author} {\bibfnamefont {L.~J.}\ \bibnamefont
  {Splitthoff}}, \bibinfo {author} {\bibfnamefont {P.}~\bibnamefont
  {Krogstrup}}, \bibinfo {author} {\bibfnamefont {B.}~\bibnamefont {{van
  Heck}}},\ and\ \bibinfo {author} {\bibfnamefont {G.}~\bibnamefont {{de
  Lange}}},\ }\href@noop {} {\bibinfo {title} {Dynamical polarization of the
  fermion parity in a nanowire {{Josephson}} junction}} (\bibinfo {year}
  {2021}),\ \Eprint {https://arxiv.org/abs/2112.01936} {arXiv:2112.01936}
  \BibitemShut {NoStop}%
\bibitem [{\citenamefont {Annunziata}\ \emph {et~al.}(2010)\citenamefont
  {Annunziata}, \citenamefont {Santavicca}, \citenamefont {Frunzio},
  \citenamefont {Catelani}, \citenamefont {Rooks}, \citenamefont {Frydman},\
  and\ \citenamefont {Prober}}]{annunziata_tunable_2010}%
  \BibitemOpen
  \bibfield  {author} {\bibinfo {author} {\bibfnamefont {A.~J.}\ \bibnamefont
  {Annunziata}}, \bibinfo {author} {\bibfnamefont {D.~F.}\ \bibnamefont
  {Santavicca}}, \bibinfo {author} {\bibfnamefont {L.}~\bibnamefont {Frunzio}},
  \bibinfo {author} {\bibfnamefont {G.}~\bibnamefont {Catelani}}, \bibinfo
  {author} {\bibfnamefont {M.~J.}\ \bibnamefont {Rooks}}, \bibinfo {author}
  {\bibfnamefont {A.}~\bibnamefont {Frydman}},\ and\ \bibinfo {author}
  {\bibfnamefont {D.~E.}\ \bibnamefont {Prober}},\ }\bibfield  {title}
  {\bibinfo {title} {Tunable superconducting nanoinductors},\ }\href
  {https://doi.org/10.1088/0957-4484/21/44/445202} {\bibfield  {journal}
  {\bibinfo  {journal} {Nanotechnology}\ }\textbf {\bibinfo {volume} {21}},\
  \bibinfo {pages} {445202} (\bibinfo {year} {2010})}\BibitemShut {NoStop}%
\bibitem [{\citenamefont {Plantenberg}(2007)}]{plantenberg_coupled_2007}%
  \BibitemOpen
  \bibfield  {author} {\bibinfo {author} {\bibfnamefont {J.~H.}\ \bibnamefont
  {Plantenberg}},\ }\href@noop {} {\emph {\bibinfo {title} {Coupled
  Superconducting Flux Qubits.}}}\ (\bibinfo  {publisher} {{s.n.}},\ \bibinfo
  {address} {{S.l.}},\ \bibinfo {year} {2007})\BibitemShut {NoStop}%
\bibitem [{\citenamefont {Clarke}\ and\ \citenamefont
  {Braginski}(2004)}]{clarke_squid_2004}%
  \BibitemOpen
  \bibinfo {editor} {\bibfnamefont {J.}~\bibnamefont {Clarke}}\ and\ \bibinfo
  {editor} {\bibfnamefont {A.~I.}\ \bibnamefont {Braginski}},\ eds.,\ \href
  {https://doi.org/10.1002/3527603646} {\emph {\bibinfo {title} {The {{SQUID
  Handbook}}: {{Fundamentals}} and {{Technology}} of {{SQUIDs}} and {{SQUID
  Systems}}}}},\ \bibinfo {edition} {1st}\ ed.\ (\bibinfo  {publisher}
  {{Wiley}},\ \bibinfo {year} {2004})\BibitemShut {NoStop}%
\bibitem [{\citenamefont {Samkharadze}\ \emph {et~al.}(2016)\citenamefont
  {Samkharadze}, \citenamefont {Bruno}, \citenamefont {Scarlino}, \citenamefont
  {Zheng}, \citenamefont {DiVincenzo}, \citenamefont {DiCarlo},\ and\
  \citenamefont {Vandersypen}}]{samkharadze_high-kinetic-inductance_2016-1}%
  \BibitemOpen
  \bibfield  {author} {\bibinfo {author} {\bibfnamefont {N.}~\bibnamefont
  {Samkharadze}}, \bibinfo {author} {\bibfnamefont {A.}~\bibnamefont {Bruno}},
  \bibinfo {author} {\bibfnamefont {P.}~\bibnamefont {Scarlino}}, \bibinfo
  {author} {\bibfnamefont {G.}~\bibnamefont {Zheng}}, \bibinfo {author}
  {\bibfnamefont {D.~P.}\ \bibnamefont {DiVincenzo}}, \bibinfo {author}
  {\bibfnamefont {L.}~\bibnamefont {DiCarlo}},\ and\ \bibinfo {author}
  {\bibfnamefont {L.~M.~K.}\ \bibnamefont {Vandersypen}},\ }\bibfield  {title}
  {\bibinfo {title} {High-{{Kinetic-Inductance Superconducting Nanowire
  Resonators}} for {{Circuit QED}} in a {{Magnetic Field}}},\ }\href
  {https://doi.org/10.1103/PhysRevApplied.5.044004} {\bibfield  {journal}
  {\bibinfo  {journal} {Physical Review Applied}\ }\textbf {\bibinfo {volume}
  {5}},\ \bibinfo {pages} {044004} (\bibinfo {year} {2016})}\BibitemShut
  {NoStop}%
\bibitem [{\citenamefont {Park}\ \emph {et~al.}(2020)\citenamefont {Park},
  \citenamefont {Metzger}, \citenamefont {Tosi}, \citenamefont {Goffman},
  \citenamefont {Urbina}, \citenamefont {Pothier},\ and\ \citenamefont
  {Yeyati}}]{park_adiabatic_2020-1}%
  \BibitemOpen
  \bibfield  {author} {\bibinfo {author} {\bibfnamefont {S.}~\bibnamefont
  {Park}}, \bibinfo {author} {\bibfnamefont {C.}~\bibnamefont {Metzger}},
  \bibinfo {author} {\bibfnamefont {L.}~\bibnamefont {Tosi}}, \bibinfo {author}
  {\bibfnamefont {M.~F.}\ \bibnamefont {Goffman}}, \bibinfo {author}
  {\bibfnamefont {C.}~\bibnamefont {Urbina}}, \bibinfo {author} {\bibfnamefont
  {H.}~\bibnamefont {Pothier}},\ and\ \bibinfo {author} {\bibfnamefont {A.~L.}\
  \bibnamefont {Yeyati}},\ }\bibfield  {title} {\bibinfo {title} {From
  adiabatic to dispersive readout of quantum circuits},\ }\href
  {https://doi.org/10.1103/PhysRevLett.125.077701} {\bibfield  {journal}
  {\bibinfo  {journal} {Physical Review Letters}\ }\textbf {\bibinfo {volume}
  {125}},\ \bibinfo {pages} {077701} (\bibinfo {year} {2020})},\ \Eprint
  {https://arxiv.org/abs/2007.05030} {arXiv:2007.05030} \BibitemShut {NoStop}%
\bibitem [{\citenamefont {Metzger}\ \emph {et~al.}(2021)\citenamefont
  {Metzger}, \citenamefont {Park}, \citenamefont {Tosi}, \citenamefont
  {Janvier}, \citenamefont {Reynoso}, \citenamefont {Goffman}, \citenamefont
  {Urbina}, \citenamefont {Yeyati},\ and\ \citenamefont
  {Pothier}}]{metzger_circuit-qed_2021-1}%
  \BibitemOpen
  \bibfield  {author} {\bibinfo {author} {\bibfnamefont {C.}~\bibnamefont
  {Metzger}}, \bibinfo {author} {\bibfnamefont {S.}~\bibnamefont {Park}},
  \bibinfo {author} {\bibfnamefont {L.}~\bibnamefont {Tosi}}, \bibinfo {author}
  {\bibfnamefont {C.}~\bibnamefont {Janvier}}, \bibinfo {author} {\bibfnamefont
  {A.~A.}\ \bibnamefont {Reynoso}}, \bibinfo {author} {\bibfnamefont {M.~F.}\
  \bibnamefont {Goffman}}, \bibinfo {author} {\bibfnamefont {C.}~\bibnamefont
  {Urbina}}, \bibinfo {author} {\bibfnamefont {A.~L.}\ \bibnamefont {Yeyati}},\
  and\ \bibinfo {author} {\bibfnamefont {H.}~\bibnamefont {Pothier}},\
  }\bibfield  {title} {\bibinfo {title} {Circuit-{{QED}} with phase-biased
  {{Josephson}} weak links},\ }\href
  {https://doi.org/10.1103/PhysRevResearch.3.013036} {\bibfield  {journal}
  {\bibinfo  {journal} {Physical Review Research}\ }\textbf {\bibinfo {volume}
  {3}},\ \bibinfo {pages} {013036} (\bibinfo {year} {2021})},\ \Eprint
  {https://arxiv.org/abs/2010.00430} {arXiv:2010.00430} \BibitemShut {NoStop}%
\bibitem [{\citenamefont {Tosi}\ \emph {et~al.}(2019)\citenamefont {Tosi},
  \citenamefont {Metzger}, \citenamefont {Goffman}, \citenamefont {Urbina},
  \citenamefont {Pothier}, \citenamefont {Park}, \citenamefont {Yeyati},
  \citenamefont {Nyg{\aa}rd},\ and\ \citenamefont
  {Krogstrup}}]{tosi_spin-orbit_2019}%
  \BibitemOpen
  \bibfield  {author} {\bibinfo {author} {\bibfnamefont {L.}~\bibnamefont
  {Tosi}}, \bibinfo {author} {\bibfnamefont {C.}~\bibnamefont {Metzger}},
  \bibinfo {author} {\bibfnamefont {M.~F.}\ \bibnamefont {Goffman}}, \bibinfo
  {author} {\bibfnamefont {C.}~\bibnamefont {Urbina}}, \bibinfo {author}
  {\bibfnamefont {H.}~\bibnamefont {Pothier}}, \bibinfo {author} {\bibfnamefont
  {S.}~\bibnamefont {Park}}, \bibinfo {author} {\bibfnamefont {A.~L.}\
  \bibnamefont {Yeyati}}, \bibinfo {author} {\bibfnamefont {J.}~\bibnamefont
  {Nyg{\aa}rd}},\ and\ \bibinfo {author} {\bibfnamefont {P.}~\bibnamefont
  {Krogstrup}},\ }\bibfield  {title} {\bibinfo {title} {Spin-orbit splitting of
  {{Andreev}} states revealed by microwave spectroscopy},\ }\href
  {https://doi.org/10.1103/PhysRevX.9.011010} {\bibfield  {journal} {\bibinfo
  {journal} {Physical Review X}\ }\textbf {\bibinfo {volume} {9}},\ \bibinfo
  {pages} {011010} (\bibinfo {year} {2019})},\ \Eprint
  {https://arxiv.org/abs/1810.02591} {arXiv:1810.02591} \BibitemShut {NoStop}%
\bibitem [{\citenamefont {{Matute-Ca{\~n}adas}}\ \emph
  {et~al.}(2022)\citenamefont {{Matute-Ca{\~n}adas}}, \citenamefont {Metzger},
  \citenamefont {Park}, \citenamefont {Tosi}, \citenamefont {Krogstrup},
  \citenamefont {Nyg{\aa}rd}, \citenamefont {Goffman}, \citenamefont {Urbina},
  \citenamefont {Pothier},\ and\ \citenamefont
  {Yeyati}}]{matute-canadas_signatures_2022}%
  \BibitemOpen
  \bibfield  {author} {\bibinfo {author} {\bibfnamefont {F.~J.}\ \bibnamefont
  {{Matute-Ca{\~n}adas}}}, \bibinfo {author} {\bibfnamefont {C.}~\bibnamefont
  {Metzger}}, \bibinfo {author} {\bibfnamefont {S.}~\bibnamefont {Park}},
  \bibinfo {author} {\bibfnamefont {L.}~\bibnamefont {Tosi}}, \bibinfo {author}
  {\bibfnamefont {P.}~\bibnamefont {Krogstrup}}, \bibinfo {author}
  {\bibfnamefont {J.}~\bibnamefont {Nyg{\aa}rd}}, \bibinfo {author}
  {\bibfnamefont {M.~F.}\ \bibnamefont {Goffman}}, \bibinfo {author}
  {\bibfnamefont {C.}~\bibnamefont {Urbina}}, \bibinfo {author} {\bibfnamefont
  {H.}~\bibnamefont {Pothier}},\ and\ \bibinfo {author} {\bibfnamefont {A.~L.}\
  \bibnamefont {Yeyati}},\ }\bibfield  {title} {\bibinfo {title} {Signatures of
  {{Interactions}} in the {{Andreev Spectrum}} of {{Nanowire Josephson
  Junctions}}},\ }\href {https://doi.org/10.1103/PhysRevLett.128.197702}
  {\bibfield  {journal} {\bibinfo  {journal} {Physical Review Letters}\
  }\textbf {\bibinfo {volume} {128}},\ \bibinfo {pages} {197702} (\bibinfo
  {year} {2022})}\BibitemShut {NoStop}%
\bibitem [{\citenamefont {Kringh{\o}j}\ \emph {et~al.}(2021)\citenamefont
  {Kringh{\o}j}, \citenamefont {Larsen}, \citenamefont {Erlandsson},
  \citenamefont {Uilhoorn}, \citenamefont {Kroll}, \citenamefont {Hesselberg},
  \citenamefont {McNeil}, \citenamefont {Krogstrup}, \citenamefont {Casparis},
  \citenamefont {Marcus},\ and\ \citenamefont
  {Petersson}}]{kringhoj_magnetic-field-compatible_2021-1}%
  \BibitemOpen
  \bibfield  {author} {\bibinfo {author} {\bibfnamefont {A.}~\bibnamefont
  {Kringh{\o}j}}, \bibinfo {author} {\bibfnamefont {T.~W.}\ \bibnamefont
  {Larsen}}, \bibinfo {author} {\bibfnamefont {O.}~\bibnamefont {Erlandsson}},
  \bibinfo {author} {\bibfnamefont {W.}~\bibnamefont {Uilhoorn}}, \bibinfo
  {author} {\bibfnamefont {J.}~\bibnamefont {Kroll}}, \bibinfo {author}
  {\bibfnamefont {M.}~\bibnamefont {Hesselberg}}, \bibinfo {author}
  {\bibfnamefont {R.}~\bibnamefont {McNeil}}, \bibinfo {author} {\bibfnamefont
  {P.}~\bibnamefont {Krogstrup}}, \bibinfo {author} {\bibfnamefont
  {L.}~\bibnamefont {Casparis}}, \bibinfo {author} {\bibfnamefont
  {C.}~\bibnamefont {Marcus}},\ and\ \bibinfo {author} {\bibfnamefont
  {K.}~\bibnamefont {Petersson}},\ }\bibfield  {title} {\bibinfo {title}
  {Magnetic-{{Field-Compatible Superconducting Transmon Qubit}}},\ }\href
  {https://doi.org/10.1103/PhysRevApplied.15.054001} {\bibfield  {journal}
  {\bibinfo  {journal} {Physical Review Applied}\ }\textbf {\bibinfo {volume}
  {15}},\ \bibinfo {pages} {054001} (\bibinfo {year} {2021})}\BibitemShut
  {NoStop}%
\bibitem [{\citenamefont {Winkler}\ \emph {et~al.}(2019)\citenamefont
  {Winkler}, \citenamefont {Antipov}, \citenamefont {{van Heck}}, \citenamefont
  {Soluyanov}, \citenamefont {Glazman}, \citenamefont {Wimmer},\ and\
  \citenamefont {Lutchyn}}]{winkler_unified_2019-1}%
  \BibitemOpen
  \bibfield  {author} {\bibinfo {author} {\bibfnamefont {G.~W.}\ \bibnamefont
  {Winkler}}, \bibinfo {author} {\bibfnamefont {A.~E.}\ \bibnamefont
  {Antipov}}, \bibinfo {author} {\bibfnamefont {B.}~\bibnamefont {{van Heck}}},
  \bibinfo {author} {\bibfnamefont {A.~A.}\ \bibnamefont {Soluyanov}}, \bibinfo
  {author} {\bibfnamefont {L.~I.}\ \bibnamefont {Glazman}}, \bibinfo {author}
  {\bibfnamefont {M.}~\bibnamefont {Wimmer}},\ and\ \bibinfo {author}
  {\bibfnamefont {R.~M.}\ \bibnamefont {Lutchyn}},\ }\bibfield  {title}
  {\bibinfo {title} {A unified numerical approach to
  semiconductor-superconductor heterostructures},\ }\href
  {https://doi.org/10.1103/PhysRevB.99.245408} {\bibfield  {journal} {\bibinfo
  {journal} {Physical Review B}\ }\textbf {\bibinfo {volume} {99}},\ \bibinfo
  {pages} {245408} (\bibinfo {year} {2019})},\ \Eprint
  {https://arxiv.org/abs/1810.04180} {arXiv:1810.04180 [cond-mat]} \BibitemShut
  {NoStop}%
\bibitem [{\citenamefont {Antipov}\ \emph {et~al.}(2018)\citenamefont
  {Antipov}, \citenamefont {Bargerbos}, \citenamefont {Winkler}, \citenamefont
  {Bauer}, \citenamefont {Rossi},\ and\ \citenamefont
  {Lutchyn}}]{antipov_effects_2018}%
  \BibitemOpen
  \bibfield  {author} {\bibinfo {author} {\bibfnamefont {A.~E.}\ \bibnamefont
  {Antipov}}, \bibinfo {author} {\bibfnamefont {A.}~\bibnamefont {Bargerbos}},
  \bibinfo {author} {\bibfnamefont {G.~W.}\ \bibnamefont {Winkler}}, \bibinfo
  {author} {\bibfnamefont {B.}~\bibnamefont {Bauer}}, \bibinfo {author}
  {\bibfnamefont {E.}~\bibnamefont {Rossi}},\ and\ \bibinfo {author}
  {\bibfnamefont {R.~M.}\ \bibnamefont {Lutchyn}},\ }\bibfield  {title}
  {\bibinfo {title} {Effects of gate-induced electric fields on semiconductor
  {{Majorana}} nanowires},\ }\href {https://doi.org/10.1103/PhysRevX.8.031041}
  {\bibfield  {journal} {\bibinfo  {journal} {Physical Review X}\ }\textbf
  {\bibinfo {volume} {8}},\ \bibinfo {pages} {031041} (\bibinfo {year}
  {2018})},\ \Eprint {https://arxiv.org/abs/1801.02616} {arXiv:1801.02616
  [cond-mat]} \BibitemShut {NoStop}%
\bibitem [{\citenamefont {Kitaev}(2001)}]{kitaev_unpaired_2001}%
  \BibitemOpen
  \bibfield  {author} {\bibinfo {author} {\bibfnamefont {A.}~\bibnamefont
  {Kitaev}},\ }\bibfield  {title} {\bibinfo {title} {Unpaired {{Majorana}}
  fermions in quantum wires},\ }\href
  {https://doi.org/10.1070/1063-7869/44/10S/S29} {\bibfield  {journal}
  {\bibinfo  {journal} {Physics-Uspekhi}\ }\textbf {\bibinfo {volume} {44}},\
  \bibinfo {pages} {131} (\bibinfo {year} {2001})},\ \Eprint
  {https://arxiv.org/abs/cond-mat/0010440} {arXiv:cond-mat/0010440}
  \BibitemShut {NoStop}%
\bibitem [{\citenamefont {Lutchyn}\ \emph {et~al.}(2010)\citenamefont
  {Lutchyn}, \citenamefont {Sau},\ and\ \citenamefont
  {Sarma}}]{lutchyn_majorana_2010}%
  \BibitemOpen
  \bibfield  {author} {\bibinfo {author} {\bibfnamefont {R.~M.}\ \bibnamefont
  {Lutchyn}}, \bibinfo {author} {\bibfnamefont {J.~D.}\ \bibnamefont {Sau}},\
  and\ \bibinfo {author} {\bibfnamefont {S.~D.}\ \bibnamefont {Sarma}},\
  }\bibfield  {title} {\bibinfo {title} {Majorana {{Fermions}} and a
  {{Topological Phase Transition}} in {{Semiconductor-Superconductor
  Heterostructures}}},\ }\href {https://doi.org/10.1103/PhysRevLett.105.077001}
  {\bibfield  {journal} {\bibinfo  {journal} {Physical Review Letters}\
  }\textbf {\bibinfo {volume} {105}},\ \bibinfo {pages} {077001} (\bibinfo
  {year} {2010})},\ \Eprint {https://arxiv.org/abs/1002.4033} {arXiv:1002.4033}
  \BibitemShut {NoStop}%
\bibitem [{\citenamefont {Oreg}\ \emph {et~al.}(2010)\citenamefont {Oreg},
  \citenamefont {Refael},\ and\ \citenamefont {{von
  Oppen}}}]{oreg_helical_2010}%
  \BibitemOpen
  \bibfield  {author} {\bibinfo {author} {\bibfnamefont {Y.}~\bibnamefont
  {Oreg}}, \bibinfo {author} {\bibfnamefont {G.}~\bibnamefont {Refael}},\ and\
  \bibinfo {author} {\bibfnamefont {F.}~\bibnamefont {{von Oppen}}},\
  }\bibfield  {title} {\bibinfo {title} {Helical liquids and {{Majorana}} bound
  states in quantum wires},\ }\href
  {https://doi.org/10.1103/PhysRevLett.105.177002} {\bibfield  {journal}
  {\bibinfo  {journal} {Physical Review Letters}\ }\textbf {\bibinfo {volume}
  {105}},\ \bibinfo {pages} {177002} (\bibinfo {year} {2010})},\ \Eprint
  {https://arxiv.org/abs/1003.1145} {arXiv:1003.1145 [cond-mat]} \BibitemShut
  {NoStop}%
\bibitem [{\citenamefont {Khalil}\ \emph {et~al.}(2012)\citenamefont {Khalil},
  \citenamefont {Stoutimore}, \citenamefont {Wellstood},\ and\ \citenamefont
  {Osborn}}]{khalil_analysis_2012}%
  \BibitemOpen
  \bibfield  {author} {\bibinfo {author} {\bibfnamefont {M.~S.}\ \bibnamefont
  {Khalil}}, \bibinfo {author} {\bibfnamefont {M.~J.~A.}\ \bibnamefont
  {Stoutimore}}, \bibinfo {author} {\bibfnamefont {F.~C.}\ \bibnamefont
  {Wellstood}},\ and\ \bibinfo {author} {\bibfnamefont {K.~D.}\ \bibnamefont
  {Osborn}},\ }\bibfield  {title} {\bibinfo {title} {An analysis method for
  asymmetric resonator transmission applied to superconducting devices},\
  }\href {https://doi.org/10.1063/1.3692073} {\bibfield  {journal} {\bibinfo
  {journal} {Journal of Applied Physics}\ }\textbf {\bibinfo {volume} {111}},\
  \bibinfo {pages} {054510} (\bibinfo {year} {2012})}\BibitemShut {NoStop}%
\bibitem [{\citenamefont {Kroll}\ \emph {et~al.}(2019)\citenamefont {Kroll},
  \citenamefont {Borsoi}, \citenamefont {{van der Enden}}, \citenamefont
  {Uilhoorn}, \citenamefont {{de Jong}}, \citenamefont {{Quintero-P{\'e}rez}},
  \citenamefont {{van Woerkom}}, \citenamefont {Bruno}, \citenamefont
  {Plissard}, \citenamefont {Car}, \citenamefont {Bakkers}, \citenamefont
  {Cassidy},\ and\ \citenamefont {Kouwenhoven}}]{kroll_magnetic_2019-2}%
  \BibitemOpen
  \bibfield  {author} {\bibinfo {author} {\bibfnamefont {J.~G.}\ \bibnamefont
  {Kroll}}, \bibinfo {author} {\bibfnamefont {F.}~\bibnamefont {Borsoi}},
  \bibinfo {author} {\bibfnamefont {K.~L.}\ \bibnamefont {{van der Enden}}},
  \bibinfo {author} {\bibfnamefont {W.}~\bibnamefont {Uilhoorn}}, \bibinfo
  {author} {\bibfnamefont {D.}~\bibnamefont {{de Jong}}}, \bibinfo {author}
  {\bibfnamefont {M.}~\bibnamefont {{Quintero-P{\'e}rez}}}, \bibinfo {author}
  {\bibfnamefont {D.~J.}\ \bibnamefont {{van Woerkom}}}, \bibinfo {author}
  {\bibfnamefont {A.}~\bibnamefont {Bruno}}, \bibinfo {author} {\bibfnamefont
  {S.~R.}\ \bibnamefont {Plissard}}, \bibinfo {author} {\bibfnamefont
  {D.}~\bibnamefont {Car}}, \bibinfo {author} {\bibfnamefont {E.~P. A.~M.}\
  \bibnamefont {Bakkers}}, \bibinfo {author} {\bibfnamefont {M.~C.}\
  \bibnamefont {Cassidy}},\ and\ \bibinfo {author} {\bibfnamefont {L.~P.}\
  \bibnamefont {Kouwenhoven}},\ }\bibfield  {title} {\bibinfo {title} {Magnetic
  field resilient superconducting coplanar waveguide resonators for hybrid
  {{cQED}} experiments},\ }\href
  {https://doi.org/10.1103/PhysRevApplied.11.064053} {\bibfield  {journal}
  {\bibinfo  {journal} {Physical Review Applied}\ }\textbf {\bibinfo {volume}
  {11}},\ \bibinfo {pages} {064053} (\bibinfo {year} {2019})},\ \Eprint
  {https://arxiv.org/abs/1809.03932} {arXiv:1809.03932} \BibitemShut {NoStop}%
\bibitem [{\citenamefont {Groth}\ \emph {et~al.}(2014)\citenamefont {Groth},
  \citenamefont {Wimmer}, \citenamefont {Akhmerov},\ and\ \citenamefont
  {Waintal}}]{groth_kwant_2014}%
  \BibitemOpen
  \bibfield  {author} {\bibinfo {author} {\bibfnamefont {C.~W.}\ \bibnamefont
  {Groth}}, \bibinfo {author} {\bibfnamefont {M.}~\bibnamefont {Wimmer}},
  \bibinfo {author} {\bibfnamefont {A.~R.}\ \bibnamefont {Akhmerov}},\ and\
  \bibinfo {author} {\bibfnamefont {X.}~\bibnamefont {Waintal}},\ }\bibfield
  {title} {\bibinfo {title} {Kwant: A software package for quantum transport},\
  }\href {https://doi.org/10.1088/1367-2630/16/6/063065} {\bibfield  {journal}
  {\bibinfo  {journal} {New Journal of Physics}\ }\textbf {\bibinfo {volume}
  {16}},\ \bibinfo {pages} {063065} (\bibinfo {year} {2014})}\BibitemShut
  {NoStop}%
\bibitem [{\citenamefont {Laeven}\ \emph {et~al.}(2020)\citenamefont {Laeven},
  \citenamefont {Nijholt}, \citenamefont {Wimmer},\ and\ \citenamefont
  {Akhmerov}}]{laeven_enhanced_2020}%
  \BibitemOpen
  \bibfield  {author} {\bibinfo {author} {\bibfnamefont {T.}~\bibnamefont
  {Laeven}}, \bibinfo {author} {\bibfnamefont {B.}~\bibnamefont {Nijholt}},
  \bibinfo {author} {\bibfnamefont {M.}~\bibnamefont {Wimmer}},\ and\ \bibinfo
  {author} {\bibfnamefont {A.~R.}\ \bibnamefont {Akhmerov}},\ }\bibfield
  {title} {\bibinfo {title} {Enhanced {{Proximity Effect}} in {{Zigzag-Shaped
  Majorana Josephson Junctions}}},\ }\href
  {https://doi.org/10.1103/PhysRevLett.125.086802} {\bibfield  {journal}
  {\bibinfo  {journal} {Physical Review Letters}\ }\textbf {\bibinfo {volume}
  {125}},\ \bibinfo {pages} {086802} (\bibinfo {year} {2020})}\BibitemShut
  {NoStop}%
\bibitem [{\citenamefont {{van Heck}}\ \emph {et~al.}(2017)\citenamefont {{van
  Heck}}, \citenamefont {V{\"a}yrynen},\ and\ \citenamefont
  {Glazman}}]{van_heck_zeeman_2017}%
  \BibitemOpen
  \bibfield  {author} {\bibinfo {author} {\bibfnamefont {B.}~\bibnamefont {{van
  Heck}}}, \bibinfo {author} {\bibfnamefont {J.~I.}\ \bibnamefont
  {V{\"a}yrynen}},\ and\ \bibinfo {author} {\bibfnamefont {L.~I.}\ \bibnamefont
  {Glazman}},\ }\bibfield  {title} {\bibinfo {title} {Zeeman and spin-orbit
  effects in the {{Andreev}} spectra of nanowire junctions},\ }\href
  {https://doi.org/10.1103/PhysRevB.96.075404} {\bibfield  {journal} {\bibinfo
  {journal} {Physical Review B}\ }\textbf {\bibinfo {volume} {96}},\ \bibinfo
  {pages} {075404} (\bibinfo {year} {2017})}\BibitemShut {NoStop}%
\bibitem [{\citenamefont {Beenakker}\ and\ \citenamefont {{van
  Houten}}(1992)}]{beenakker_resonant_1992-1}%
  \BibitemOpen
  \bibfield  {author} {\bibinfo {author} {\bibfnamefont {C.~W.~J.}\
  \bibnamefont {Beenakker}}\ and\ \bibinfo {author} {\bibfnamefont
  {H.}~\bibnamefont {{van Houten}}},\ }\bibfield  {title} {\bibinfo {title}
  {Resonant {{Josephson}} current through a quantum dot}\ }(\bibinfo {year}
  {1992})\ pp.\ \bibinfo {pages} {175--179},\ \Eprint
  {https://arxiv.org/abs/cond-mat/0111505} {arXiv:cond-mat/0111505}
  \BibitemShut {NoStop}%
\bibitem [{\citenamefont {Park}\ and\ \citenamefont
  {Yeyati}(2017)}]{park_andreev_2017}%
  \BibitemOpen
  \bibfield  {author} {\bibinfo {author} {\bibfnamefont {S.}~\bibnamefont
  {Park}}\ and\ \bibinfo {author} {\bibfnamefont {A.~L.}\ \bibnamefont
  {Yeyati}},\ }\bibfield  {title} {\bibinfo {title} {Andreev spin qubits in
  multichannel {{Rashba}} nanowires},\ }\href
  {https://doi.org/10.1103/PhysRevB.96.125416} {\bibfield  {journal} {\bibinfo
  {journal} {Physical Review B}\ }\textbf {\bibinfo {volume} {96}},\ \bibinfo
  {pages} {125416} (\bibinfo {year} {2017})},\ \Eprint
  {https://arxiv.org/abs/1707.04273} {arXiv:1707.04273} \BibitemShut {NoStop}%
\bibitem [{\citenamefont {B\'eri}\ \emph {et~al.}(2008)\citenamefont {B\'eri},
  \citenamefont {Bardarson},\ and\ \citenamefont {Beenakker}}]{beri2008}%
  \BibitemOpen
  \bibfield  {author} {\bibinfo {author} {\bibfnamefont {B.}~\bibnamefont
  {B\'eri}}, \bibinfo {author} {\bibfnamefont {J.~H.}\ \bibnamefont
  {Bardarson}},\ and\ \bibinfo {author} {\bibfnamefont {C.~W.~J.}\ \bibnamefont
  {Beenakker}},\ }\bibfield  {title} {\bibinfo {title} {Splitting of andreev
  levels in a josephson junction by spin-orbit coupling},\ }\href
  {https://doi.org/10.1103/PhysRevB.77.045311} {\bibfield  {journal} {\bibinfo
  {journal} {Phys. Rev. B}\ }\textbf {\bibinfo {volume} {77}},\ \bibinfo
  {pages} {045311} (\bibinfo {year} {2008})}\BibitemShut {NoStop}%
\bibitem [{\citenamefont {Katsaros}\ \emph {et~al.}(2020)\citenamefont
  {Katsaros}, \citenamefont {Kuku{\v c}ka}, \citenamefont {Vuku{\v s}i{\'c}},
  \citenamefont {Watzinger}, \citenamefont {Gao}, \citenamefont {Wang},
  \citenamefont {Zhang},\ and\ \citenamefont {Held}}]{katsaros_zero_2020}%
  \BibitemOpen
  \bibfield  {author} {\bibinfo {author} {\bibfnamefont {G.}~\bibnamefont
  {Katsaros}}, \bibinfo {author} {\bibfnamefont {J.}~\bibnamefont {Kuku{\v
  c}ka}}, \bibinfo {author} {\bibfnamefont {L.}~\bibnamefont {Vuku{\v
  s}i{\'c}}}, \bibinfo {author} {\bibfnamefont {H.}~\bibnamefont {Watzinger}},
  \bibinfo {author} {\bibfnamefont {F.}~\bibnamefont {Gao}}, \bibinfo {author}
  {\bibfnamefont {T.}~\bibnamefont {Wang}}, \bibinfo {author} {\bibfnamefont
  {J.-J.}\ \bibnamefont {Zhang}},\ and\ \bibinfo {author} {\bibfnamefont
  {K.}~\bibnamefont {Held}},\ }\bibfield  {title} {\bibinfo {title} {Zero
  {{Field Splitting}} of {{Heavy-Hole States}} in {{Quantum Dots}}},\ }\href
  {https://doi.org/10.1021/acs.nanolett.0c01466} {\bibfield  {journal}
  {\bibinfo  {journal} {Nano Letters}\ }\textbf {\bibinfo {volume} {20}},\
  \bibinfo {pages} {5201} (\bibinfo {year} {2020})}\BibitemShut {NoStop}%
\bibitem [{\citenamefont {Newville}\ \emph {et~al.}(2021)\citenamefont
  {Newville}, \citenamefont {Otten}, \citenamefont {Nelson}, \citenamefont
  {Ingargiola}, \citenamefont {Stensitzki}, \citenamefont {Allan},
  \citenamefont {Fox}, \citenamefont {Carter}, \citenamefont {Micha{\l}},
  \citenamefont {Osborn}, \citenamefont {Pustakhod}, \citenamefont {Lneuhaus},
  \citenamefont {Weigand}, \citenamefont {Glenn}, \citenamefont {Deil},
  \citenamefont {Mark}, \citenamefont {Hansen}, \citenamefont {Pasquevich},
  \citenamefont {Foks}, \citenamefont {Zobrist}, \citenamefont {Frost},
  \citenamefont {Beelen}, \citenamefont {Stuermer}, \citenamefont {Azelcer},
  \citenamefont {Hannum}, \citenamefont {Polloreno}, \citenamefont {Nielsen},
  \citenamefont {Caldwell}, \citenamefont {Almarza},\ and\ \citenamefont
  {Persaud}}]{newville_lmfitlmfit-py_2021}%
  \BibitemOpen
  \bibfield  {author} {\bibinfo {author} {\bibfnamefont {M.}~\bibnamefont
  {Newville}}, \bibinfo {author} {\bibfnamefont {R.}~\bibnamefont {Otten}},
  \bibinfo {author} {\bibfnamefont {A.}~\bibnamefont {Nelson}}, \bibinfo
  {author} {\bibfnamefont {A.}~\bibnamefont {Ingargiola}}, \bibinfo {author}
  {\bibfnamefont {T.}~\bibnamefont {Stensitzki}}, \bibinfo {author}
  {\bibfnamefont {D.}~\bibnamefont {Allan}}, \bibinfo {author} {\bibfnamefont
  {A.}~\bibnamefont {Fox}}, \bibinfo {author} {\bibfnamefont {F.}~\bibnamefont
  {Carter}}, \bibinfo {author} {\bibnamefont {Micha{\l}}}, \bibinfo {author}
  {\bibfnamefont {R.}~\bibnamefont {Osborn}}, \bibinfo {author} {\bibfnamefont
  {D.}~\bibnamefont {Pustakhod}}, \bibinfo {author} {\bibnamefont {Lneuhaus}},
  \bibinfo {author} {\bibfnamefont {S.}~\bibnamefont {Weigand}}, \bibinfo
  {author} {\bibnamefont {Glenn}}, \bibinfo {author} {\bibfnamefont
  {C.}~\bibnamefont {Deil}}, \bibinfo {author} {\bibnamefont {Mark}}, \bibinfo
  {author} {\bibfnamefont {A.~L.~R.}\ \bibnamefont {Hansen}}, \bibinfo {author}
  {\bibfnamefont {G.}~\bibnamefont {Pasquevich}}, \bibinfo {author}
  {\bibfnamefont {L.}~\bibnamefont {Foks}}, \bibinfo {author} {\bibfnamefont
  {N.}~\bibnamefont {Zobrist}}, \bibinfo {author} {\bibfnamefont
  {O.}~\bibnamefont {Frost}}, \bibinfo {author} {\bibfnamefont
  {A.}~\bibnamefont {Beelen}}, \bibinfo {author} {\bibnamefont {Stuermer}},
  \bibinfo {author} {\bibnamefont {Azelcer}}, \bibinfo {author} {\bibfnamefont
  {A.}~\bibnamefont {Hannum}}, \bibinfo {author} {\bibfnamefont
  {A.}~\bibnamefont {Polloreno}}, \bibinfo {author} {\bibfnamefont {J.~H.}\
  \bibnamefont {Nielsen}}, \bibinfo {author} {\bibfnamefont {S.}~\bibnamefont
  {Caldwell}}, \bibinfo {author} {\bibfnamefont {A.}~\bibnamefont {Almarza}},\
  and\ \bibinfo {author} {\bibfnamefont {A.}~\bibnamefont {Persaud}},\ }\href
  {https://doi.org/10.5281/ZENODO.598352} {\bibinfo {title} {Lmfit/lmfit-py:
  1.0.3}},\ \bibinfo {howpublished} {Zenodo} (\bibinfo {year}
  {2021})\BibitemShut {NoStop}%
\end{thebibliography}%

\end{document}


\preprint{APS/123-QED}

\title{Supplementary information: Microwave spectroscopy of interacting Andreev spins}
\author{J.\,J. Wesdorp\textsuperscript{1}}\email{j.j.wesdorp@tudelft.nl}
\author{A. Vaartjes\textsuperscript{1}}
\author{L. Gr{\"u}nhaupt\textsuperscript{1}}
\author{T. Laeven\textsuperscript{2}}
\author{S. Roelofs\textsuperscript{1}}
\author{L.\,J. Splitthoff\textsuperscript{1}}
\author{M. Pita-Vidal\textsuperscript{1}}
\author{A. Bargerbos\textsuperscript{1}}
\author{D. J. van Woerkom\textsuperscript{2}}
\author{L. P. Kouwenhoven\textsuperscript{1}}
\author{C. K. Andersen\textsuperscript{1}}
\author{B. van Heck\textsuperscript{2}}
\author{G. de Lange\textsuperscript{2}}\email{gijs.delange@microsoft.com}
\affiliation{\textsuperscript{1}QuTech and Kavli Institute of Nanoscience, Delft University of Technology, 2628 CJ, Delft, The Netherlands \\
\textsuperscript{2}Microsoft Quantum Lab Delft, 2628 CJ, Delft, The Netherlands}
\author{F. J. Matute-Cañadas\textsuperscript{4}}
\author{A. Levy Yeyati\textsuperscript{4}}
\affiliation{\textsuperscript{4} Departamento de Física Teórica de la Materia Condensada, Condensed Matter Physics Center (IFIMAC) and Instituto Nicolás Cabrera, Universidad Autónoma de Madrid, 28049 Madrid, Spain}
\author{P. Krogstrup\textsuperscript{3}}
\affiliation{\textsuperscript{3} Center for Quantum Devices, Niels Bohr Institute, University of Copenhagen
and Microsoft Quantum Materials Lab Copenhagen, Denmark}
\date{\today}
\maketitle
\tableofcontents
\newpage

\section{Methods}
\begin{figure}[b]
   \includegraphics{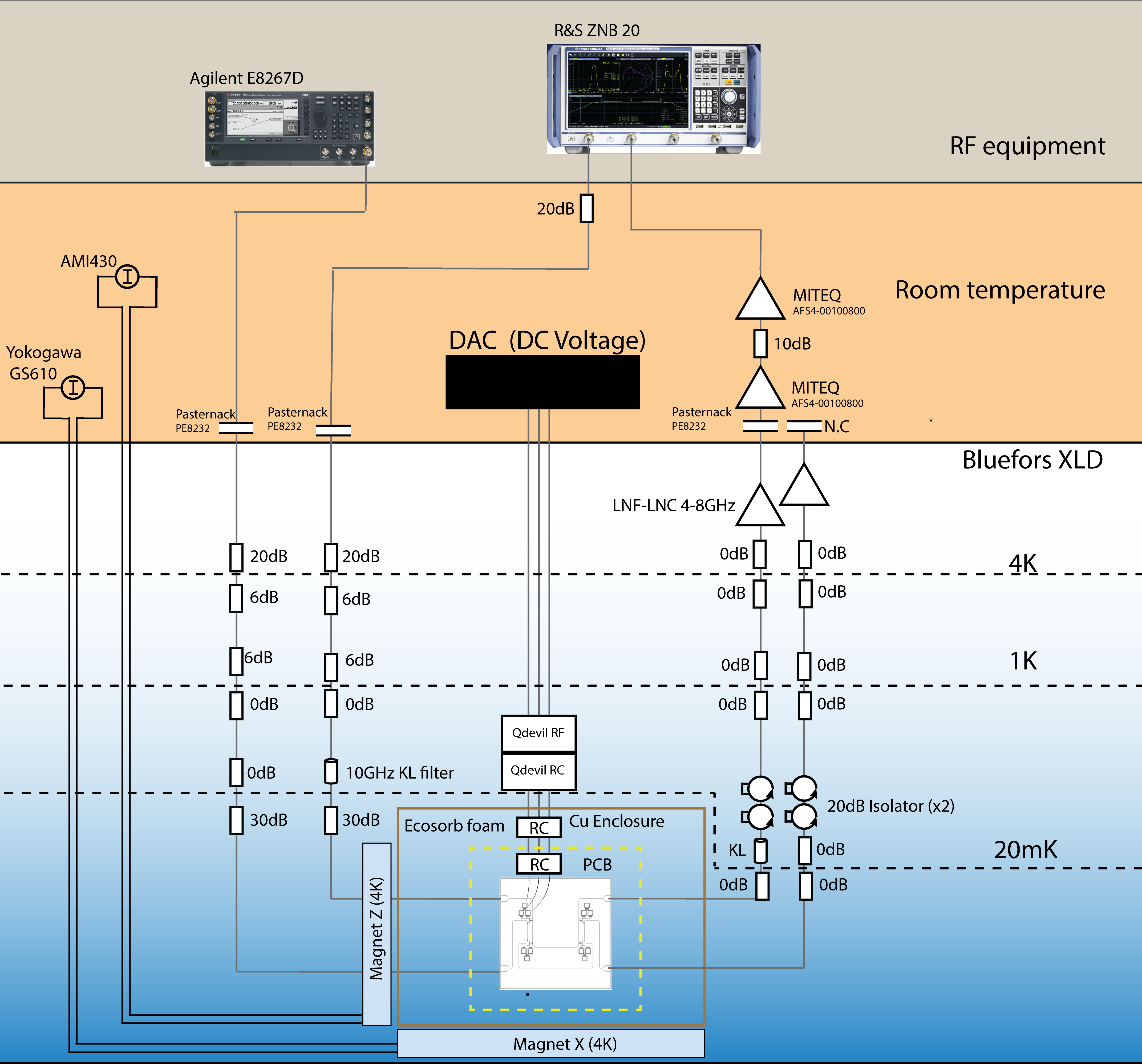}
   \caption{\label{fig:wiring-diagram} Wiring diagram of the experimental setup.}
\end{figure}

The sample is fabricated using the methods described in~\cite{wesdorp_dynamical_2021}.
Since this is the same device as used in~\cite{wesdorp_dynamical_2021}, we refer to the supplementary information there for detailed device images, a fit of the resonator quality factor at zero field and targeted coupling strength to the Andreev bound states (ABS).  A wiring diagram of the setup is shown in~\cref{fig:wiring-diagram}. Measurements were performed at a temperature of $\approx \SI{20}{\milli\kelvin}$ in a Bluefors XLD dilution refrigerator. A magnetic field is applied using a 6-1-1 Tesla vector magnet thermally anchored at the 4K stage. The Z-axis (\SI{6}{\tesla}) is controlled with a large AMI430 current source (\SI{60}{\ampere} max). For the X and Y axis we used smaller Yokogawa GS210 (\SI{200}{\milli\ampere} max) and GS610 (\SI{3}{\ampere} max) sources respectively in order to get finer flux control and more precise alignment with the out-of-plane (Y) field. This limited the maximum field to $\sim\SI{45}{\milli\tesla}$ ($\SI{3}{\milli\tesla}$) in the X (Y) direction. 

\subsection{Resonator frequency targeting}
We targeted the resonator frequencies using simulations in AWR Microwave office, that uses the method of moments to solve quasi-3D geometries. The simulation consists of a stack of materials infinite in the z-plane and finite in the x-y plane.
The stack consisted of \SI{500}{\micro\meter} silicon with a dielectric constant of $\epsilon_\mathrm{r}=11.7\epsilon_0$, where $\epsilon_0$ is the vacuum permittivity,  followed by a metal with the circuit design, and finally a layer of vacuum. To emulate the kinetic inductance of the metal, we simulated kinetic inductance as a fixed inductance per square in the thin-film approximation.
The kinetic inductance per square was estimated from normal state resistance measurements of the feedline using a dirty superconductor approximation~\cite{annunziata_tunable_2010}.
We found a normal state sheet resistance of the NbTiN film $R_\square=\SI{87}{\ohm}$ and used a critical temperature $T_\mathrm{c}=\SI{10.6}{\kelvin}$ found in similar devices to obtain an estimate of $L_{k/\square}=\SI{11.3}{\pico\henry}$. Using these simulations, we reproduced the fundamental mode of the resonator at \SI{4.82}{\giga\hertz} when using $L_\mathrm{k/\square}=\SI{14.5}{\pico\henry}$.
Using the same settings, the first higher harmonic was found to be the self-resonance of the inductor at \SI{28.5}{\giga\hertz}, making the lumped-element resonators very suitable for spectroscopy in the frequency range of interest. 

\subsection{Linearity of flux-phase in gradiometric SQUID}\label{sec:sup:gradiometer}
To verify that we can measure the un-distorted phase dispersion of ABS, we model the phase difference over the junction $\varphi$ as a function of the externally applied magnetic flux $\Phi$.
The relation should be linear and limits the maximum allowed shunt-inductance in the SQUID loop design. We use a procedure that calculates $\varphi(\Phi)$ assuming a sinusoidal current phase relation (CPR) with typical critical current strength as this remains approximately valid in the case of a few channel-wire with a modified CPR.
\begin{figure}[h!]
  \centering
   \includegraphics{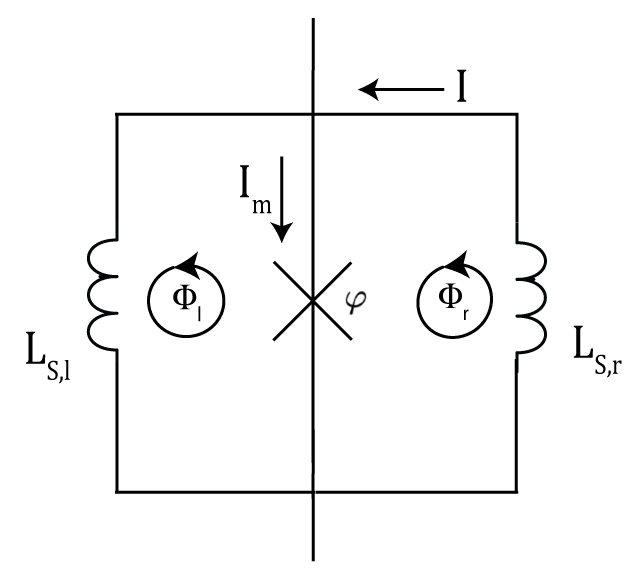}
   \caption{Sketch of the gradiometer SQUID (top view)} \label{fig-sup-gradiometer-schematic}
 \end{figure}

 We begin by a change of variables from the current flowing through the left loop $I_\mathrm{l}$ and right loop $I_\mathrm{r}$ to the total outer loop current $I$ and the middle branch current $I_\mathrm{m}$. 
  \begin{align}
I_\mathrm{r} = I + \frac{1}{2}I_\mathrm{m} \\
I_\mathrm{l} = I - \frac{1}{2}I_\mathrm{m}
\end{align}
The first step is to quantize the flux in the loops. See e.g. Ref~\cite{plantenberg_coupled_2007} and Ref.~\cite{clarke_squid_2004} Sections 2.3 and 2.1.
\begin{align}
\frac{2\pi}{\phi_0}\left[\phi_\mathrm{r} - L_\mathrm{r}I_\mathrm{r} + MI_\mathrm{l} \right] + \varphi = 2\pi n_\mathrm{r} \label{eq:sup-gradiometer-left-loop} \\
\frac{2\pi}{\phi_0}\left[\phi_\mathrm{l} - L_\mathrm{l}I_\mathrm{l} + MI_\mathrm{r} \right] - \varphi = 2\pi n_\mathrm{l}  \label{eq:sup-gradiometer-right-loop}
\end{align}
Here, $\phi_0$ is the flux quantum, $\phi_\mathrm{r}, \phi_\mathrm{l}$ the externally applied flux in each of the sub-loops. $L_\mathrm{r}, L_\mathrm{l}$ correspond to the total inductance, kinetic and geometric, in the right loop and left loop respectively and $I_\mathrm{r}, I_\mathrm{l}$ are the currents inside each of the loops. $M$ is the mutual inductance between the loops.

Adding~Eq. \eqref{eq:sup-gradiometer-left-loop} and~Eq. \eqref{eq:sup-gradiometer-right-loop} gives a condition for the total loop flux $\phi_\mathrm{E} = \phi_\mathrm{l} + \phi_\mathrm{r}$ while the middle branch drops out
\begin{equation}
\phi_\mathrm{r} + \phi_\mathrm{l} - \phi_\mathrm{w} = \left(n_\mathrm{r} + n_\mathrm{l}\right)\phi_0
\end{equation}
with $\phi_\mathrm{w} = I\left(L_\mathrm{r} + L_\mathrm{l} - 2M\right) - I_\mathrm{m}\left(\frac{L_\mathrm{l}-L_\mathrm{r}}{2}\right)$ the total flux coming from the current in the big loop. Taking the difference of the loop equations results in an expression for the JJ phase difference $\varphi$ as a function of external fluxes
\begin{equation}
\frac{\pi}{\phi_0}\left[ \phi_\mathrm{r}-\phi_\mathrm{l} + I\left(L_\mathrm{l}-L_\mathrm{r}\right) - I_\mathrm{m}\left(\frac{L_\mathrm{r} + L_\mathrm{l}}{2} + M\right)\right] + \varphi  = \pi\left(n_\mathrm{r} - n_\mathrm{l}\right)
\end{equation}
Replacing the current in the big loop $I$ using the sum of the flux quantization equations and combining that into the difference of the flux quantization equations results in an expression for $\varphi$ which only depends on itself, the CPR $I_\mathrm{m}(\varphi)$ and known factors
\begin{equation}
  \begin{aligned}
&\varphi + I_\mathrm{m}\left(\varphi\right)\frac{\pi}{\phi_0}\left[\frac{\left(L_\mathrm{l} - L_\mathrm{r}\right)^2}{2\left(L_\mathrm{r}+L_\mathrm{l}-2M\right)} - \frac{1}{2}\left(L_\mathrm{r} + L_\mathrm{l}\right) - M \right] = \\
 &\pi\left(n_\mathrm{r}-n_\mathrm{l}\right) - \frac{\pi}{\phi_0}\left(\phi_\mathrm{r}-\phi_\mathrm{l}\right) - \frac{\pi}{\phi_0}\frac{L_\mathrm{l}-L_\mathrm{r}}{L_\mathrm{r}+L_\mathrm{l} - 2M}\left(\phi_0\left(n_\mathrm{r}+n_\mathrm{l}\right) - \phi_\mathrm{r} - \phi_\mathrm{l}\right)
  \end{aligned}
\end{equation}
We can solve this equation numerically assuming a sinusoidal CPR $I_\mathrm{m}(\varphi)=\sin\varphi$  and get the phase drop over the junction as a function of applied external field as a result. Using the above model, for a critical current $I_\mathrm{c}=\SI{10}{\nano\ampere}$ and the inductances specified in the circuit, we have investigated the relation and remain well in the linear regime. Non-linearities start appearing at $I_\mathrm{c}\sim\SI{100}{\nano\ampere}$, much bigger than what we estimate the nanowire to have based on the change in \fnull\ of the resonator (see discussion in main text near Fig. 1).

\FloatBarrier
\subsection{Magnetic field alignment procedure}
\label{sec:field_alignment}
\FloatBarrier
\begin{figure}[h]
     \centering
     \includegraphics[width=\textwidth]{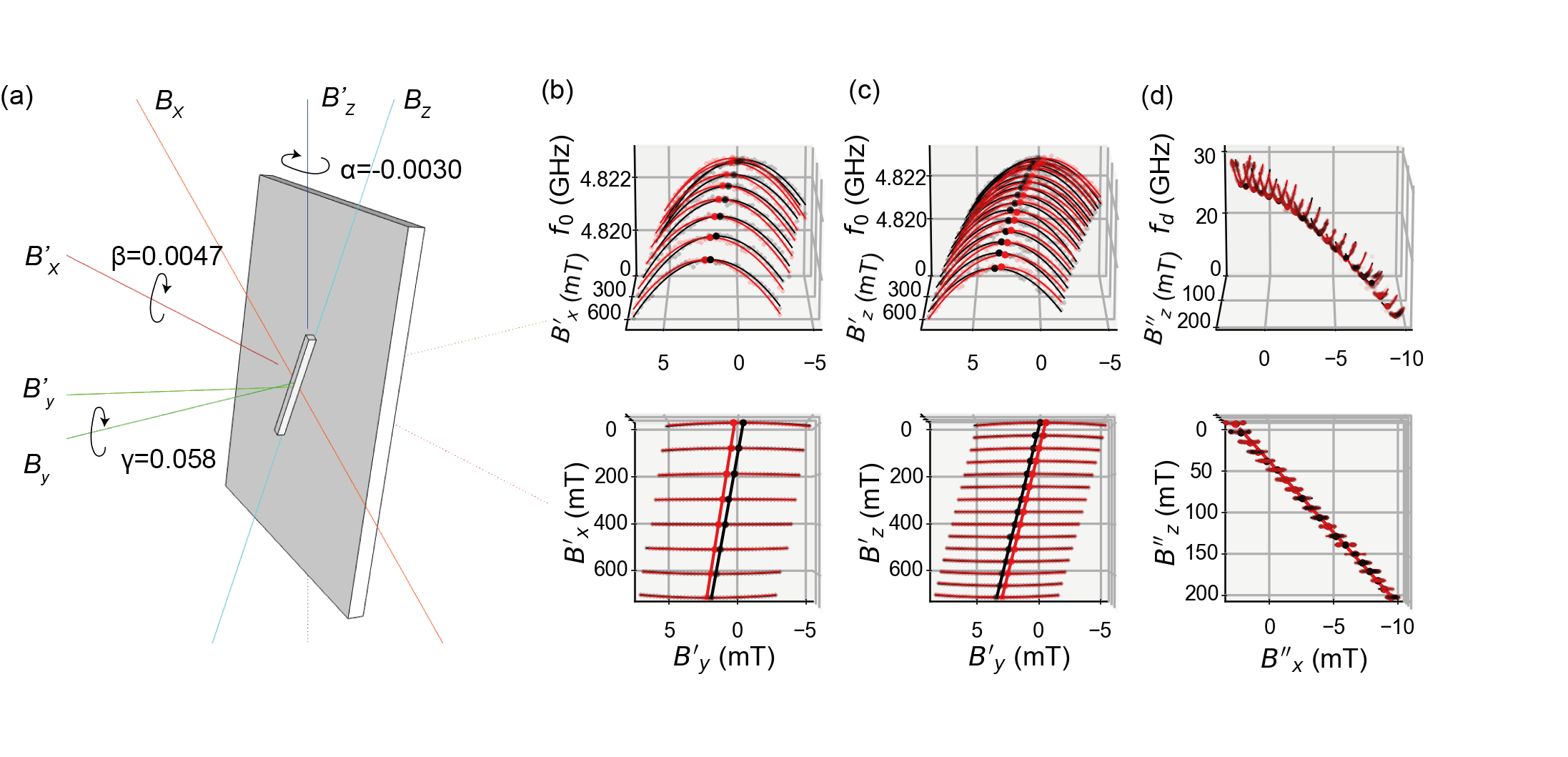}
    \caption{\textbf{Magnetic field alignment. a)} $\alpha$ and $\beta$ define the chip rotation and $\gamma$ describes the rotation of the nanowire on the chip plane. The rotation angles are exaggerated for clarity. 
    \textbf{b)} $\alpha$ alignment (front and top view). The maxima of the resonance frequency parabolas indicate a zero $B_y$ component. Black and red lines, indicating forward and backward sweeps, show a slight hysteresis. The slope of the linear fits through the maxima determine the rotation angle around the $z$-axis: $\alpha=-0.0030 \pm 0.0002$.
    \textbf{c)} Determination of the rotation angle around the $x$-axis, $\beta=0.0047 \pm 0.00006$ 
    \textbf{d)} Here, minima of an even ABS transition in two-tone spectroscopy are used to determine the nanowire angle $\gamma$. Linear fits give $\gamma = 0.058 \pm 0.0002$. The minimas are extracted by the procedure described in appendix \ref{sec:data-extraction}. Note that the hysteresis in the magnet and on-chip trapped flux due to circulating currents and vortices cause a larger effective uncertainty than quoted here based on the fit uncertainty alone.}
    \label{fig:chip-alignment}
\end{figure}

Here we explain the alignment of the magnetic field with the nanowire coordinate system. We denote the raw field vector as $\vec{B}'$ and the aligned coordinate vector as $\vec{B}$.
Figure \ref{fig:chip-alignment} (a) shows a schematic of the misaligned chip with respect to the coordinate system of the magnetic field. The rotation of the nanowire with respect to the magnet coordinate system is defined by three (extrinsic) Euler angles: $\alpha$, $\beta$ and $\gamma$. $\alpha$ and $\beta$ determine the orientation of the whole chip, whereas $\gamma$ defines the on-chip rotation of the nanowire. The Euler rotation matrices are given by

\begin{equation}
    R_z(\alpha) = 
    \begin{pmatrix}
     \cos(\alpha) & -\sin(\alpha) & 0\\
     \sin(\alpha) & \cos(\alpha) & 0\\
     0 & 0 & 1
    \end{pmatrix} \hskip 0.3cm
    R_x(\beta) = 
    \begin{pmatrix}
     1 & 0 & 0\\
     0 & \cos(\beta) & -\sin(\beta) \\
     0 & \sin(\beta) & \cos(\beta)
    \end{pmatrix} \hskip 0.3cm
    R_y(\gamma) = 
    \begin{pmatrix}
     \cos(\gamma) & 0 & \sin(\gamma)\\
     0 & 1 & 0 \\
     -\sin(\gamma) & 0 & \cos(\gamma)\\
    \end{pmatrix}.
\end{equation}

We transform from $\vec{B'}$ to $\vec{B}$ as follows
\begin{equation}
\vec{B}=
R_y(\gamma)R_x(\beta)R_z(\alpha)
\vec{B'}
\label{eq:transformation}, 
\end{equation}
or in matrix form:
\begin{equation}
\begin{pmatrix}
 B_x \\ B_y \\ B_z
\end{pmatrix}
 = 
    \begin{pmatrix}
     \cos(\alpha)\cos(\gamma) + \sin(\gamma)\sin(\beta)\sin(\alpha) & -\cos(\gamma)\sin(\alpha)+\sin(\gamma)\sin(\beta)\cos(\alpha) & \sin(\gamma)\cos(\beta) \\
     \cos(\beta)\sin(\alpha) & \cos(\alpha)\cos(\beta) & -\sin(\beta) \\
     -\sin(\gamma)\cos(\alpha) + \cos(\gamma)\sin(\beta)\sin(\alpha) & \sin(\gamma)\sin(\alpha) +  \cos(\gamma)\sin(\beta)\cos(\alpha) & \cos(\gamma)\cos(\beta)
    \end{pmatrix}
    \begin{pmatrix}
    B'_x \\ B'_y \\ B'_z
    \end{pmatrix}
    \label{eq:rotation_matrix}
\end{equation}

Due to increase of Cooper-pair breaking rate~\cite{samkharadze_high-kinetic-inductance_2016-1, annunziata_tunable_2010}, the kinetic inductance of a thin-film resonator increases with out-of-plane magnetic field more strongly than with in-plane magnetic field. This causes the resonance frequency to decrease parabolically with an increasing out-of-plane magnetic field, and we can use the maximum of this parabola to determine where the out-of-plane field is zero. 
In order to avoid additional changes in the resonance frequency due to SQUID oscillations, we go to gate voltages where the wire is pinched-off and measure the resonance frequency as a function of the (out-of-plane) field $B'_y$. This measurement is repeated for increasing steps of $B'_x$ (for $\alpha$) or $B'_z$ (for $\beta$), such that for each step a parabola is measured. At the maxima of the parabola's, $B_y=0$.
By applying the rotation matrix in Eq. (\ref{eq:transformation}), we can calculate $\alpha$ and $\beta$:

\begin{equation}
    B_y = \cos(\beta)\sin(\alpha) B'_x + \cos(\alpha)\cos(\beta)B'_y - \sin(\beta) B'_z = 0
\end{equation}
For the $\alpha$ measurement, where $B'_z=0$, this leads to
\begin{equation}
    - \frac{B'_y}{B'_x} = \frac{\sin(\alpha)}{\cos(\alpha)} \approx \alpha
\end{equation}
Likewise, for $\beta$
\begin{equation}
    \frac{B'_y}{B'_z} = \frac{\sin(\beta)}{\cos(\alpha)\cos(\beta)} \approx \beta
\end{equation}
To determine $\gamma$ we align the field vector $\vec{B'}$ with the chip plane first:
\begin{equation}
    \vec{B''} = R_x(\beta)R_z(\alpha) \vec{B'}.
    \label{eq:pre-alignment}
\end{equation}
Then, the last rotation is simply given by $R_y(\gamma)$
\begin{equation}
    \vec{B} = R_y(\gamma)\vec{B''}.
    \label{eq:gamma_1}
\end{equation}

Since the last Euler angle $\gamma$ determines the rotation of the nanowire in the plane of the superconducting circuit, we can no longer use the previous strategy to determine the orientation. Instead, we make use of the SQUID-loop, by measuring two-tone spectroscopy data and keeping track of the even ABS minima, as we increase $B''_{z}$. Due to the pre-alignment step in Eq. (\ref{eq:pre-alignment}), $B_y=B''_y=0$. Therefore, the flux shift of an ABS transition is caused by the $B_x$ component of $B''_{z}$. Filling in Eq.~(\ref{eq:gamma_1}) on the minima leads to

\begin{equation}
    B_x = \cos(\beta)B''_x + \sin(\gamma) B''_z = c
\end{equation}
where $c$ is a constant. To first order, $\gamma$ can be expressed as 
\begin{equation}
    \gamma = - \frac{B''_{ \mathrm{x}}}{B''_\mathrm{z}} + c,
\end{equation}
which is found by extracting the slope of a linear fit through the ABS minima.  \\[1.5\baselineskip]
With all the angles known, we can calculate the raw magnetic field vector $\vec{B'}$ necessary to create the desired field $\vec{B}$ in the rotated coordinate system via the inverse of the rotation matrix in Eq.~(\ref{eq:rotation_matrix}).

\FloatBarrier
\newpage
\FloatBarrier
\section{Supporting datasets}
\subsection{Gate dependence of the spectrum}
\begin{figure}[h]
    \centering
    \includegraphics{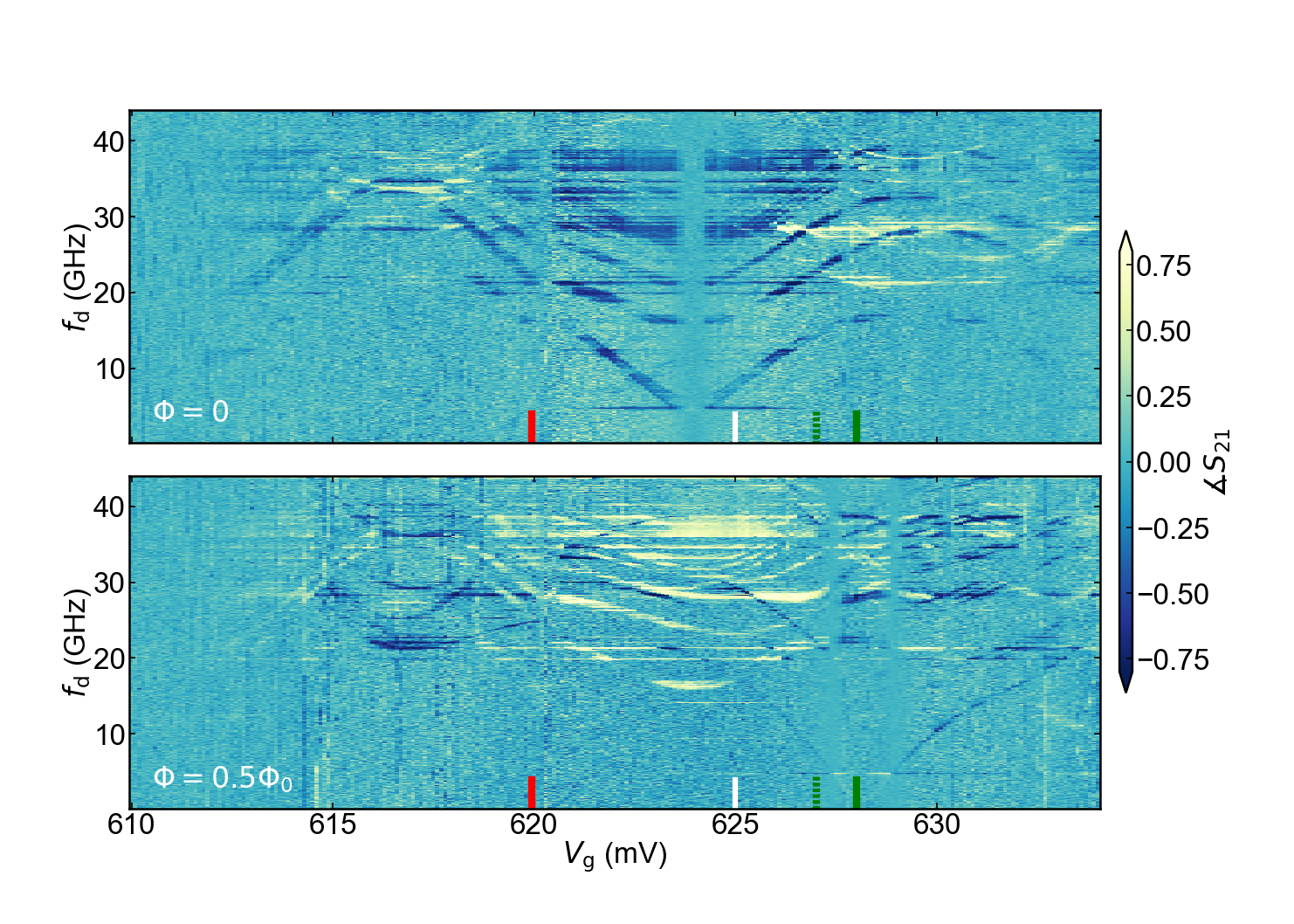}
    \caption{Gate dependence of the Andreev spectrum. Taken at external Flux: (a) $\Phi=0$ and (b) $\Phi=\Phi_0/2$. Vertical bars indicate gate positions of \cref{fig-field_sweep_supplement} (red), Fig. 3 in the main text, Fig. \ref{fig-field_sweep_supplement} \& \ref{fig:spiders} (white) and Fig. 4 of the main text, \ref{fig:selective_spectroscopy} \& \ref{fig:sup:z_field_high_tau} (green). Note that the green bar is indicative only and is not placed at \SI{627}{\milli\volt} (full bar) where the high transparancy data was measured, but is put at \SI{628}{\milli\volt} (dashed bar) as the data at high transparency was measured a month later after a gate jump occurred, and this gate sweep is from before the jump. The jump shifted the spectrum by $\approx$\SI{1}{\milli\volt} and caused additional small changes in the ABS energies at the shifted gate settings; gate jumps in general happened on timescales of weeks to months. }
    \label{fig:gate_dependence}
\end{figure}

Figure \ref{fig:gate_dependence} shows the transition spectrum as a function of the gate voltage at $\Phi\approx0, \Phi_0/2$. The gate voltage tunes the transparency of the ABS. We show flux and field dependence at various points in gate space varying from low to high transparancy later in this Section. For $\Phi=0$, the even transitions (bright lines) are at a maximum and the odd transitions (dark lines) at a minimum. For $\Phi=\Phi_0/2$ the opposite is true, which can be seen in the flux dependence in figure Fig. 2. The colors for even and odd parity are reversed at $\Phi=0$, since the sign of the shift is related to the curvature of the ABS energies, which switches sign~\cite{park_adiabatic_2020-1, metzger_circuit-qed_2021-1}. Similar to Ref.~\cite{tosi_spin-orbit_2019}, we observe opposite dispersion of even and odd transitions as a function of gate voltage. \\\\

\FloatBarrier
\newpage
\subsection{\Bz\ dependence at \Vg~=~\SI{619.95}{\milli\volt} (low transparency)}
We show the spectrum at~\SI{619.95}{\milli\volt} where the ABS have low transparency, both as a function of flux at zero field, and as a function of parallel field at $\Phi=0.5\Phi_0$. Here, the slope of the linearly increasing odd transition corresponds to an effective g-factor of 3.7. This dataset was used to calculate the alignment angle $\gamma$ in \cref{sec:field_alignment}.  

\begin{figure}[h]
   \includegraphics{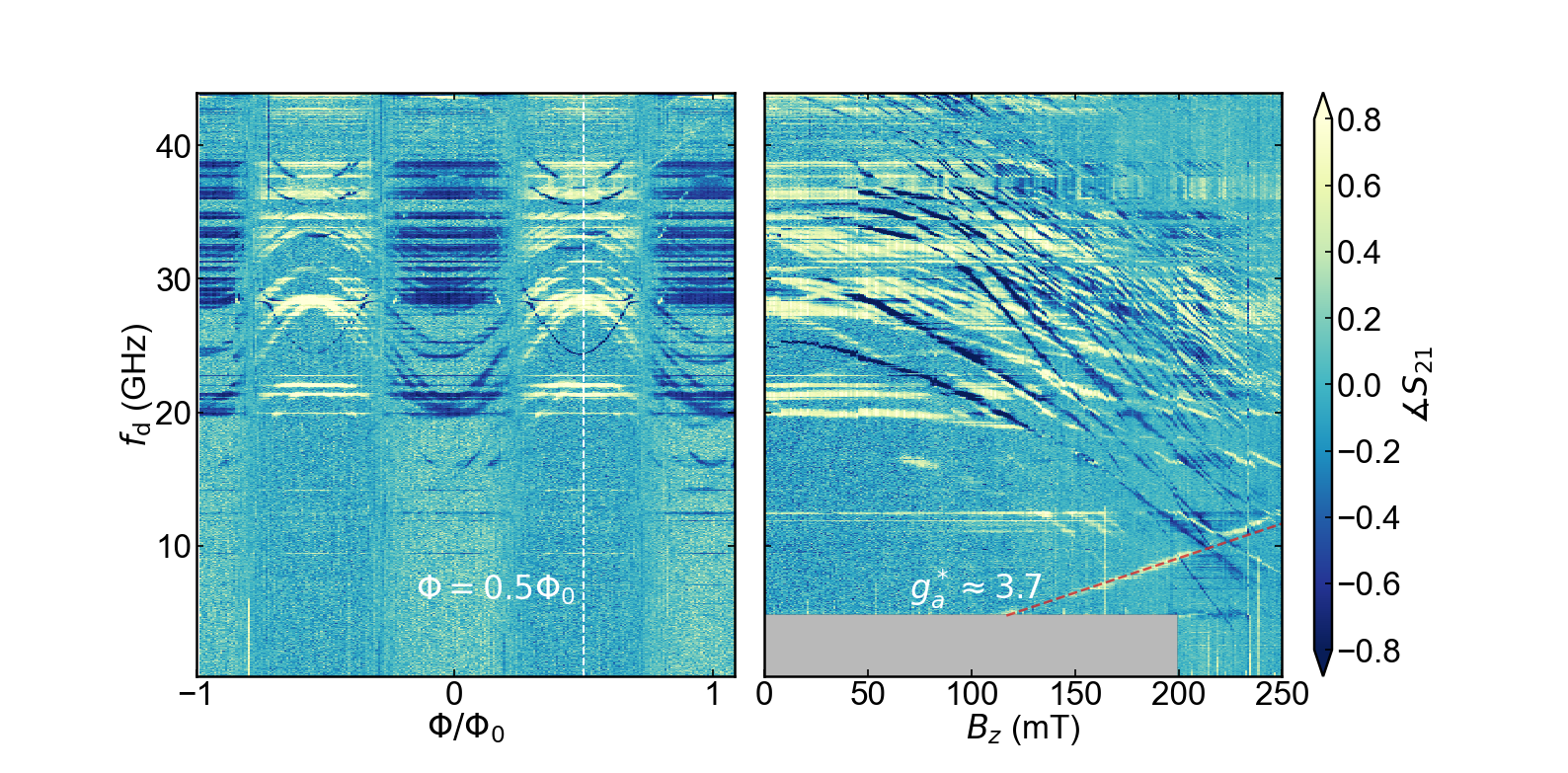}
   \caption{Parallel field dependence at \Vg~=~\SI{619.95}{\milli\volt}. (a) Zero field spectrum vs. flux (b) Spectrum as a function of parallel field, taken at constant flux $\Phi=0.5\Phi_0$. No data was taken in the grey area.}
   
   \label{fig-field_sweep_supplement}
\end{figure}

\clearpage
\subsection{Additional data at the gate voltage of Fig. 3 : \Vg~=~\SI{625}{\milli\volt} (medium transparency)}
\subsubsection{Field dependence of odd transitions}
\begin{figure}[h]
    \centering
    \includegraphics{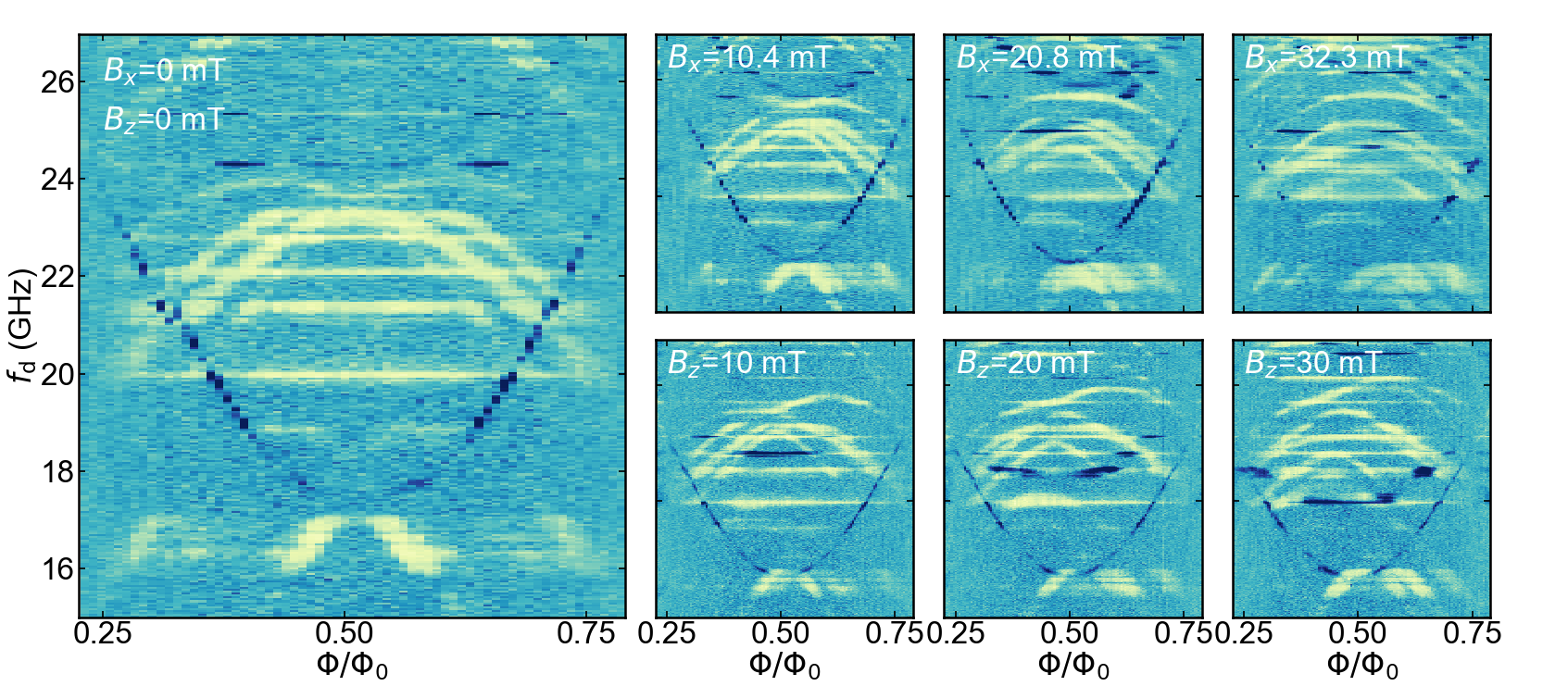}
    \caption{Field dependence of the odd transitions, at the same gate voltage as figure 2, 3. We observe asymmetric splitting when applying parallel field \Bz, which is opposite from the symmetric splitting observed in Ref.~\cite{tosi_spin-orbit_2019}. This could indicate that the effective spin-orbit field is not fully along the $y$-axis as expected for a wire with the transport direction along the $z$-axis, but has a component along the $z$-axis as well. }
    \label{fig:spiders}
\end{figure}

\clearpage
\subsubsection{Flux dependence during parallel field sweep}
\begin{figure}[h]
    \centering
    \includegraphics{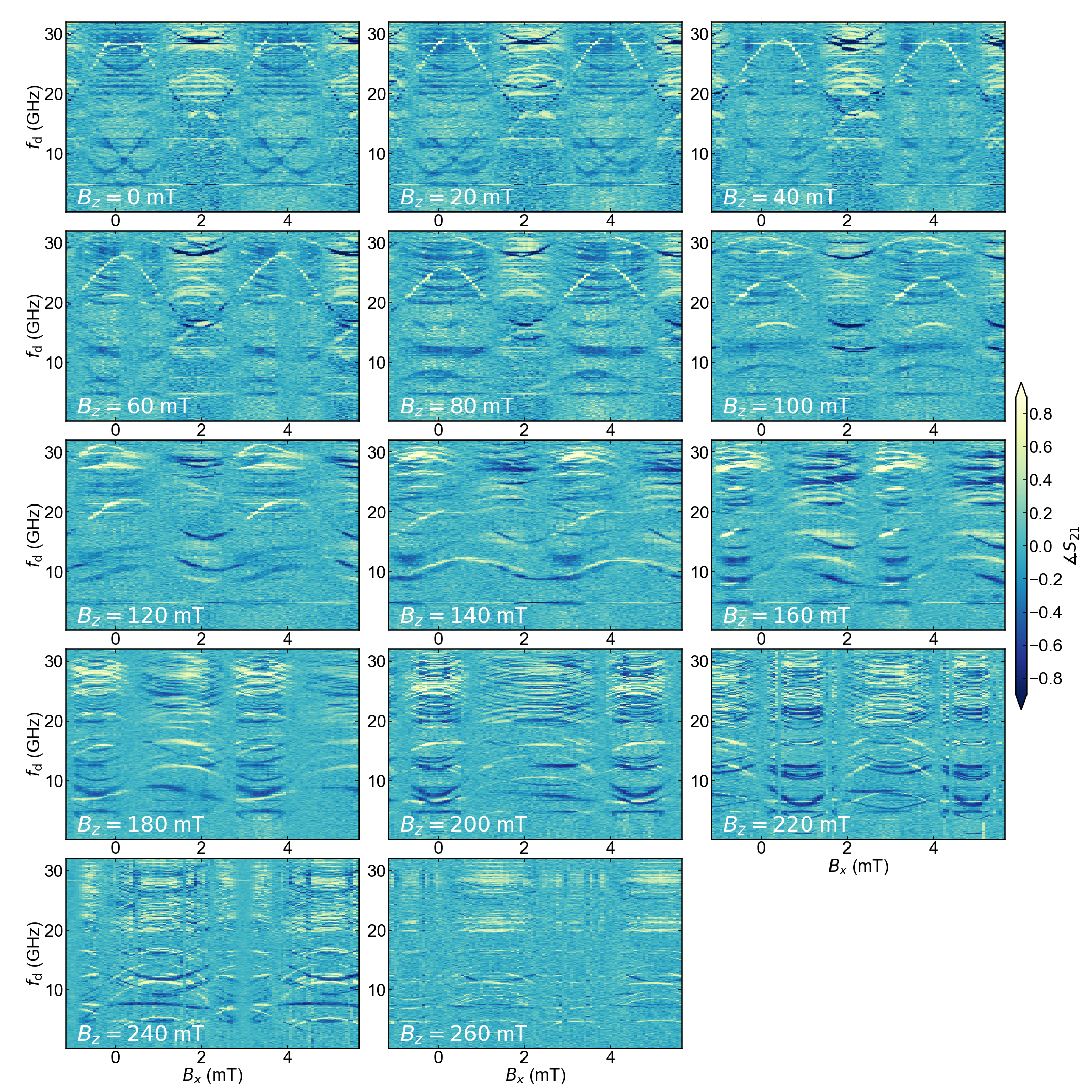}
    \caption{Flux dependence at the same gate as in Fig. 3 in the main text. Note that the \Tzerotilde\ transition keeps its minimum at \Bx~=~\SI{2}{\milli\tesla} up to \Bz~=~\SI{160}{\milli\tesla}, demonstrating the correct alignment, possibly compensating for some linear AJE, of the magnetic field used in Fig. 3 in the main text. }
    \label{fig:flux_dependence_field}
\end{figure}

\clearpage
\subsubsection{Additional flux dependence on spin-split even transitions}

\begin{figure}[h!]
    \centering
    \includegraphics{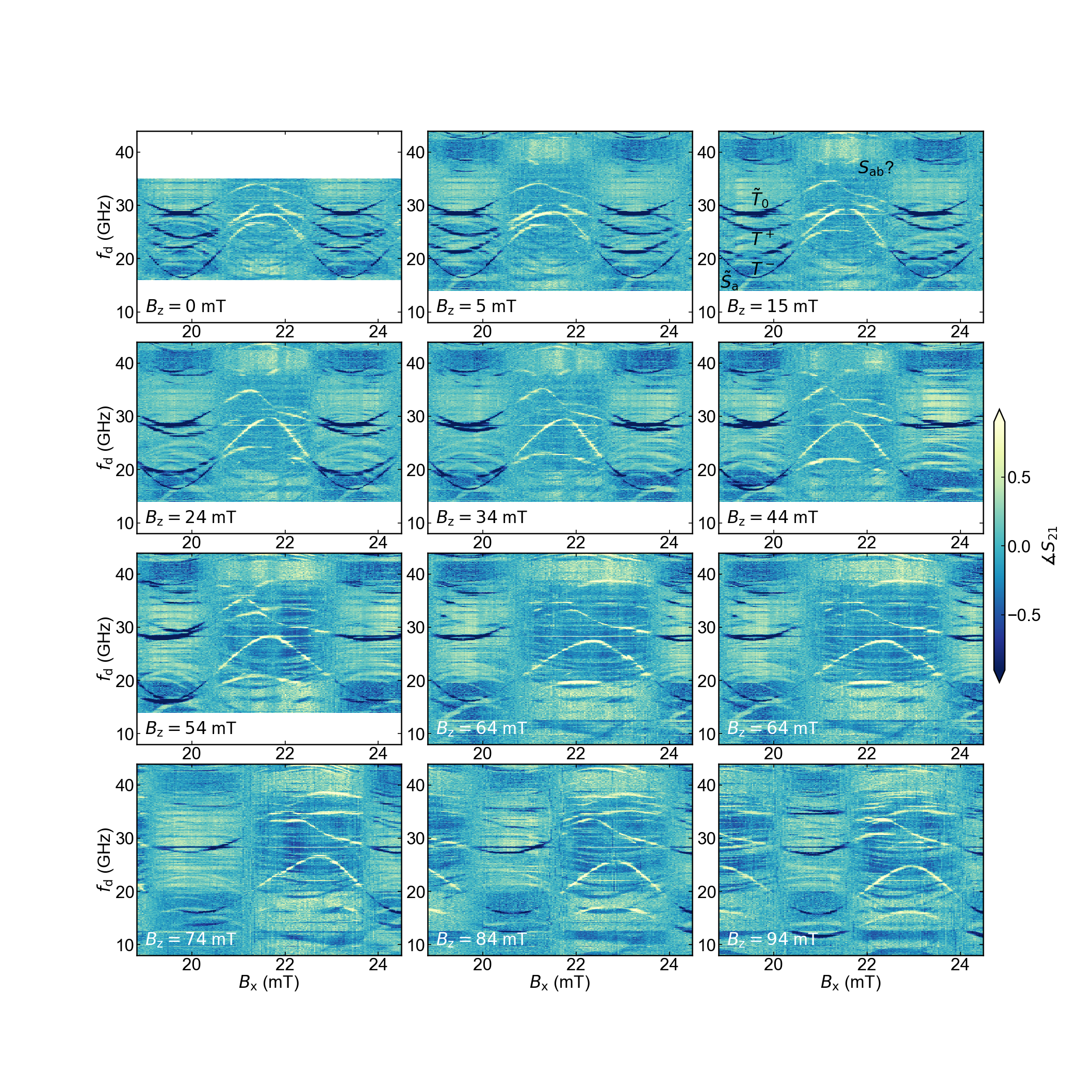}
    \caption{Flux dependence of hybridized singlet and triplet transitions at the same gate used in Fig. 3 of the main text: $V_\mathrm{g}=\SI{625}{\milli\volt}$ versus parallel magnetic field. This data was taken at \SI{10}{\dBm} higher drive power and finite $B_\mathrm{x}=\SI{22}{\milli\tesla}$ to increase visibility of the \Tmin\ and \Tplus\ transitions. 
    This data was measured to demonstrate the phase-dispersion of the triplet states and explore whether there are no other transitions than \Tmin, \Tplus\ visible between the lowest even transition identified as \Sa\ and the higher transition labeled as \Sb\ without interactions. The non-interacting model would require two more transitions visible with similar dispersion, c.f. Fig.3 (c) in the main text and the discussion in \cref{sec:non-interacting-sims}. Additionally, hints of a camel-back shaped transition become visible above \Tzerotilde\ which could correspond to \Sab\ similar to what was observed in a recent work~\cite{matute-canadas_signatures_2022} (see labels at \Bz~=~\SI{15}{\milli\tesla}). Note that at higher fields the transitions get distorted, possibly due to hybridization with other manifolds.}
    \label{fig:flux_dependence_even_transitions_field}
\end{figure}

\subsubsection{Estimate of supercurrent change due to the triplet state}
\begin{figure}[h!]
    \centering
    \includegraphics{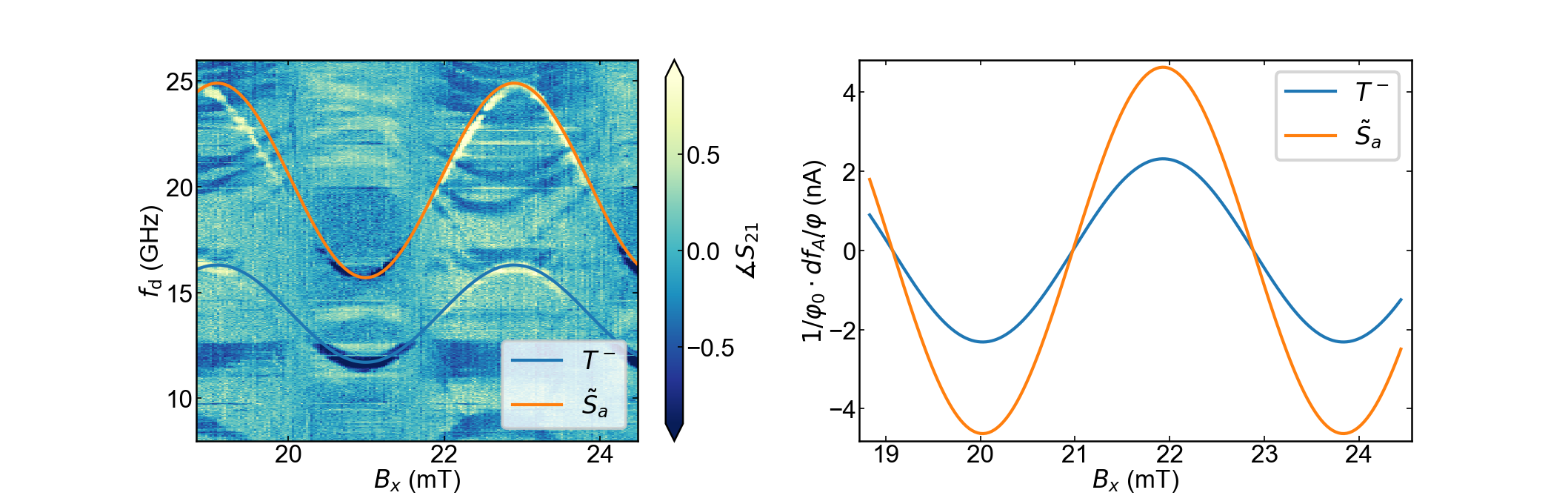}
    \caption{Estimate of the supercurrent change when exciting the triplet and singlet at $B_z=\SI{94}{\milli\tesla}$, where the two transitions are clearly visible and \Tmin\ is already split below \Satilde\ (the last panel in \cref{fig:flux_dependence_even_transitions_field}). To provide a rough estimate of the supercurrent carried by the triplet state \Tmin, we manually approximate the the phase-dispersion of the even transitions (left panel) with $f_A=A*\sin(\varphi+b)+c$ resulting in $A=\SI{2.3}{\giga\hertz}$, $c=\SI{14}{\giga\hertz}$ for \Tmin and subsequently calculate the resulting supercurrent through the derivative $1/\varphi_0*df_A/d\varphi$ with $\varphi_0=\hbar/2e$ (right panel). Note that the \Satilde\ transition seems very well approximated by using a twice as large amplitude $2A$ but $c=\SI{20}{\giga\hertz}$. This could be due to a much larger dispersion of the lower manifold compared to the higher manifold, see discussion in Ref.~\cite{matute-canadas_signatures_2022} }
    \label{fig:flux_dependence_triplet_supercurrent}
\end{figure}

\clearpage

\subsection{Figure 4 full dataset : Phase dependence at \Vg~=~\SI{627}{\milli\volt} (high transparency)}
 \begin{figure}[h]
    \includegraphics{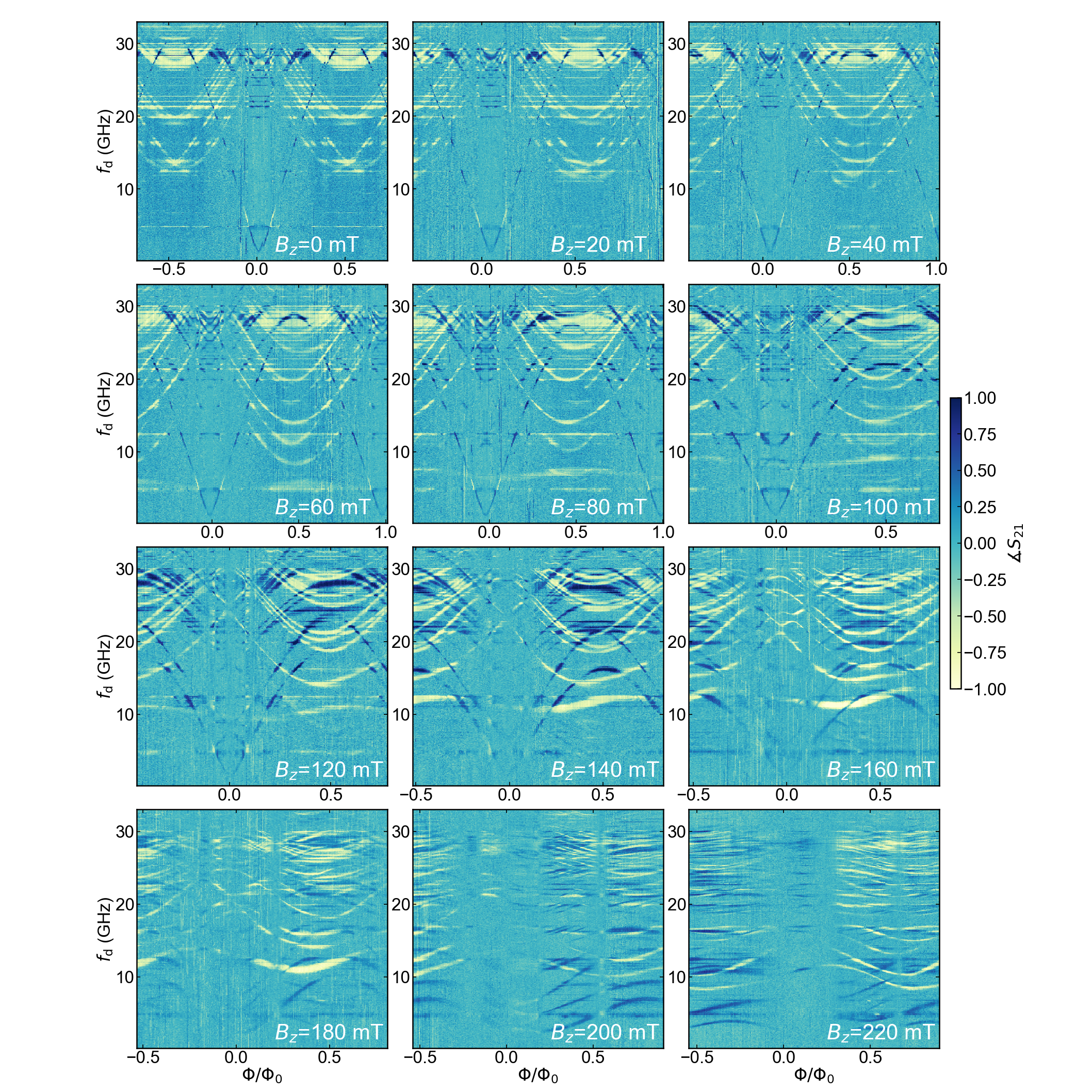}
    \caption{Full dataset corresponding to Fig. 4 in the main text. From \SI{60}{\milli\tesla} onwards, we see a bright, flat line coming up, which is the intra doublet transition \Da.  Measurement taken at $V_\mathrm{g}=\SI{627}{\milli\volt}$ versus \Bz. $B_\mathrm{y}$ is set to zero, raw $B_\mathrm{x}'$ is used for flux control. 
    In the even parity manifold, \Sa\ remains approximately constant near $\Phi=\Phi_0/2$ while reducing at $\Phi=0$ due to interaction with other transitions. The spectra also show a strongly dispersing even triplet transition \Tmin\ come down in field faintly in the beginning but more apparent at higher fields.}
    \label{fig:sup:z_field_high_tau}
\end{figure}

\clearpage
\subsubsection{Parity selective-spectroscopy at \Bz~=~\SI{100}{\milli\tesla}}
In the main text, the junction parity was implied from the sign of the dispersive shift. The presence of crossings between the ABS and the resonator mode can result in sign-changes of the shift~\cite{metzger_circuit-qed_2021-1}. In order to make sure that we correctly labeled the parity of the transitions we have performed parity selective spectroscopy using the techniques of Ref.~\cite{wesdorp_dynamical_2021} at \Bz~=~\SI{100}{\milli\tesla}.

\begin{figure}[h]
\includegraphics[width=0.5\textwidth]{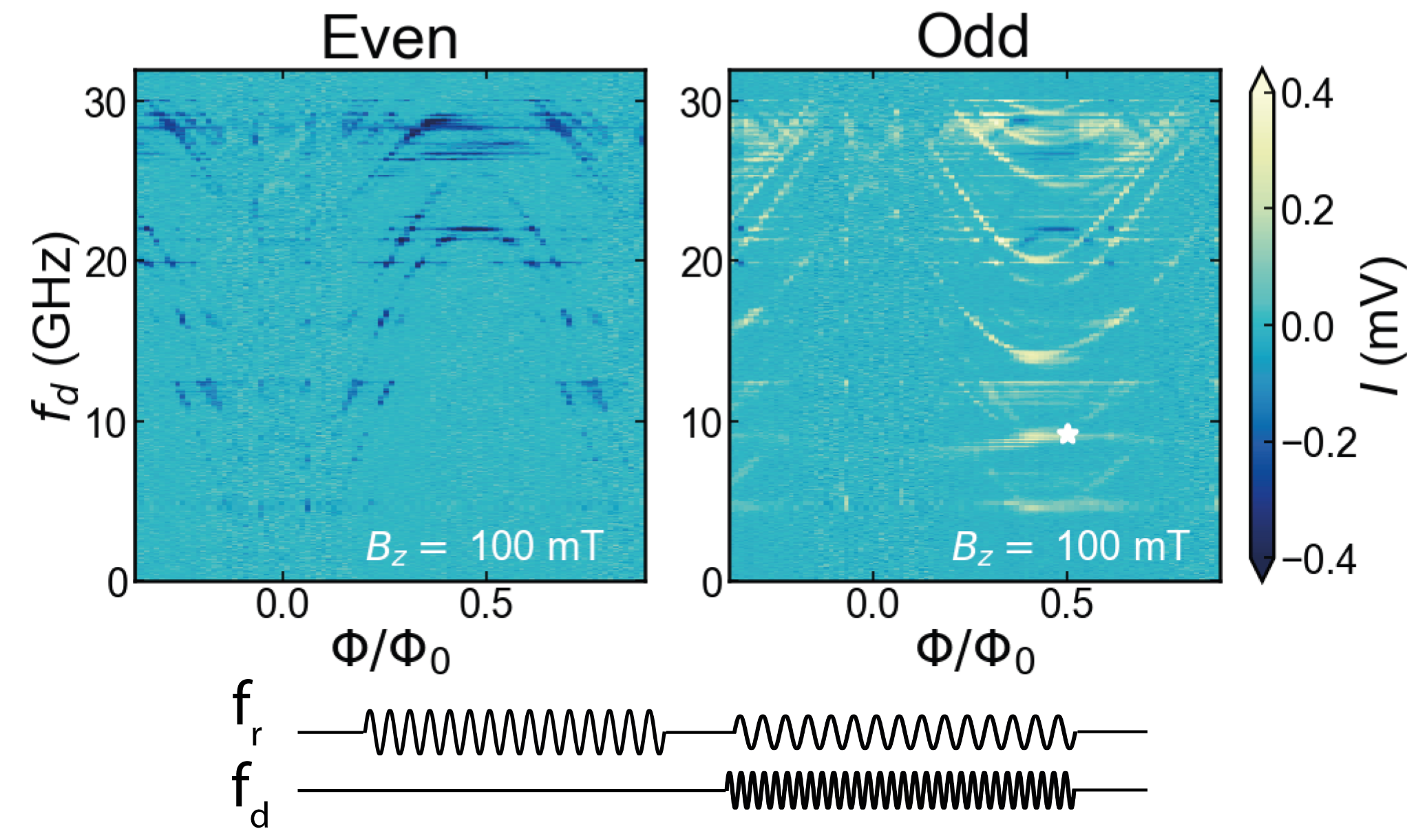}
\caption{\label{fig:selective_spectroscopy} Parity selective spectroscopy at the gate voltage of Fig.4 at \SI{100}{\milli\tesla} confirming the odd parity nature of the intra-doublet transition. (a) Spectra at \SI{100}{\milli\tesla} after post-selection on initial junction parity using the pulse sequence in (b). The pulse sequence consists of a \SI{20}{\micro\second} readout pulse that measures the junction parity followed by a \SI{20}{\micro\second} spectroscopy pulse conditioned on the outcome of the first pulse. To increase contrast of the parity selection, we thresholded the parity selection keeping roughly 20\% of the data. This was measured with the setup described in~\cite{wesdorp_dynamical_2021} }
\end{figure}

\clearpage

\subsection{Extended \Bz\ dependence up to \SI{1}{\tesla}}
\begin{figure}[h]
    \centering
    \includegraphics{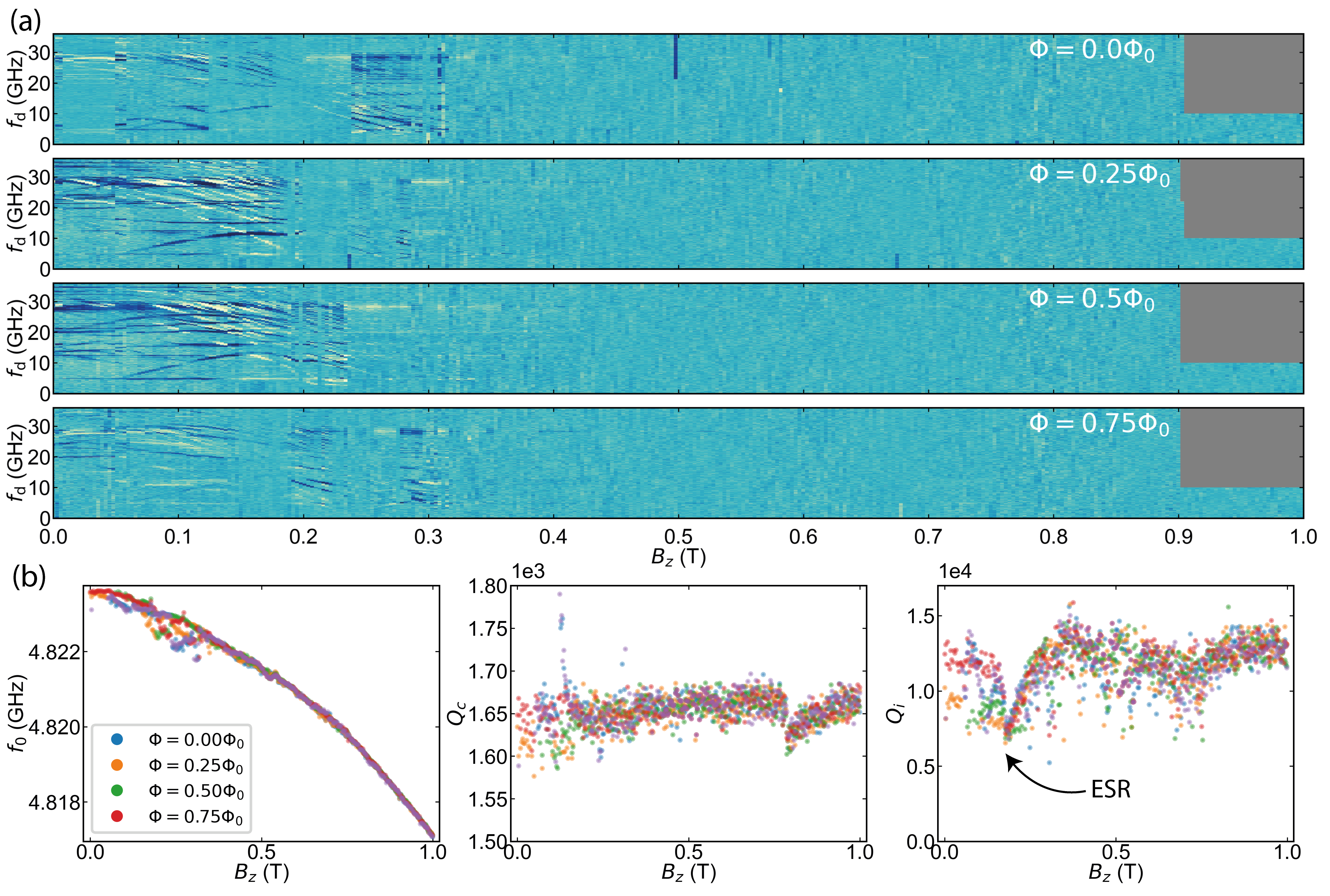}
    \caption{Extended \Bz\ dependence up to \SI{1}{\tesla} \textbf{(a)} Parallel field sweeps at four different starting external flux points, to see the maximum field where a signal remains in spectroscopy. No revival is observed after the gap closing at \SI{325}{\milli\tesla} due to the orbital effect. Grey regions denote regions where no data is measured.
    \textbf{(b)} The resonance frequency \fnull\ extracted from a fit for the different flux points and field during the two-tone sweeps. Fits with a large $\chi^2$ are discarded in order to filter out points where the ABS avoided cross with the resonator. \fnull\ shows a parabolic decline with increasing parallel field \Bz due to a Cooper-pair breaking increase in kinetic inductance\cite{samkharadze_high-kinetic-inductance_2016-1, kroll_magnetic_2019-2}. $Q_\mathrm{c}$ stays fixed as a function of $B_z$ and $Q_\mathrm{i}$ stays $>10$k, except at points where crossings occur. There is however a clear dip in $Q_\mathrm{i}$ visible corresponding to the electron-spin-resonance frequency (2$\mu_\mathrm{B}$~\Bz~=~h\fnull).  }
    \label{fig:field_compatibility}
\end{figure}

An extended field sweep could show a revival of the spectrum if the gap was limited by the orbital effect~\cite{kringhoj_magnetic-field-compatible_2021-1, winkler_unified_2019-1, antipov_effects_2018} or by a transition to a topological phase~\cite{kitaev_unpaired_2001, lutchyn_majorana_2010, oreg_helical_2010}.
At positive back-gate voltages, which is the case here, the hybrid states can form ring-like shapes due to surface accumulation~\cite{winkler_unified_2019-1}. Destructive interference then occurs at roughly half integer flux quanta threading through the nanowire cross-section with diameter $d$, i.e. $\frac{1}{4}\pi d B_\mathrm{z} = \Phi_0/2$~\cite{kringhoj_magnetic-field-compatible_2021-1}.  
In~\cref{fig:field_compatibility}.(a) we perform an aligned parallel field sweep for different initial fluxes to test the limiting field where we can observe the spectrum without being affected by a shift in phase due to the AJE or flux-jumps. We observe a crash of the spectrum at $\approx \SI{300}{\milli\tesla}$ and extract $d=\SI{65}{\nano\meter}$ close to the \SI{80}{\nano\meter} nanowire diameter. The sweep is continued up to \SI{1}{\tesla} as one would expect a revival around $\Phi_0\sim$\SI{600}{\milli\tesla} if limited by orbital effects of a single state. We do not observe this, which can be the case if the many visible ABS have different effective wavefunction cross-sections~\cite{winkler_unified_2019-1}.
Another cause for the low visibility at high-fields could be a wrong choice of readout point due to ABS crossing the resonator.

\cref{fig:field_compatibility}.(b) shows \fnull, coupling quality factor $Q_c$ and internal quality factor $Q_\mathrm{i}$ as a function of \Bz\ during the sweeps. For each \Bz, these quantities are extracted by fitting resonator traces to an asymmetric resonator model measured in transmission \cite{khalil_analysis_2012}. At least up to \Bz =\SI{1}{\tesla}, we do not observe a significant trend in the resonator's quality factors. Fluctuations may be caused by avoided crossings between ABS transitions and the resonator. \cref{fig:field_compatibility}.(b) shows the loss of visibility in the spectrum around $B_\mathrm{z}=\Phi_0/2$, is not related to limited field compatibility of the resonator. Furthermore, we observe an ESR dip in $Q_\mathrm{i}$ at \Bz~=~$hf_0/(|g_\mathrm{e}|\mu_\mathrm{B})$~=~\SI{172}{\milli\tesla} with $g_\mathrm{e}$ the electron g-factor where the resonator frequency matches the surface electron spin-flip frequency. 

\FloatBarrier

\section{Supporting data $\varphi_0$-effect}
\FloatBarrier
\subsection{Data extraction procedure}
\label{sec:data-extraction}
To obtain the minima and maxima in Fig. 5 of the main text and panel \textbf{d} of \cref{fig:chip-alignment}, we performed the following data extraction protocol. This example is for finding a maximum. 

\begin{enumerate}
    \item Make an initial guess of the coordinates $\left(\Phi, f_\mathrm{d}\right)$ of the minima and maxima. Then, construct a guess parabola $p$: a parabola of which the maximum is located at $\left(\Phi_\mathrm{max}, f_\mathrm{d,max}\right)$.
    \item Select a region of interest around the guess parabola: $\Phi \in \large[\Phi_\mathrm{max} - 0.2 \Phi_0$; $\Phi_\mathrm{max} + 0.2 \Phi_0  \large] \cap f_d \in \large[p(\Phi)-\SI{0.8}{\giga\hertz}$;  $p(\Phi)+\SI{0.8}{\giga\hertz} \large]$. In this region, we save the phase points exceeding a threshold as raw datapoints. We set the threshold to the 95th percentile of the selected region's phase data. For each flux point $\Phi$, at most one datapoint is extracted by taking the median of the raw datapoints. The standard deviation $\sigma$ is calculated for each flux value as well.
    \item Fit a second order polynomial through the extracted datapoints, weighted by the inverse variance 1/$\sigma^2$.
    \item Subsequently, we identify the coordinates of the maximum of the fit as an estimate of the transition line maximum. 
\end{enumerate}

\begin{figure}[h!]
  \centering
   \includegraphics{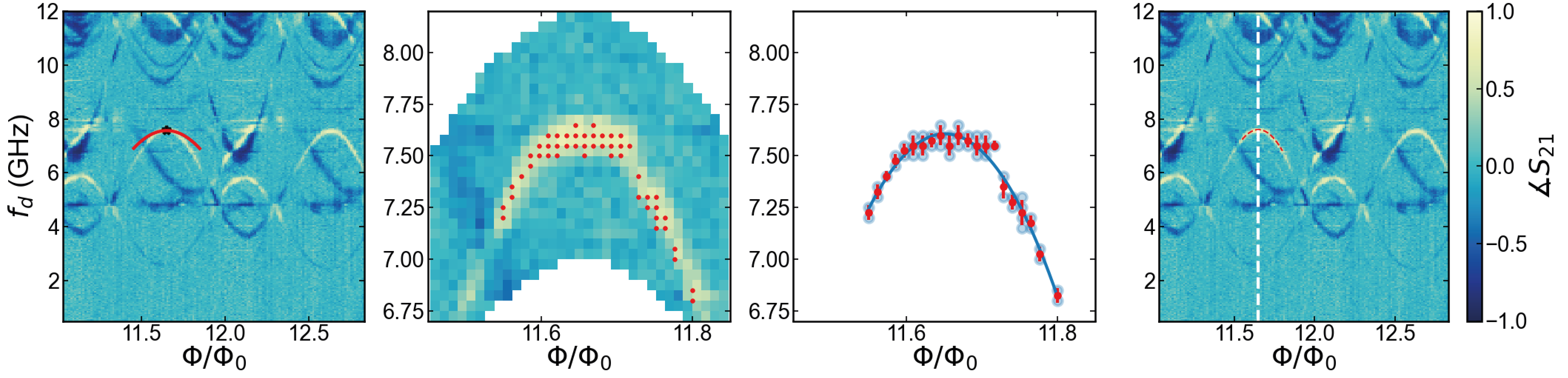}
   \caption{Data extraction steps. (a) Construct guess parabola. (b) Select region of interest, and extract datapoints with a threshold. (c) Fit a parabola through the extracted datapoints. (d) Identify the vertex of the parabola as the maximum.} \label{fig-sup-data-extraction}
 \end{figure}

\FloatBarrier
\subsection{Identification of parity of the transitions at high field for $\varphi_0$ sweeps}

\begin{figure}[h]
   \includegraphics{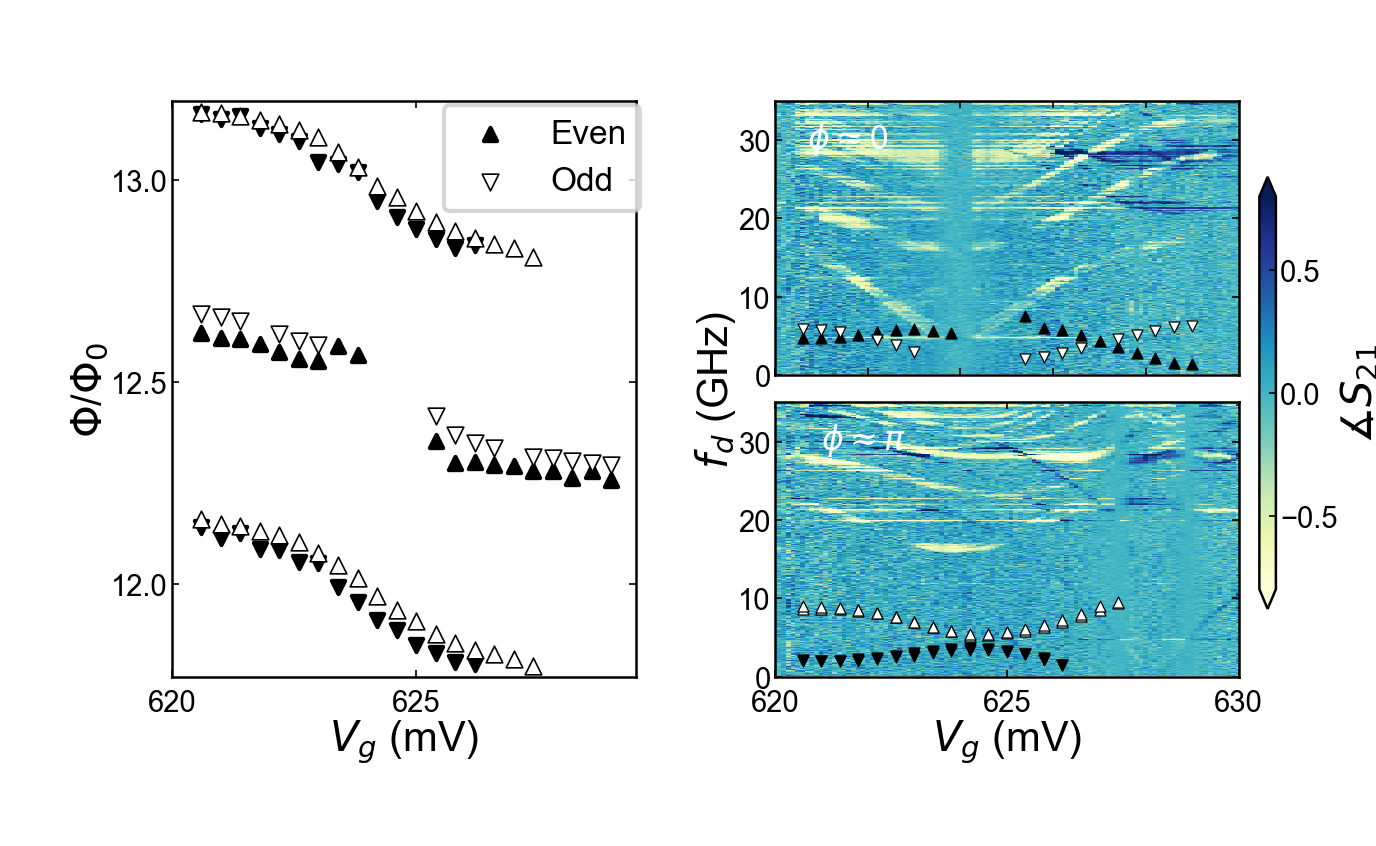}
   \caption{Even and odd labeling. (a) Scatterplot of extracted minima ($\triangledown$) and maxima ($\triangle$) for an odd (white) and even (black) transition in the two-tone dataset of Fig. 5 of the main text. For both parities, we observe a similar downward trend, indicating a negative phase shift as a function of the gate voltage.
   (b)
   Comparison of the extracted datapoints with the gate two-tone scans at zero field. The white datapoints qualitatively follow the behaviour of the yellow lines, and the black points behave oppositely, like the blue lines, justifying the labeling of the white triangles as odd and the black triangles as even.}
   
   \label{fig-even_vs_odd}
\end{figure}

To extract the anomalous phase $\varphi_0$ from the two tone spectra in Fig. 5 of the main text, we followed the extrema of a single transition line as a function of gate voltage. Figure \ref{fig-even_vs_odd} (a) shows the anomalous phase shift of a transition line of both parities, where the white datapoints are used in the main text. We observe similar  trends for both even and odd parity.
To identify states of even and odd parity at the high magnetic field setting used here, we compare the gate dependence of the y-coordinate (\fd) of the extracted maxmima and minima for both parities with the gate spectrum at zero field, see~\cref{fig-even_vs_odd}. Although the transition frequencies are significantly lowered due to the reduction of $\Delta$, the even and odd traces still peak and dip at the same gate points as same parity transitions at \Bz~=~0. From this we concluded that the transition traced in the main text have odd-parity. We therefore tracked an even-parity state as well, which gave a similar $\varphi_0$ shift as the odd state. 

\FloatBarrier
\subsection{Symmetry breaking in two-tone spectrum}

\begin{figure}[h]
   \includegraphics{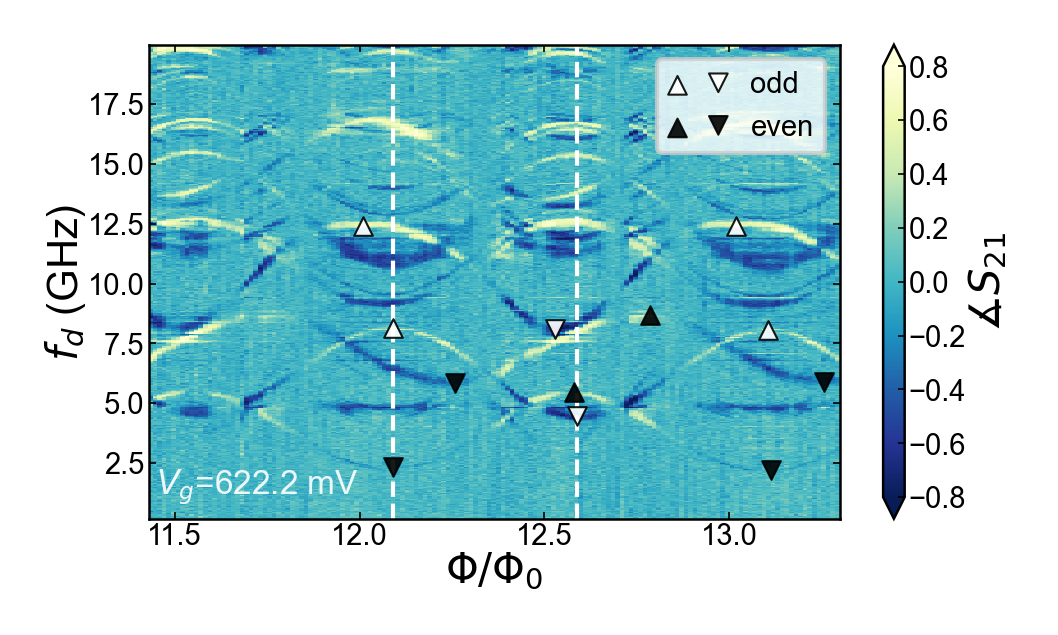}
   \caption{Symmetry breaking. Zoom-in of one of the panels of Fig. 5(a) of the main text. Black and white triangles are placed on the extrema of even and odd transitions, respectively indicating strong asymmetry between different ABS levels.} \label{fig:symmetry_breaking}
\end{figure}

Figure \ref{fig:symmetry_breaking} shows one panel of Fig. 5 (a) in the main text. At \Vg=\SI{622.2}{\milli\volt}, \Bz=\SI{220}{\milli\tesla} and \Bx=\SI{45}{\milli\tesla}, highlighting that there is not only a global shift of the whole spectrum, but that the Andreev transitions are also strongly shifted with respect to each other and that there is no symmetry axis to be found at $\varphi=0,\pi$. Note that this observation alone rules out most trivial explanations for the AJE.

\FloatBarrier
\clearpage

\subsection{$\varphi_0$ under field-reversal}

We have performed measurements of the $\varphi_0$-effect under reversal of the field vector ($B_x=\SI{45}{\milli\tesla}$, $B_z=\SI{225}{\milli\tesla}$), both for the spectroscopy and squid oscillations. This helps verify that indeed there is no phase shift due to a non-linear flux-phase relation (see~\cref{sec:sup:gradiometer}), which could in turn be caused by a large shunt inductance relative to the junction inductance. 
\begin{figure}[h]
   \includegraphics{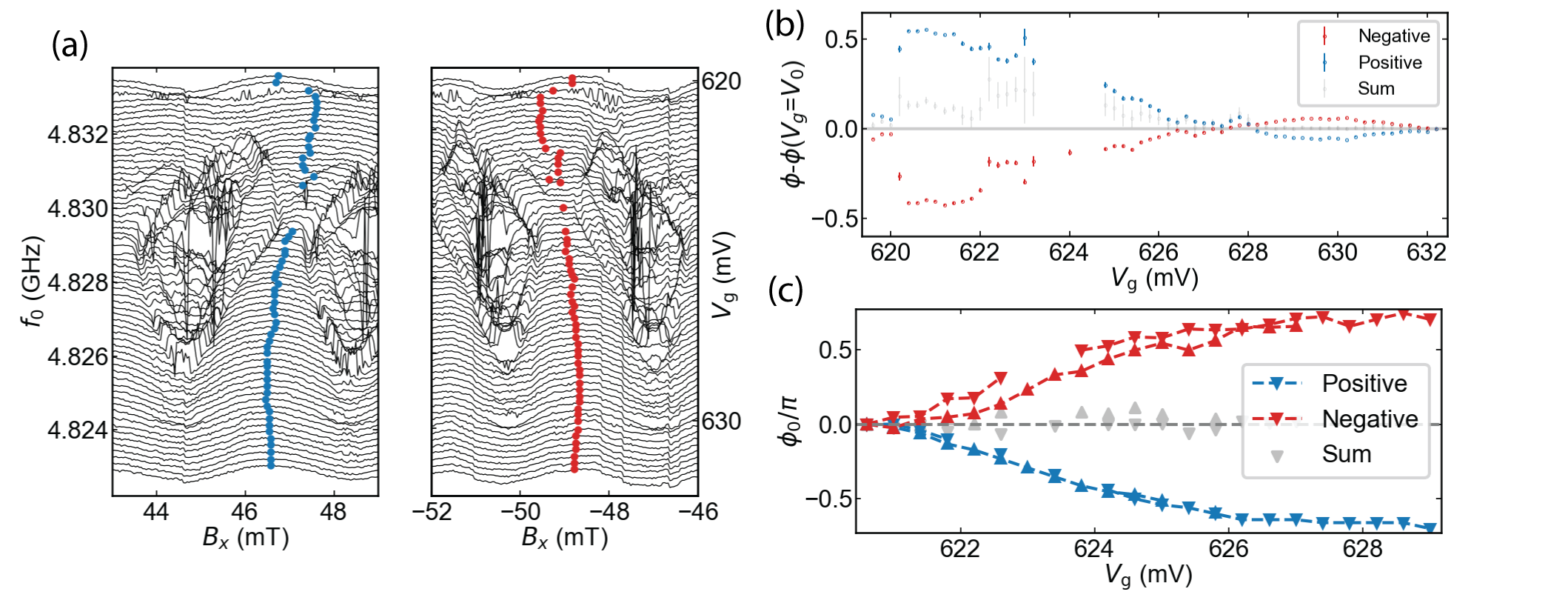}
   \caption{$\varphi_0$ effect under reversal of the field vector. \textrm{(a)} SQUID oscillations for both field directions. The vertical straight features are flux-jumps due to vortices in vicinity of the resonator, which appear as vertical lines because the gate is swept as the fast-axis. \textrm{(b)} Extracted phase shift from the SQUID oscillations for positive and negative fields. \textrm{(c)} Extracted $\varphi_0$-effect in two-tone spectroscopy under field reversal.} \label{fig:field_reversed_phi_0}
\end{figure}
\clearpage
\subsection{Field dependence of the $\varphi_0$-effect}
We investigate the field-dependence of the anomalous Josephson effect by reducing the magnitude of the $\vec{B}$ starting from the vector shown in Fig. 5 of the main text ($B_x=\SI{45}{\milli\tesla}$, $B_z=\SI{225}{\milli\tesla}$), while keeping the angle the same. We measure two-tone spectra, keeping track of the lowest visible transition. At 0 field, we do not find a significant phase-shift as expected, while for increasing field we observe a steep transition to a large phase shift. The data presented here resulted from tracing the transitions by hand, as the crowding of the spectra and lower resolution did not allow for the numerical methods presented above. To allow for verification, we have provided collections of images of the spectra with the traced maxima/minima in the data-repository. At some gate and field values data is missing because the transition was no longer visible in spectroscopy. 

\begin{figure}[h]
   \includegraphics{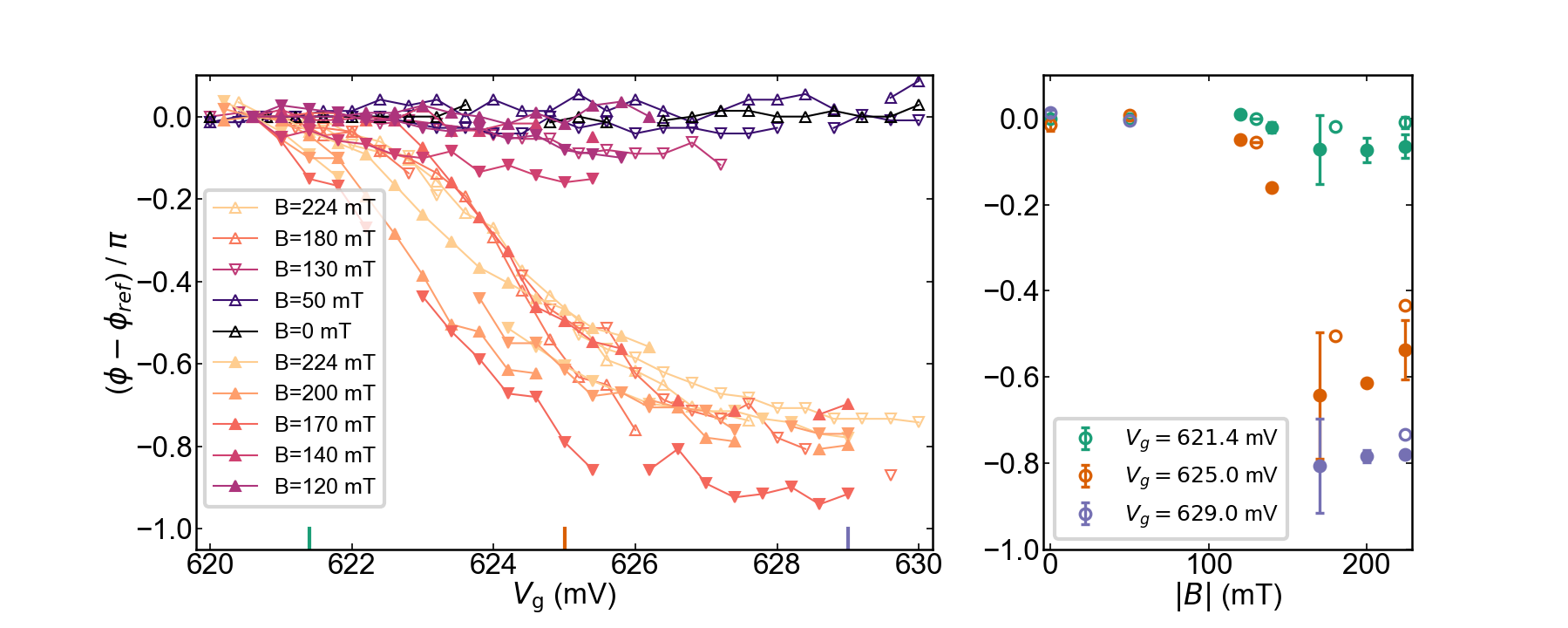}
   \caption{\textbf{(a)}Gate dependence of the anomalous phase for increasing magnetic field strength applied along the vector defined in Fig. 5 of the main text. We plot the combined data of two sweeps tracing the extrema separated in time, indicated by open or closed triangles respectively. Upward triangles indicate tracing of a minimum and downward of a maximum. \textbf{(b)} Relative phase shifts (average of minima and maxima) with respect to $V_\mathrm{g}=\SI{620}{\milli\volt}$ for the gate values indicated by the same colored line-cuts in (a) versus field strength.} \label{fig-phi_0_mag}
\end{figure}

\subsection{Extracted $g$-factor of the lowest ABS manifold versus \Vg}

In an attempt to find a cause for the gate-dependence of the AJE, we have performed parallel field scans similar to Fig.3 (b) of the main text for the gate range where we measured the AJE.
By extracting the slope of the spin-flip transition when visible we can extract the effective $g$-factor of the lowest ABS manifold. 
We found a strong gate-dependence of the effective $g$-factor of the lowest ABS manifold, correlated with the anomalous phase shift. This is consistent with the scaling of the AJE magnitude with an effective Zeeman term.

\begin{figure}[h]
   \includegraphics{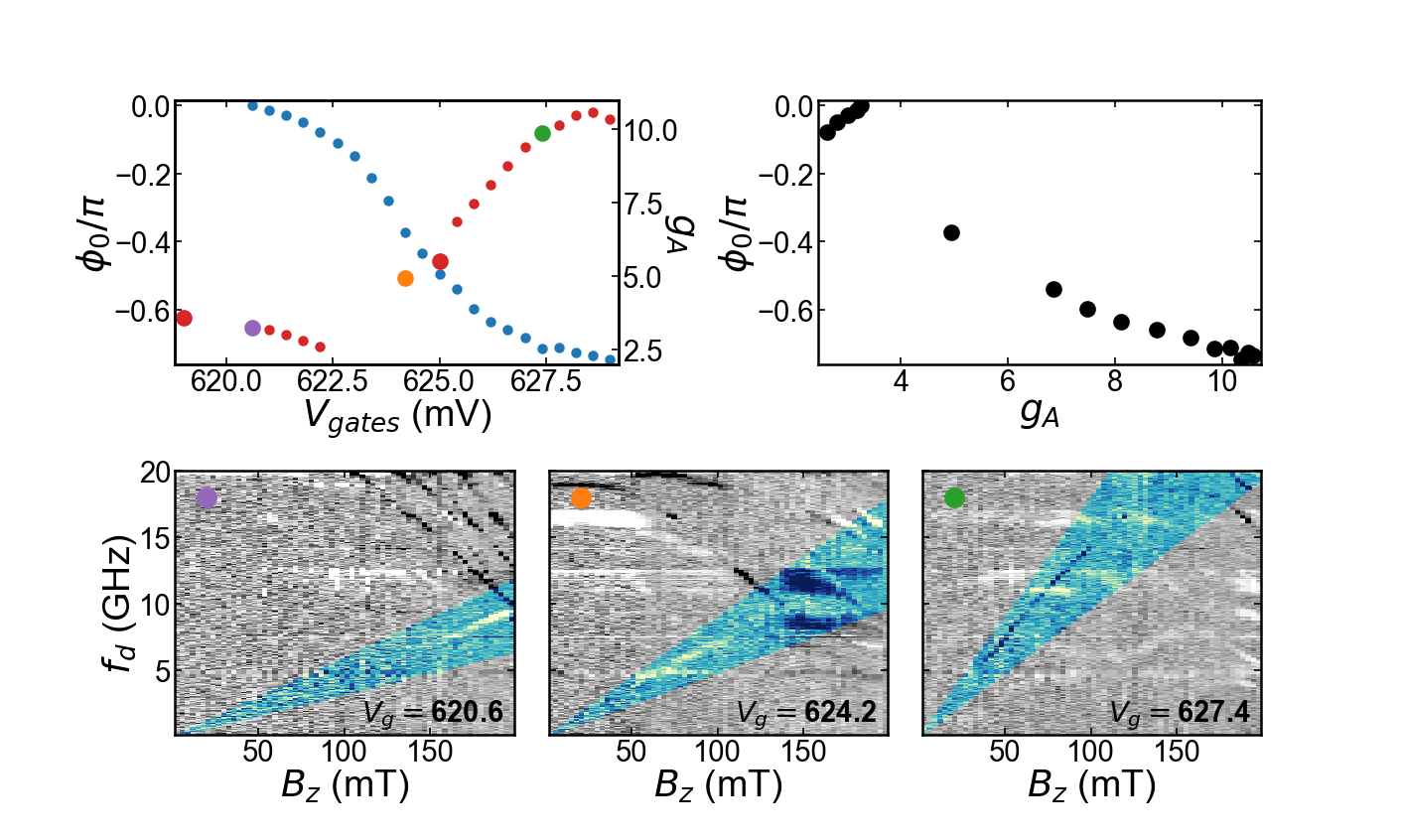}
   \caption{\textbf{Extracted g-factor of the lowest ABS level pair.} (a) $\varphi_0$ Of two-tone spectroscopy shown in main text plotted versus $g_\mathrm{a}^*$. (b) Correlation between $\varphi_0$ and $g_\mathrm{a}^*$. (c) Example parallel field sweeps that were used to extract $g_\mathrm{a}^*$ of the lowest doublet. Full dataset is provided in the data-repository. } \label{fig-g-factor-correlation}
\end{figure}
\FloatBarrier

\newpage
\newpage
\clearpage
\section{Theoretical modeling}
\subsection{Non-interacting tight-binding simulations}\label{sec:non-interacting-sims}
Tight-binding simulations of the Josephson junction were performed using the Kwant package~\cite{groth_kwant_2014}, and in particular adapting code from Ref.~\cite{laeven_enhanced_2020}. We use the following two-dimensional Hamiltonian including Rashba spin-orbit coupling, Zeeman effect and a position dependent barrier potential

\begin{equation}
\begin{aligned}
    H= &
    \left(\frac{\hbar^2\mathbf{k}^2}{2m} - E_\mathrm{F}) - e\phi(\mathbf{r})\right)\sigma_0 \\
    & + \alpha(k_z\sigma_x - k_x\sigma_z)\\
    & + \mathbf{E_\mathrm{Z}}\cdot \mathbf{\sigma}
\end{aligned}
\end{equation}
From numerically solving the tight-binding Hamiltonian we obtain the BdG eigenenergies which give all required information as we assume no exchange interaction is present.
We first take only the positive eigenvalues, due to the degeneracy induced by PH symmetry,
$\{E_{i,\sigma}\}$,
where $i$ stands for the manifold index and $\sigma$ the pseudo-spin direction.
The low-energy Hamiltonian then is a sum over the Andreev levels~\cite{van_heck_zeeman_2017}
\begin{equation}
H = \sum_{i,\sigma}E_{i,\sigma} (c_{i,\sigma}^\dagger c_{i,\sigma} - \frac{1}{2})
\end{equation}

\begin{figure}[h]
     \centering
     \includegraphics{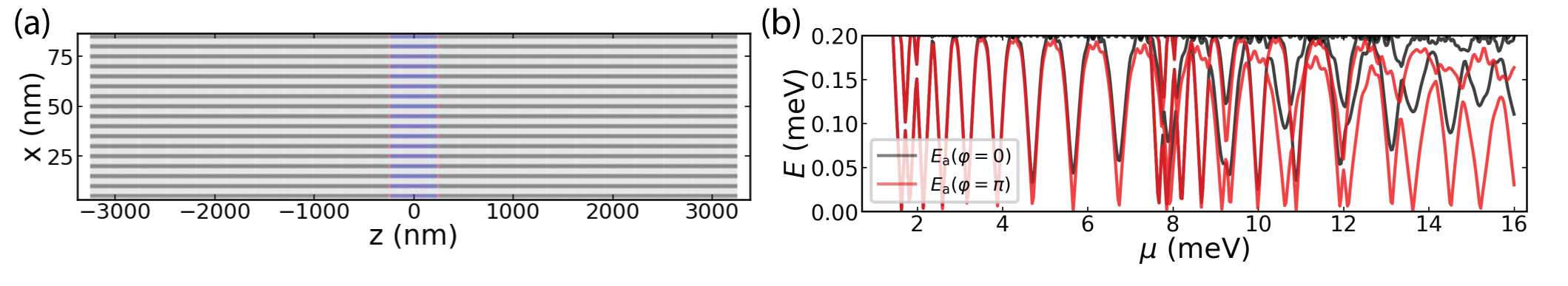}
    \caption{Kwant model. \textbf{(a)} Tight binding grid used for the simulations. \textbf{(b)} Chemical potential sweep using the grid in (a).}
    \label{fig:kwant-tight-binding}
\end{figure}

We simulate the junction using a 2D-grid with lattice constant of \SI{5}{\nano\meter}. The grid consists of a \SI{100}{\nano\meter} wide and \SI{3000}{\nano\meter} long superconducting lead, where $\Delta\not=0$ on each side  of a \SI{500}{\nano\meter} long junction, where $\Delta=0$, and the rest of the parameters are kept equal in both sections. 
To approach the data qualitatively, we add two three side wide barriers at the left and right edge of the Josephson junction to emulate the effect of a reduction in the maxima of the ABS due to confinement~\cite{beenakker_resonant_1992-1}. In the device, the reduction of screening in the uncovered Josephson junction and possible Al-InAs interface barriers can cause these potential barriers. The resulting tight-binding grid is shown in~\cref{fig:kwant-tight-binding} together with a chemical potential sweep illustrating the first and second subband entry. Fabry-Perot oscillations are visible in the transparency due to the strong confinement. 
Settings used here are, $\alpha=\SI{40}{\milli\electronvolt\nano\meter}$, $g=15$, $\Delta=\SI{0.2}{\milli\electronvolt}$, $m=0.0023m_e$ with $m_e$ is the electron mass and the potential barrier height is set to \SI{13}{\milli\electronvolt}.

We first attempt to gain qualitative insight into the ABS spectrum by searching for a chemical potential that approximately matches the two lowest manifolds in the experimental data. We thus fine-tune to a "gate" point at $\mu=\SI{9.20}{\milli\electronvolt}$ right after the second subband-entry where we have two ABS that disperse similarly in phase, see~\cref{fig:kwant-spectrum}(a).
Here we perform a parallel-field dependence and observe the two ABS manifolds spin-split in field~\cref{fig:kwant-spectrum}(b), with different effective g-factors similar to the data. Next we will stay at this point and investigate the microwave absorption.

\subsubsection{Matrix elements}
We follow Ref.~\cite{van_heck_zeeman_2017} to calculate the matrix elements of the current operator that give the transition spectrum.
From diagonalizing $H_\mathrm{BdG}$ we get a set of eigenvalues $+E_m, -E_m$ and with corresponding eigenstates $\Psi_m, \mathcal{P}\Psi_m$, where $\mathcal{P}$ denotes the particle-hole operator. 
We can thus evaluate the current matrix elements $j_{n,m} = \langle \Psi_n|I_A|\Psi_m\rangle$. The diagonal elements give the contribution of the supercurrent of a single ABS level. The off-diagonal elements are useful when considering possible microwave transitions. 
We can use linear response theory to calculate the finite-temperature microwave susceptibility of the junction and thus get the absorption spectrum. 
This results in~\cite{van_heck_zeeman_2017}
\begin{equation}
\chi(\omega) = \sum_{n \geq m} |j_{n,\mathcal{P}m}|^2 \delta(\omega-(E_m+E_n))\left[1-f(E_m)-f(E_n)\right]+ \sum_{n \geq m} |j_{n,m}|^2 \delta(\omega-(E_n-E_m))\left[f(E_m)-f(E_n)\right]
\end{equation}
where $j_{n,\mathcal{P}m}$ is defined as $\langle \Psi_n|I_A|\mathcal{P}\Psi_m\rangle$ and the Fermi-Dirac function at finite temperature T is defined as
\begin{equation}
f(E) = \frac{1}{e^{(E-\mu )/k_B T}-1}
\end{equation}
where we take $\mu=0$
\begin{figure}[t!]
     \centering
     \includegraphics{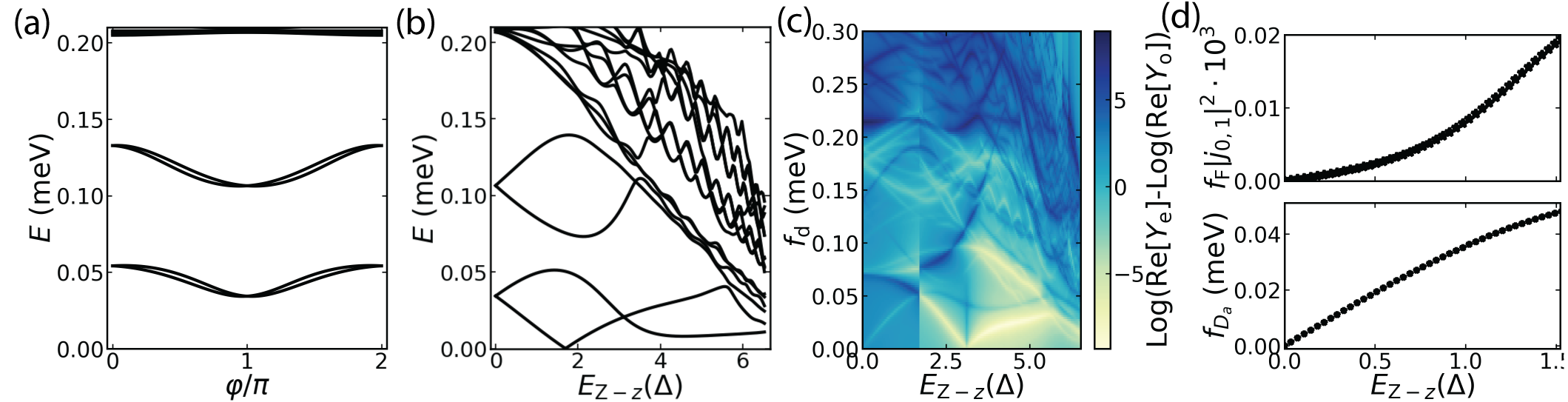}
    \caption{Tight binding simulations at $\mu=\SI{9.20}{\milli\electronvolt}$ with the same settings as used in Fig.\ref{fig:kwant-tight-binding}.  
    \textbf{(a)} Phase dependence of ABS energies at \Bz=0. Two manifolds are visible inside the gap, spin-split due to spin-orbit interaction.
    \textbf{(b)} Field evolution at $\varphi=\pi$ of the ABS energies.
    \textbf{(c)} Linear response illustrating the resulting spectrum versus \Bz\ at an effective temperature of \SI{20}{\micro\electronvolt}. The vertical line is due to a ground state fermion parity switch which swaps the assignment between even and odd for the lowest ABS.  
    \textbf{(d)} Matrix element $j_{0,1}$ multiplied by the odd parity fermi-factor $f_\mathrm{F}= [f(E_0)-f(E_1)]$ and transition frequency of \Da\ at the same effective temperature. 
    \label{fig:kwant-spectrum}}
\end{figure}

\begin{figure}[t!]
     \centering
     \includegraphics{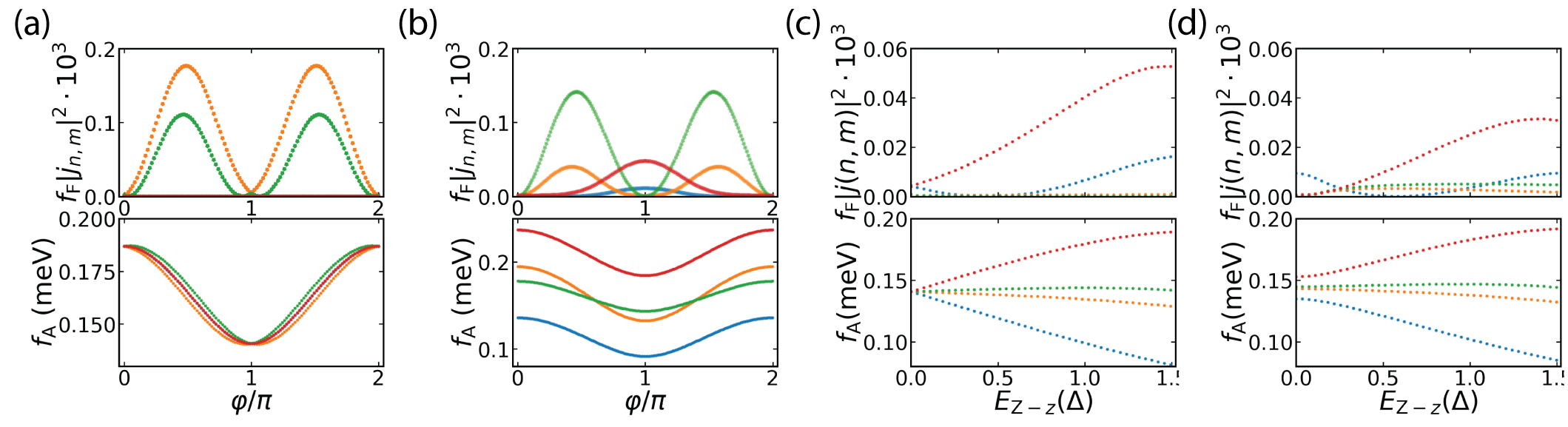}
    \caption{Matrix elements of even $\{M_\mathrm{ab}\}$ transitions at $\mu=\SI{9.20}{\milli\electronvolt}$ multiplied by the even-parity Fermi-factor $f_\mathrm{F}=1-f(E_m)-f(E_n)$. The transitions (bottom panels) are colored with the same color as their corresponding matrix elements (top panels). 
    \textbf{(a)} Phase dependence at \Bz~=~0. Here, only \Tzero and \Sab\ have a finite matrix element.
    \textbf{(b)} Phase dependence at $E_{\mathrm{Z}-z}=1.2\Delta$. This illustrates that at $\varphi=\pi$, \Tmin\ and \Tplus\ have a finite matrix element while \Tzero\ and \Sab\ have zero matrix element there. 
    \textbf{(c)} Field dependence at $\varphi=\pi$ without perpendicular field. At $\varphi=\pi$, \Sab, \Tzero\ transitions have a zero matrix element.
    \textbf{(d)} Field dependence at $\varphi=\pi$ at finite \Bx ($E_{\mathrm{Z}-x} = 0.1\Delta$). Here the \Sab, \Tzero\ get a finite matrix element as well. 
    \label{fig:kwant-matrix-even}}
\end{figure}

In~\cref{fig:kwant-spectrum}(c) we show the microwave absorption versus field, similar to Fig. 3 of the main text. Note that we see qualitatively similar features: the odd (yellow) \Da\ transition coming up, the even \Sa\ and the mixed transitions \Tmin, \Tplus, dispersing in field. Additionally, we do not observe the $\ket{0}\rightarrow\ket{\uparrow_\mathrm{a}\downarrow_\mathrm{b}},\ket{0}\rightarrow\ket{\downarrow_\mathrm{a}\uparrow_\mathrm{b}}$ transitions.

To explore this further, we investigate the current operator matrix elements. The odd transition matrix element is shown in~\cref{fig:kwant-spectrum}(d) and increases with field as expected~\cite{park_andreev_2017, van_heck_zeeman_2017}.
The even-parity matrix elements are shown in~\cref{fig:kwant-matrix-even}. Here we see that only at $\varphi=\pi$ the elements for the $\ket{0}\rightarrow\ket{\uparrow_\mathrm{a}\downarrow_\mathrm{b}},\ket{0}\rightarrow\ket{\downarrow_\mathrm{a}\uparrow_\mathrm{b}}$ are zero and thus the transitions should appear when measuring the phase-dependence as we did in ~\cref{fig:flux_dependence_even_transitions_field}. Furthermore, we show the matrix elements versus field, and see that indeed only the \Tmin, \Tplus, matrix elements become finite at $\varphi=\pi$, while the others stay zero. This is lifted by applying a small \Bx, where now also $\ket{0}\rightarrow\ket{\uparrow_\mathrm{a}\downarrow_\mathrm{b}},\ket{0}\rightarrow\ket{\downarrow_\mathrm{a}\uparrow_\mathrm{b}}$ obtain a finite matrix element at $\varphi=\pi$. Thus we would have expected to see $\ket{0}\rightarrow\ket{\uparrow_\mathrm{a}\downarrow_\mathrm{b}},\ket{0}\rightarrow\ket{\downarrow_\mathrm{a}\uparrow_\mathrm{b}}$ transitions in the data.
This is by no means a complete data set as there are many parameters to vary, but we have investigated similar spectra at a few other chemical potential points -- including for  junctions without the double barrier close to the first sub-band entry, where the manifolds disperse oppositely and behave like the analytical expressions shown in~\cite{park_andreev_2017} -- and found qualitatively similar conclusions.
The visibility of the even-mixed transitions heavily depended on the exact chemical potential setting, and often they were not visible at all.  

\subsubsection{Field directions for a single ABS manifold}
\begin{figure}[t!]
     \centering
     \includegraphics{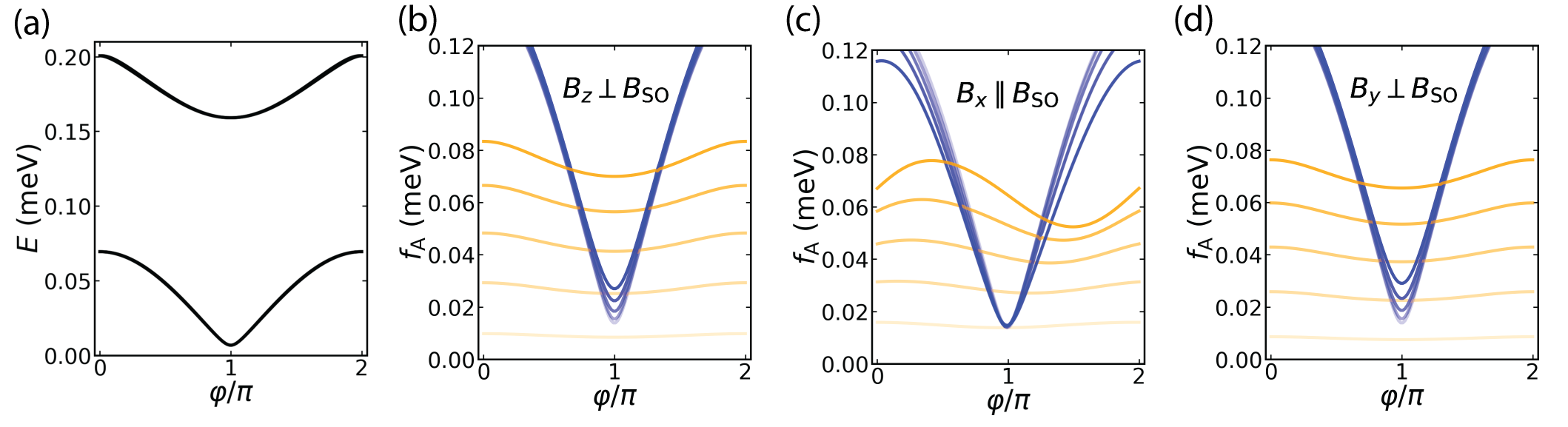}
    \caption{ 
    Relative shift in minimum of \Da\ transition compared to the \Sa\ transition at high transparency ($\mu=\SI{13.12}{\milli\electronvolt}$) versus different field directions.  
    \textbf{(a)} BdG energies at $\mu=\SI{13.12}{\milli\electronvolt}$. We observe two nearly degenerate manifolds. 
    \textbf{(b) - (d)} Transition spectra of the lowest ABS manifold illustrating both the \Sa\ (blue) and \Da\ (yellow) transition for increasing field strength and 3 different field directions. Darker colors indicate higher fields. The Zeeman energy used in (b), (d) ranges from $E_\mathrm{Z}=0.2\rightarrow2\Delta$ in equal steps. For (c), the parallel field direction, the Zeeman energy ranges from $E_\mathrm{Z}=0.1\rightarrow0.5\Delta$
    \label{fig:kwant-high-tau-field-direction-dependence}}
\end{figure}

We now shift towards a chemical potential where the lowest ABS manifold is more transparant, to qualitatively match the situation in Fig. 4 of the main text. In~\cref{fig:kwant-high-tau-field-direction-dependence}, we investigate the dispersion of the odd spin-flip transition \Da\ for three magnetic field directions relative to the effective spin-orbit field. We observe that the case displayed in~\cref{fig:kwant-high-tau-field-direction-dependence}(c) is most similar to the data where $B\parallel B_\mathrm{SO}$ as for the other directions there is no relative shift of the minimum of \Da\ with respect to the minimum of \Sa. Thus, the data observed in Fig.4 of the main text and~\cref{fig:sup:z_field_high_tau} is consistent with a component of the spin-orbit along the wire axis, i.e. the $z$-axis. Note that we have not investigated whether a random disorder potential can change this picture.  
\subsection{Minimal model including exchange interaction}
Here, we supply additional information about the minimal model including interaction presented in Eq.(1) in the main text to provide insight on the effect of each ingredient.
The aim in this minimal model is to keep tunable the parameters that are relevant to reproduce the qualitative hierarchy of the lines (spin-orbit, magnetic field and Coulomb interaction) but to avoid an explicit description of other features such as the size of the junction, its transparency, the chemical potential or the superconducting pairing. To this end, we consider the basis of two lowest spinless bare ABSs manifolds \(i=a,b\), with energies \(E_i\) implicitly taking into account the latter features of the junction, and then introduce the spin-orbit, the magnetic field and the Coulomb interaction within this basis. 

To describe the spin-orbit coupling, we project it on the bare ABSs, writing the spin-orbit Hamiltonian as:
\begin{equation}
    \tilde{H}_{SO} = \sum_{i\sigma i'\sigma'} \bra{i\sigma}H_{SO}\ket{i'\sigma'} \gamma^\dagger_{i\sigma}\gamma_{i'\sigma'}\,,
\end{equation}
where $\gamma^\dagger_{i\sigma}$ creates a quasiparticle  with spin \(\sigma\) in manifold \(i\), and \(H_{SO} = \alpha /\hbar \left( p_z \sigma_x - p_x \sigma_z  \right) \) is described with a 2D Rashba model, \(z\) being the direction parallel to the nanowire.
Imposing time reversal symmetry on \(\tilde{H}_{SO}\), we get the relations \(\bra{i\sigma}H_{SO}\ket{i'\sigma} = \bra{i\bar{\sigma}}H_{SO}\ket{i'\bar{\sigma}}^*\) and \(\bra{i\sigma}H_{SO}\ket{i'\bar{\sigma}} = -\bra{i\bar{\sigma}}H_{SO}\ket{i'\sigma}^*\), where it has been used \(\mathcal{T}c\gamma_{i,\sigma}\mathcal{T}^{-1}=c^*\sigma\gamma_{i,\bar{\sigma}}\). Thus, we can write:
\begin{equation}	
	\bar{H}_{SO} =   i\alpha_\parallel \left(  \gamma^\dagger_{a\uparrow}\gamma_{b\downarrow} +  \gamma^\dagger_{a\downarrow}\gamma_{b\uparrow} \right) 
	-
	i\alpha_\perp \left(  \gamma^\dagger_{a\uparrow}\gamma_{b\uparrow} -  \gamma^\dagger_{a\downarrow}\gamma_{b\downarrow} \right) + \textrm{H.c.}\,, 
\end{equation}
where \(\alpha_{\perp (\parallel)} \in \mathbb{R}\) are treated as fitting parameters but can be related with the spatial profile of the basis wavefunctions \(i\alpha_{\perp (\parallel)} = \mp(\alpha/\hbar) \bra{a} p_{x (z)} \ket{b}\). A more general spin-orbit Hamiltonian can be devised just imposing time reversal symmetry, resulting in the non-diagonal terms \( \alpha_s \gamma^\dagger_{a\uparrow}\gamma_{b\uparrow} + \alpha_s^* \gamma^\dagger_{a\downarrow}\gamma_{b\downarrow} + \alpha_d \gamma^\dagger_{a\uparrow}\gamma_{b\downarrow} - \alpha_d^*\gamma^\dagger_{a\downarrow}\gamma_{b\uparrow} + H.c.\), with \(\alpha_{s,d}\in\mathbb{C}\). We checked that this coupling provides qualitatively similar results.

Assuming that these basis wavefunctions are separable in \(\ket{i} = \ket{\phi_{iz}}\ket{\phi_{ix}}\), we expect the transverse part \(i\alpha_{\perp} =  - (\alpha/\hbar) \braket{\phi_{az}|\phi_{bz}} \bra{\phi_{ax}} p_{x}\ket{\phi_{bx}}\) to be negligible in situations with almost spatially symmetric transverse confining potentials (\(V(x) \approx V(-x)\)) where both manifolds share the same channel. This would happen because both \(\ket{\phi_{ax}}\) and \(\ket{\phi_{bx}}\) would have approximately the same spatial parity around \(x=0\), and thus \(\bra{\phi_{ax}} p_{x}\ket{\phi_{bx}} \approx 0\).
However, our case likely corresponds to a multichannel situation, given the moderate length of the junction and the existence of multiple intermanifold single QP transitions visible in the spectroscopy over \(\varphi\).
Thus, each ABSs manifold would arise from a different transverse channel, enabling a finite \(\alpha_\perp\). Still, this separation is not complete because of the anticrossings, which are a consequence of a non-zero \(\alpha_\parallel \propto \braket{\phi_{ax}|\phi_{bx}}\).
Nevertheless, for simplicity, in order to estimate a lower bound for the spin-orbit strength we use the case where the manifolds \(1,2\) strictly correspond to different channels.
This yields \( |\bra{\phi_{ax}} p_{x} \ket{\phi_{bx}}| \le \hbar C/W\), with \(W\) the width of the nanowire and \(C\) a factor of order \({\sim}1\) that depends on the confinement potential.
Then, using the fitted parameter \(\alpha_\perp{\approx}5\)GHz and \(|\braket{\phi_{az}|\phi_{bz}}|\le1 \), we get \(\alpha = \alpha_\perp \hbar /( \braket{\phi_{az}|\phi_{bz}} \bra{\phi_{ax}} p_{x} \ket{\phi_{bx}}) \ge 0.02 W \) meVnm (\(W\) in nm).
This provides a lower bound of \({\sim} 2\) meVnm for \(W{\sim}100\)nm.

\begin{figure}[h]
   \includegraphics{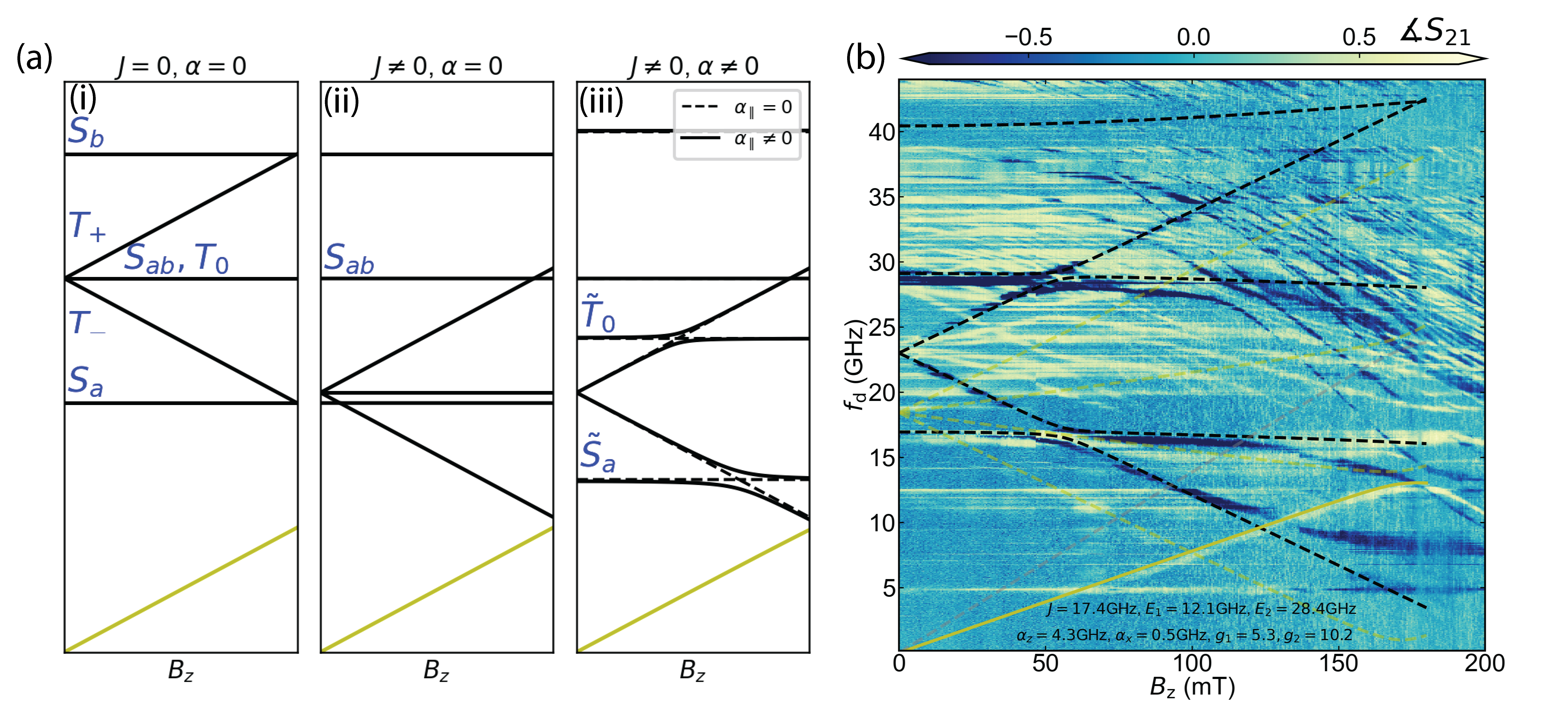}
   \caption{\textbf{(a)} Minimal model with example parameters to illustrate the effect of each ingredient separately. (i) the spectrum without exchange interaction and equal g-factors, (ii) the spectrum with exchange interaction, but zero spin-orbit interaction, (iii) the spectrum both with spin-orbit and exchange interaction. This motivates the sketch made in Fig 2.d in the main text. Unitless parameter chosen are:  $E_a=1.2$, $E_b=2.4$, $\alpha_{\perp}=0.5, \alpha_{||}=0.1, g^*_a=g^*_b=0.8,J=0.85$.
   \textbf{(b)} Fit of the data using real units as shown in Fig.3(c) of the main text overlayed on the data. \label{fig-sup-interacting-fits}}
 \end{figure}

It must be highlighted that even if the actual specific microscopic origin of the spin orbit effective parameters is arguable, their presence is necessary in order to reproduce the structure of the even transitions and their anticrossings.
Note that the spin-orbit interaction must be taken into account even though we are considering the case $\varphi=\pi$, where it does not affect the single-particle Andreev levels degeneracy due to Kramers' theorem~\cite{beri2008}, since we are interested in the two-particle states as well.
Nevertheless, the predicted unbroken degeneracy of \(T_{\pm}\) at zero field (Fig. \ref{fig-sup-interacting-fits}(a)iii) suggests that it should be improved to account for the full splitting in \(\varphi=\pi\) that is sometimes observed in other spectra, for example introducing a third manifold, which allows to avoid that degeneracy.

The behaviour of ABSs energies with small Zeeman fields depends on the interplay between the spin orbit, the transmission of the junction and the chemical potential~\cite{van_heck_zeeman_2017}.
In our simple model, we introduce effective g-factors for the bare ABSs as fitting parameters.

Finally, to describe the Coulomb interaction we use a ferromagnetic effective exchange. The Coulomb interaction in the nanowire is strongly screened by the environment and its free charges, rendering the interaction approximately local. However, the form of the interaction in the subspace of the ABSs is more convoluted because of their spatial overlap. In the states with 2 QPs, due to this overlap, the interaction leads to an effective ferromagnetic coupling between the quasiparticle spins, as in the case of Hund's rule in atomic physics. 

The full Hamiltonian is projected over the states with 1 and 2 quasiparticles, resulting in the matrices 
\begin{equation}
	\left( H \right)_{1QP} = \begin{pmatrix}
		E_a+g^*_a B_z &0 &-i\alpha_\perp &i\alpha_\parallel \\
		0 &E_a-g^*_a B_z &i\alpha_\parallel &i\alpha_\perp \\
		i\alpha_\perp &-i\alpha_\parallel &E_b+g^*_b B_z &0 \\
		-i\alpha_\parallel &-i\alpha_\perp &0 &E_b-g^*_b B_z
	\end{pmatrix},
\end{equation}
in the basis \(\mathcal{B}_{1QP}=\left\lbrace 
\ket{1\uparrow},\ket{1\downarrow},
\ket{2\uparrow},\ket{2\downarrow} \right\rbrace \), where the constant term \(-\frac{3J}{8}\) has been neglected, and 
\begin{equation}
	\left( H \right)_{2QP} {=} \begin{pmatrix}
		2E_a &0 &i\alpha_\parallel &i\sqrt{2}\alpha_\perp &-i\alpha_\parallel &0 \\
		0 &E_a{+}E_b &0 &(g^*_a{-}g^*_b)B_z &0 &0 \\
		-i\alpha_\parallel &0 &E_a{+}E_b{-}J{+}(g^*_a{+}g^*_b)B_z &0 &0 &-i\alpha_\parallel\\
		-i\sqrt{2}\alpha_\perp &(g^*_a{-}g^*_b)B_z &0 &E_a{+}E_b{-}J &0 &-i\sqrt{2}\alpha_\perp\\
		i\alpha_\parallel &0 &0 &0 &E_a{+}E_b{-}J{-}(g^*_a{+}g^*_b)B_x &i\alpha_\parallel\\
		0 &0 &i\alpha_\parallel &i\sqrt{2}\alpha_\perp &-i\alpha_\parallel &2E_b
	\end{pmatrix},
\end{equation}
in the basis \(\mathcal{B}_{2QP}=\left\lbrace 
\ket{S_a},\ket{S_{ab}},
\ket{T_+},\ket{T_0},\ket{T_-},\ket{S_b} \right\rbrace \).

In order to visualize the effect of each ingredient, we plot in Fig. \ref{fig-sup-interacting-fits} the evolution with magnetic field in 3 situations. In Fig. \ref{fig-sup-interacting-fits}(a.i), without $J$ and $\alpha$, in Fig. \ref{fig-sup-interacting-fits}(a.ii) with $J$ without $\alpha$ , and in Fig. \ref{fig-sup-interacting-fits}(a.iii) with $J$ and $\alpha$. The combination of SO and interaction acts as an effective anisotropic exchange~\cite{katsaros_zero_2020}.
Note that the ordering of the lines depends on the relative strenghts of $J$ and the energies $E_a$ and $E_b$, therefore with stronger $J$ it is possible to have \Sa above \Tzero and thus flip \Satilde\ and \Tzerotilde. This is actually the case in the fit of the data in the main text in Fig.3, but since they are strongly hybridized, we kept the notation where \Satilde is the lowest state for clarity. 
An overlay of the simple fit using the LMFIT package~\cite{newville_lmfitlmfit-py_2021} shown in Fig.3(c) of the main text is shown in Fig. \ref{fig-sup-interacting-fits}(b). For the fit, we extracted the positions of the four lowest even transitions and the lowest odd interband transitions at zero field. We then let LMFIT determine the energies $E_a$, $E_b$, $J$ and $\alpha_\perp$ by fitting both the odd and even parity diagonalization simultaneously. The g-factors are extracted directly based on the discussion in the main text and $\alpha_{||}$ is set to the size of the avoided crossing in the data. 

\bibliography{bib/main_mac.bib}